\documentclass[preprint,notoc]{JHEP3}
\usepackage{epsfig}

\def\Vind{V^{\rm induced}}
\def\eslt{\not\!\!{E_T}}
\def\eslt{E_T^{\rm miss}}
\def\emiss{\not\!\!{E}}
\def\to{\rightarrow}

\def\bi{\begin{itemize}}
\def\ei{\end{itemize}}
\def\te{\tilde e}

\def\tu{\tilde u}

\def\tb{\tilde b}

\def\tst{\tilde t}
\def\ttau{\tilde \tau}
\def\tmu{\tilde \mu}
\def\tg{\tilde g}

\def\tell{\tilde\ell}
\def\tq{\tilde q}
\def\tB{\widetilde B}
\def\tw{\widetilde W}
\def\tz{\widetilde Z}
\def\alt{\stackrel{<}{\sim}}
\def\agt{\stackrel{>}{\sim}}
\def\be{\begin{equation}}  
\def\ee{\end{equation}}  
\def\CM{\cal M}
\title{Exploring the BWCA (Bino-Wino Co-Annihilation) \\
Scenario for Neutralino Dark Matter}
\author{Howard Baer$^a$, Tadas Krupovnickas$^b$, 
Azar Mustafayev$^a$, Eun-Kyung Park$^a$, Stefano Profumo$^{a,c}$ and
Xerxes Tata$^d$\\
$^a$Department of Physics, Florida State University Tallahassee, 
FL 32306, USA\\
$^b$High Energy Theory Group, Brookhaven National Laboratory, 
Upton, NY 11973, USA\\
$^c$Division of Physics, Mathemathics and Astronomy, California Institute of Technology, Mail Code 106-38, Pasadena, CA 91125, USA\\
$^d$Department of Physics and Astronomy, University of Hawaii,
Honolulu, HI 96822, USA\\
E-mail: \email{baer@hep.fsu.edu},\email{tadas@quark.phy.bnl.gov},
\email{mazar@hep.fsu.edu},\email{epark@hep.fsu.edu},
\email{profumo@caltech.edu},\email{tata@phys.hawaii.edu}}

\preprint{\vbox{FSU-HEP-050817, BNL-HET-05/24, UH-511-1079-05}} 
\abstract{
In supersymmetric models with non-universal gaugino masses, it is 
possible to have opposite-sign $SU(2)$ and $U(1)$ gaugino mass terms. 
In these models,
the gaugino eigenstates experience little mixing so that the lightest
SUSY particle remains either pure bino or pure wino. The neutralino
relic density can only be brought
into accord with the WMAP measured value when bino-wino
co-annihilation (BWCA) acts to enhance the dark matter annihilation rate.
We map out parameter space regions and mass spectra which are characteristic
of the BWCA scenario. Direct and indirect 
dark matter detection rates are shown to be typically very low. 
At collider experiments, the BWCA scenario is typified by a small mass gap
$m_{\tz_2}-m_{\tz_1}\sim 20-80$ GeV, so that tree level two body decays
of $\tz_2$ are not allowed. However,
in this case the second lightest neutralino has an 
enhanced loop decay branching fraction to photons. While the photonic
neutralino decay signature looks difficult to extract at the 
Fermilab Tevatron, it should  lead to distinctive events 
at the CERN LHC and at a linear $e^+e^-$ collider.
}


\begin{document}

\section{Introduction}
\label{sec:intro}
In 
$R$-parity conserving supergravity models 
a stable neutralino ($\tz_1$) is the 
lightest supersymmetric particle (LSP) over a large part of the
parameter space of the model. A neutralino LSP
is generally considered an excellent candidate to comprise the bulk
of the cold dark matter (CDM) in the universe.
The relic density of neutralinos in supersymmetric models can be 
calculated by solving the Boltzmann equation for the
neutralino number density\cite{griest}. 
The central part of the calculation is to
evaluate the thermally averaged neutralino annihilation and 
co-annihilation cross section times velocity. 
The computation requires evaluating many thousands of Feynman diagrams.
Several computer codes are now publicly\cite{rdcodes,isared} 
available to evaluate the
neutralino relic density $\Omega_{\tz_1}h^2$. 

From its analysis of the anisotropies in the cosmic microwave background
radiation, the WMAP collaboration has inferred that
the CDM density of the universe 
is given by\cite{wmap}, 
\begin{equation}
\Omega_{CDM}h^2=0.113\pm 0.009 .
\end{equation}
Since the dark matter could well be composed of several components,
strictly speaking, the WMAP measurement only implies an upper limit on
the density of any single dark matter candidate. Nevertheless, even this
upper bound 
imposes a tight constraint on all models that contain such candidate
particles, and in particular,
on supersymmetric models with a conserved $R$-parity \cite{wmapcon}.

Many analyses have been recently performed in the context of the
paradigm minimal supergravity model\cite{msugra} (mSUGRA), which is
completely specified by the parameter set, 
$$m_0,\ m_{1/2},\ A_0,\ \tan\beta \ {\rm and} \ sign(\mu ). $$ The
mSUGRA model assumes that the minimal supersymmetric model (MSSM) is
valid between the mass scales $Q=M_{GUT}$ and $Q=M_{weak}$. A common
value $m_0$ ($m_{1/2}$) (($A_0$)) is assumed for all scalar mass
(gaugino mass) ((trilinear soft SUSY breaking)) parameters at
$Q=M_{GUT}$, and 
$\tan\beta$ is
the ratio of vacuum expectation values of the two Higgs fields that
give masses to the up and down type fermions.
The magnitude of the superpotential Higgs mass term $\mu$, but not its
sign, is fixed so as to reproduce the observed $Z$ boson mass.  The
values of couplings and other model parameters renormalized at the weak
scale can be computed via renormalization group (RG) evolution from
$Q=M_{GUT}$ to $Q=M_{weak}$. Once these weak scale parameters that are
relevant to phenomenology are obtained, sparticle masses and mixings may
be computed, and the associated relic density of neutralinos can be
determined.

In most of the allowed mSUGRA parameter space, 
the relic density $\Omega_{\tz_1}h^2$
turns out to be considerably larger than the WMAP value. Consistency
with WMAP thus implies that neutralinos should be able to annihilate
very efficiently. In the mSUGRA model,
the annihilation rate is enhanced in 
just the following regions of parameter space, where the sparticle masses
and/or the neutralino composition assume special forms.
\begin{itemize}
\item The bulk region occurs at low values of $m_0$ and
$m_{1/2}$\cite{haim,bulk}.  In this region, neutralino annihilation is
enhanced by $t$-channel exchange of relatively light sleptons. The bulk
region, featured prominently in many early analyses of the relic
density, has been squeezed from below by the LEP2 bound on the chargino
mass $m_{\tw_1}>103.5$ GeV and the measured value of the branching
fraction $B(b\to s\gamma)$, and from above by the tight bound from WMAP.
\item The stau co-annihilation region  occurs at low $m_0$ for
almost any $m_{1/2}$ value where $m_{\ttau_1}\simeq m_{\tz_1}$. The
staus, being charged, can annihilate rapidly so that
$\ttau_1\tz_1$  co-annihilation processes
that maintain $\tz_1$ in thermal equilibrium with $\ttau_1$, serve to reduce
the relic density of neutralinos \cite{stau}.
\item The hyperbolic branch/focus point (HB/FP) region at large $m_0\sim$ 
several TeV, where $|\mu|$ becomes small, and neutralinos efficiently 
annihilate via their higgsino components\cite{hb_fp}.
This is the case of mixed higgsino dark matter (MHDM).
\item The $A$-annihilation funnel occurs at large $\tan\beta$ values
when $2m_{\tz_1}\sim m_A$ and neutralinos can efficiently annihilate
through the relatively broad $A$ and $H$ Higgs resonances\cite{Afunnel}.
\end{itemize}
In addition, a less prominent light Higgs $h$ annihilation corridor occurs at
low $m_{1/2}$\cite{drees_h} 
and a top squark co-annihilation region occurs at 
particular $A_0$ values when $m_{\tst_1}\simeq m_{\tz_1}$\cite{stop}.

Many analyses have also been performed for gravity-mediated SUSY
breaking models with non-universal soft terms. Non-universality of soft
SUSY breaking (SSB) scalar masses can,  1. pull one or more scalar masses to
low values so that ``bulk'' annihilation via $t$-channel exchange of
light scalars can occur\cite{nmh,nuhm},  2. they can bring in new near
degeneracies of various sparticles with the $\tz_1$ so that new
co-annihilation regions open up\cite{auto,nuhm,sp}, 3.
bring the value of $m_A$ into accord with $2m_{\tz_1}$ so that funnel
annihilation can occur\cite{ellis,nuhm}, or 4. they can pull the value
of $\mu$ down so that higgsino annihilation can
occur\cite{ellis,drees2,nuhm}.  It is worth noting that these
general mechanisms for increasing the neutralino annihilation rate can 
all occur in the mSUGRA model.  Moreover, in all these cases the
lightest neutralino is either bino-like, or a bino-higgsino mixture.

If non-universal gaugino masses are allowed, then qualitatively new
possibilities arise that are not realized 
in the mSUGRA model\cite{ibanez,gunion,dermisek,models}. 
One case, that of mixed wino dark matter (MWDM), has been addressed in a
previous paper\cite{mwdm}.  In this case, as the weak scale value of
$SU(2)$ gaugino mass $M_2({\rm weak})$ is lowered from its mSUGRA value,
keeping the hypercharge gaugino mass $M_1({\rm weak})$ fixed, the wino
component of $\tz_1$ continuously increases until it becomes dominant
when $M_2({\rm weak}) < M_1({\rm weak})$ (assuming $|\mu|$ is
large). The $\tz_1\tw_{1,2}W$ coupling becomes large when $\tz_1$
becomes wino-like, resulting in enhanced $\tz_1\tz_1\to W^+W^-$
annihilations.  Moreover, co-annihilations with the lightest chargino
and with the next-to-lightest neutralino help to further suppress the
LSP thermal relic abundance. Indeed, if the wino component of the
neutralino is too large, this annihilation rate is very big and the
neutralino relic density falls well below the WMAP value.

A qualitatively different case arises in supersymmetric models if the
SSB gaugino masses $M_1$ and $M_2$ are of {\it opposite
  sign}.\footnote{The sign of the gaugino mass under RG evolution is
  preserved at the one loop level.}  As we
will see below, the transition from a bino-like $\tz_1$ to a wino-like
$\tz_1$ is much more abrupt as $-M_2({\rm weak})$ passes through
$M_1({\rm weak})$.  Opposite sign masses for $SU(3)$ relative to $SU(2)$
and $U(1)$ gaugino mass parameters are well known to arise in the
anomaly-mediated SUSY breaking (AMSB) model \cite{amsb}. An opposite
sign between bino and wino masses, which is of interest to us here, can
arise in supersymmetric models with a non-minimal gauge kinetic function
(GKF).  In supergravity Grand Unified Theories (GUT), the GKF $f_{AB}$
must transform as the symmetric product of two adjoints of the GUT
group.  In minimal supergravity, the GKF transforms as a singlet. In
$SU(5)$ SUGRA-GUT models, it can also transform as a 24, 75 or 200
dimensional representation\cite{anderson}, while in $SO(10)$ models it
can transform as 1, 54, 210 and 770 dimensional
representations\cite{chamoun,nath}.  Each of these non-singlet cases
leads to unique predictions for the ratios of GUT scale gaugino masses,
though of course (less predictive) combinations are also possible.  The
GUT scale and weak scale ratios of gaugino masses are listed in Table
\ref{tab:one} for these non-singlet representations of the GKF.  If the
GKF transforms as a linear combination of these higher dimensional
representations, then essentially arbitrary gaugino masses are allowed.
In this report, we will adopt a phenomenological approach as in
Ref. \cite{mwdm}, and regard the three MSSM gaugino masses as
independent parameters, with the constraint that the neutralino relic
density should match the WMAP measured value.  However, in this paper,
we will mainly address the special features 
that arise when the $SU(2)$ and $U(1)$ gaugino masses have opposite sign.
\begin{table}
\begin{center}
\begin{small}
\begin{tabular}{|c|c|ccc|ccc|}
\hline \multicolumn{2}{|c|} {} & \multicolumn{3}{c|} {$M_{\rm GUT}$} &
\multicolumn{3}{c|}{$M_Z$} \cr \hline group & $F_{h}$ & $M_3$ & $M_2$ &
$M_1$ & $M_3$ & $M_2$ & $M_1$ \cr \hline $SU(5)$ & ${\bf 1}$ & $1$
&$\;\; 1$ &$\;\;1$ & $\sim \;6$ & $\sim \;\;2$ & $\sim \;\;1$ \cr
$SU(5)$ & ${\bf 24}$ & $2$ &$-3$ & $-1$ & $\sim 12$ & $\sim -6$ & $\sim
-1$ \cr $SU(5)$ & ${\bf 75}$ & $1$ & $\;\;3$ &$-5$ & $\sim \;6$ & $\sim
\;\;6$ & $\sim -5$ \cr $SU(5)$ & ${\bf 200}$ & $1$ & $\;\; 2$ & $\;10$ &
$\sim \;6$ & $\sim \;\;4$ & $\sim \;10$ \cr $SO(10)\to G_{442}$ & ${\bf
54}$ & $1$ & $\;\; -1.5$ & $\;-1$ & $\sim \;3$ & $\sim \;\;-1.3$ & $\sim
\;-1$ \cr $SO(10)\to SU(2)\times SO(7)$ & ${\bf 54}$ & $1$ & $\;\; -7/3$
& $\;1$ & $\sim \;3$ & $\sim \;\;-2.1$ & $\sim \;0.42$ \cr $SO(10)\to
H_{51}$ & ${\bf 210}$ & $1$ & $\;\; 1$ & $\;-96/25$ & $\sim \;3$ & $\sim
\;\;0.88$ & $\sim \;-1.6$ \cr \hline
\end{tabular}
\end{small}
\smallskip
\caption{Relative gaugino mass parameters at $Q=M_{\rm GUT}$ and 
their relative values evolved to $Q=M_Z$
in various possible $F_{h}$ irreducible representations
in $SU(5)$ and $SO(10)$ SUSY GUTs with a non-minimal GKF. Here, $G_{442}
= SU(4)\times SU(2)\times SU(2)$, and $H_{51}$ denotes the flipped
$SU(5)\times U(1)$ symmetry group. }
\label{tab:one}
\end{center}
\end{table}

Much work has already been done on evaluating the relic density in
models with gaugino mass non-universality.  Prospects for direct and
indirect detection of DM have also been studied. 
Griest and Roszkowski first pointed out that a wide
range of relic density values could be obtained by abandoning gaugino
mass universality\cite{gr}.  A specific form of gaugino mass
non-universality occurs in AMSB models mentioned above, where the
gaugino masses are proportional to the $\beta$-functions of the
corresponding low energy gauge groups: $M_1:M_2:M_3\sim 3:1:-10$. In
this case the $\tz_1$ is almost a pure wino and so can annihilate very
efficiently, resulting in a very low thermal relic density of
neutralinos.  This led Moroi and Randall\cite{moroi} to suggest that the
decay of heavy moduli to wino-like neutralinos in the early universe
could account for the observed dark matter density.  Corsetti and Nath
investigated dark matter relic density and detection rates in models
with non-minimal $SU(5)$ GKF and also in O-II string
models\cite{cor_nath}. Birkedal-Hanson and Nelson showed that a GUT
scale ratio $M_1/M_2\sim 1.5$ would bring the relic density into accord
with the measured CDM density via MWDM, and also presented direct
detection rates\cite{birkedal}.  Bertin, Nezri and Orloff studied the
variation of relic density and the enhancements in direct and indirect
DM detection rates as non-universal gaugino masses are
varied\cite{nezri}. Bottino {\it et al.} performed scans over
independent weak scale parameters to show variation in indirect DM
detection rates, and noted that neutralinos as low as 6 GeV are
allowed\cite{bottino}. Belanger {\it et al.} have recently presented
relic density plots in the $m_0\ vs.\ m_{1/2}$ plane for a variety of
universal and non-universal gaugino mass scenarios, and showed that
large swaths of parameter space open up when the $SU(3)$ gaugino mass
$M_3$ becomes small \cite{belanger}: this is primarily because the value
of $|\mu|$ reduces with the corresponding value of $M_3$, resulting in
an increased higgsino content of the neutralino. Mambrini and Mu\~noz, and
also Cerdeno and Mu\~noz, examined direct and indirect detection rates for
models with scalar and gaugino mass non-universality\cite{munoz}.  Auto
{\it et al.}\cite{auto} proposed non-universal gaugino masses to reconcile
the predicted relic density in models with Yukawa coupling unification
with the WMAP result.  Masiero, Profumo and Ullio exhibit the relic
density and direct and indirect detection rates in split supersymmetry
where $M_1$, $M_2$ and $\mu$ are taken as independent weak scale
parameters with ultra-heavy squarks and sleptons\cite{mpu}.
 
The main purpose of this paper is to examine the phenomenology of
SUSY models with non-universal gaugino masses and examine their impact
upon the cosmological relic density of DM and its
prospects for detection in direct and indirect detection experiments,
and finally for direct detection of sparticles at the Fermilab Tevatron, 
the CERN LHC and at the 
future international $e^+e^-$ linear collider (ILC). Towards this end, 
we will adopt a model with GUT scale parameters
including universal scalar masses, but with
independent $SU(2)$ and $U(1)$ gaugino masses which can be of
{\it opposite-sign}. Indeed while some of the earlier studies with
non-universal gaugino masses mentioned in the previous paragraph do
allow for negative values of $M_1/M_2$, we are not aware of a systematic
exploration of this part of parameter space. 
For the most part, we  
adjust the gaugino masses until the relic density matches
the central value determined by WMAP.
Whereas the case of {\it same-sign} gaugino masses allows consistency
with WMAP via both bino-wino mixing and bino-wino co-annihilations (the MWDM scenario), 
the opposite sign case admits essentially no mixing between
the bino and wino gaugino components.
Agreement with the WMAP value can be attained if the LSP is bino-like,
and the wino mass $M_2 \simeq -M_1$ at the weak scale, so that
bino-wino co-annihilation (BWCA) processes come into play
and act to reduce the bino relic density to acceptable values.

The BWCA scenario leads to a number of distinct 
phenomenological consequences. For direct 
and indirect DM search experiments (except when sfermions are also very light),
very low detection rates are expected in the BWCA
scenario because $SU(2)_L\times U(1)_Y$ gauge invariance precludes
couplings of the bino to gauge bosons. 
Regarding collider searches, the BWCA scenario yields relatively low
$\tz_2 -\tz_1$ mass gaps, so that two-body tree level neutralino decays
are not kinematically allowed.  However, when $m_0\alt .5-1$ TeV, the
loop-induced radiative decay $\tz_2\to\tz_1\gamma$ is enhanced, and can
even be the dominant $\tz_2$ decay mode. This may give rise to unique
signatures involving isolated photon plus jet(s) plus lepton(s) plus
$\eslt$ events at the Fermilab Tevatron and the CERN LHC hadron
colliders.  At the ILC, $\tz_1\tz_2$ production can
lead to ``photon plus nothing'' events, while $\tz_2\tz_2$ production
can lead to diphoton plus missing energy events at large rates.  It
would be interesting to examine if these signals can be separated from
SM backgrounds involving neutrinos, $e^+e^- \gamma$ and multiple gamma
processes where the electron and positron, or some of the photons, are
lost down the beam pipe.  Potentially, the energy spectrum of the signal
$\gamma$ may allow for the extraction of $m_{\tz_2}$ and $m_{\tz_1}$ via
photon energy spectrum endpoint measurements.

The remainder of this paper is organized as follows.
In Sec. \ref{sec:pspace}, we present the parameter space for the 
BWCA scenario, and show the spectrum of sparticle masses which 
are expected to occur. We also discuss some fine points of the BWCA
relic density analysis.
In Sec. \ref{sec:ddet}, we show rates for direct and indirect 
detection of DM in the BWCA scenario. 
These rates are expected to be below detectable levels unless 
some sfermions are very light or additional annihilation mechanisms
can be active.
In Sec. \ref{sec:z2z1g}, we present expectations for the radiative
neutralino decay $\tz_2\to\tz_1\gamma$ in BWCA parameter space.
In Sec. \ref{sec:col}, we examine the implications of the BWCA scenario
for the Fermilab Tevatron, CERN LHC and the ILC.
In Sec. \ref{sec:conclude}, we present our conclusions. In the Appendix
we adapt the idea of integrating out heavy degrees of freedom, familiar
in quantum field theory, to quantum mechanics, and use the results to
obtain simplified expressions for neutralino masses in the large $|\mu|$
limit where the higgsinos can be ``integrated out''. 

\section{Sparticle mass spectrum in the BWCA scenario}
\label{sec:pspace}

It is well known that a pure bino LSP can annihilate rapidly enough to
give the observed relic density only if scalars are sufficiently light,
as for instance in the so-called bulk region of the mSUGRA model. Our
goal here is to explore SUGRA models with universal values of high scale
SSB scalar masses and $A$-parameters, but without the assumption of
universality on gaugino masses. In models without gaugino mass
universality, the annihilation rate of a bino LSP may be increased in
several ways, including
\begin{enumerate}
\item  by increasing the higgsino content of the LSP, which may be achieved
by decreasing the gluino mass relative to the 
electroweak gaugino masses\cite{belanger};
\item by increasing the wino content of the LSP, by reducing the ratio
  $M_1/M_2$ as in the MWDM scenario which has been the subject of many
  investigations\cite{birkedal,mwdm}; 
\item by allowing co-annihilations between highly pure bino-like and wino-like
  states with comparable physical masses. We dub this the bino-wino
  coannihilation (BWCA) scenario. 
\end{enumerate}

The MWDM scenario is realized when the $SU(2)$ gaugino mass parameter
$M_2({\rm weak})\equiv M_2(M_{weak})$ approaches the 
$U(1)$ gaugino mass parameter $M_1({\rm weak})$. 
Since the $\tz_1$ in MWDM can annihilate very efficiently into $W^+W^-$
pairs via its wino component, care must be taken to ensure that this
wino component is not so large that the relic density falls {\it below}
its WMAP value.\footnote{We emphasize that, while a neutralino relic
density smaller than the WMAP value is not excluded, in this paper we
confine ourselves to scenarios that accommodate this value.} The reader
can easily check that if the electroweak gaugino masses are equal at the
{\it weak scale}, the photino state $$\tilde{\gamma} \equiv
\cos\theta_W\tilde{B}+\sin\theta_W\tilde{W}, $$ is an exact mass
eigenstate of the tree level neutralino mass matrix, and has a mass
equal to the common weak scale gaugino mass, $M$ \cite{wss}. Since the
photino can readily annihilate to $W^+W^-$ pairs via chargino exchange
in the $t$ channel, we would expect that it gives a relic density that
is smaller than the WMAP value: in other words, the wino content of the
$\tz_1$ must be {\it smaller} than $\sin\theta_W$ in order to obtain the
WMAP value of the relic density. We will see shortly that this is indeed
the case.

In the MWDM scenario, treating the
gaugino-higgsino mixing entries in the neutralino mass matrix, whose
scale is set by $M_W$, as a
perturbation, the signed tree level gaugino
masses for the case $M_1({\rm weak})=M_2({\rm weak})=M$ are given (to second order) by,
\begin{eqnarray} 
M \ \ ({\rm photino})\hspace{4cm}  \nonumber \\
M+ \frac{1}{2} M_Z^2 \left[\frac{(\sin\beta-\cos\beta)^2}{M-\mu}+
\frac{(\sin\beta+\cos\beta)^2}{M+\mu}\right] \ \ ({\rm zino}) .
\label{eq:photinozino}
\end{eqnarray}
If instead of exact degeneracy between $M_1({\rm weak})$ and $M_2({\rm weak})$, 
we have $M_1({\rm weak})=M-\delta$ and $M_2({\rm weak})=M+\delta$, where $\delta$ is as
small or comparable to the gaugino-higgsino mixing entries, these
eigenvalues change to,
\begin{equation}
M+\frac{1}{2}[g^2+g'^2\mp\xi]v_{\rm pert}, 
\end{equation}
where
\begin{equation}
v_{\rm
pert}=\frac{1}{2}\frac{M_Z^2}{g^2+g'^2}
\left[\frac{(\sin\beta-\cos\beta)^2}{M-\mu}+
\frac{(\sin\beta+\cos\beta)^2}{M+\mu}\right], \nonumber
\end{equation}
and 
\begin{equation}
\xi^2 = (g^2+g'^2)^2 + 4(g^2-g'^2)\frac{\delta}{v_{\rm pert}} 
+ 4\frac{\delta^2}{v_{\rm pert}^2}\;. \nonumber
\end{equation}
Clearly, these eigenvalues reduce to the masses (\ref{eq:photinozino})
of the photino and zino states when $\delta \to 0$.

The generic case where the differences between $M_1({\rm weak})$,
$M_2({\rm weak})$ and $\pm\mu$ are all much larger than the
gaugino-higgsino mixing entries is much simpler to treat since we do not
have to worry about degeneracies as in the $M_1({\rm weak})\simeq M_2({\rm weak})$ 
case that we have just discussed. In the absence of
gaugino-higgsino mixing, the bino and neutral wino are the gaugino mass
eigenstates.  Then, again treating the gaugino-higgsino mixing entries
in the neutralino mass matrix as a perturbation, we see that mixing
between wino and bino states occurs only at {\it second} order in
$M_W/\CM$, where $\CM$ denotes a generic mass difference between the
``unperturbed'' higgsino and gaugino mass eigenvalues. This is in sharp
contrast to the MWDM scenario where even the ``unperturbed'' gaugino
states are strongly mixed because of the degeneracy of the eigenvalues
when $M_1({\rm weak})=M_2({\rm weak})$.  It is simple to show that the
signed masses of the bino-like and wino-like neutralino eigenstates take
the form,
\begin{eqnarray}
M_1 + \frac{1}{2} M_Z^2\sin^2\theta_W
\left[\frac{(\sin\beta-\cos\beta)^2}{M_1-\mu}+
\frac{(\sin\beta+\cos\beta)^2}{M_1+\mu}\right] \ ({\rm bino}), \nonumber \\
M_2 +\frac{1}{2} M_Z^2\cos^2\theta_W
\left[\frac{(\sin\beta-\cos\beta)^2}{M_2-\mu}+
\frac{(\sin\beta+\cos\beta)^2}{M_2+\mu}\right]  \ ({\rm wino}),
\label{eq:binowino}
\end{eqnarray}
where $M_1$ and $M_2$ in (\ref{eq:binowino}) denote their values at the
weak scale.  Note that for the special case $M_1({\rm weak})=-M_2({\rm weak})$, 
the  formulae (\ref{eq:binowino}) apply even though the
physical masses of the two lighter neutralinos may be very close to one
another.  The main purpose of our discussion
of the approximate masses and mixing patterns is that these will enable
us to better understand the differences in phenomenology in the MWDM
scenario, where $M_1({\rm weak})$ is slightly smaller than $M_2({\rm weak})$, 
and the BWCA scenario where $M_1({\rm weak}) \simeq -M_2({\rm weak})$.


Before proceeding to explore other sparticle masses in these scenarios,
we first check that the qualitative expectations for neutralino mixing
patterns and relic density discussed above are indeed realized by
explicit calculation. Toward this end, we adopt the subprogram Isasugra,
which is a part of the Isajet v 7.72$^\prime$ event generator
program\cite{isajet}.\footnote{Isajet 7.72$^{\prime}$ is Isajet v7.72
with the subroutine SSM1LP modified to yield chargino and neutralino 
mass diagonalization all at a common scale $Q=\sqrt{m_{\tz_1}m_{\tz_2}}$.}
Isasugra allows the user to obtain sparticle
masses for a wide variety of GUT scale non-universal soft SUSY breaking
terms. The sparticle mass spectrum is generated using 2-loop MSSM
renormalization group equations (RGE) for the evolution of all couplings
and SSB parameters.  An iterative approach is used to evaluate the
supersymmetric spectrum.  Electroweak symmetry is broken radiatively, so
that the magnitude, but not the sign, of the superpotential $\mu$
parameter is determined.  The RG-improved 1-loop effective potential is
minimized at an optimized scale to account for the most important 2-loop
effects. Full 1-loop radiative corrections are incorporated for all
sparticle masses. To evaluate the neutralino relic density, we adopt the
IsaRED program\cite{isared}, which is based on CompHEP\cite{comphep} to
compute the several thousands of neutralino annihilation and
co-annihilation Feynman diagrams. Relativistic thermal averaging of the
cross section times velocity is performed\cite{gg}. The parameter space
we consider is given by 
\be m_0,\ m_{1/2},\ A_0,\ \tan\beta ,\ sign (\mu
),\ M_1\ {\rm or}\ M_2 , 
\ee 
where we take either $M_1$ or $M_2$ to be
free parameters (renormalized at $Q=M_{GUT}$), and in general not equal
to $m_{1/2}$. When $M_1$ is free, we maintain $M_2=m_{1/2}$, while when
$M_2$ is free, we maintain $M_1=m_{1/2}$.

We show our first results in Fig. \ref{fig:rd_bw1}, where we take
$m_0=m_{1/2}=300$ GeV, with $A_0=0$, $\tan\beta =10$, $\mu >0$ with
$m_t=178$ GeV. We plot the neutralino relic density $\Omega_{\tz_1}h^2$
in frame {\it a}) versus variation in the $U(1)$ gaugino mass parameter
$M_1$. For $M_1=300$ GeV, we are in the
mSUGRA case, and $\Omega_{\tz_1}h^2=1.3$, so that this model would be
strongly excluded by the WMAP measurement.  For smaller values of $M_1$,
the bino-like neutralino becomes lighter and two dips occur in the
neutralino relic density.  These correspond to the cases where
$2m_{\tz_1}\simeq m_h$ and $M_Z$ as one moves towards decreasing $M_1$,
{\it i.e.} one has either light Higgs $h$ or $Z$ resonance
annihilation\footnote{For values of $|M_1|\alt 100$ GeV, we have checked
that the contribution of $\Gamma (Z\to\tz_1\tz_1)$ is always below
limits from LEP on the invisible width of the $Z$ boson.}. 
Instead, as $M_1$ increases past its mSUGRA value, the
$\tz_1$ becomes increasing wino-like and the relic density agrees with
the WMAP value at $M_1\sim 490$ GeV, where the MWDM scenario is realized
\cite{mwdm}. For yet larger values of $M_1$, the wino content of $\tz_1$
is so large that the relic density falls below the WMAP value.
Turning to negative values of $M_1$, we see that as $M_1$
starts from zero and becomes increasingly {\it negative}, the $Z$
and $h$ poles are again encountered. The relic density is again much larger
than the WMAP bound until it begins decreasing for $M_1<-400$ GeV.  At
$M_1\simeq -480$ GeV, the relic density is again in accord with the
WMAP value, while for more negative values of $M_1$, the relic density
is too low, so that some other form of CDM or non-thermal production 
of neutralinos would be needed to account for the WMAP measurement.
\FIGURE[!htb]{
\epsfig{file=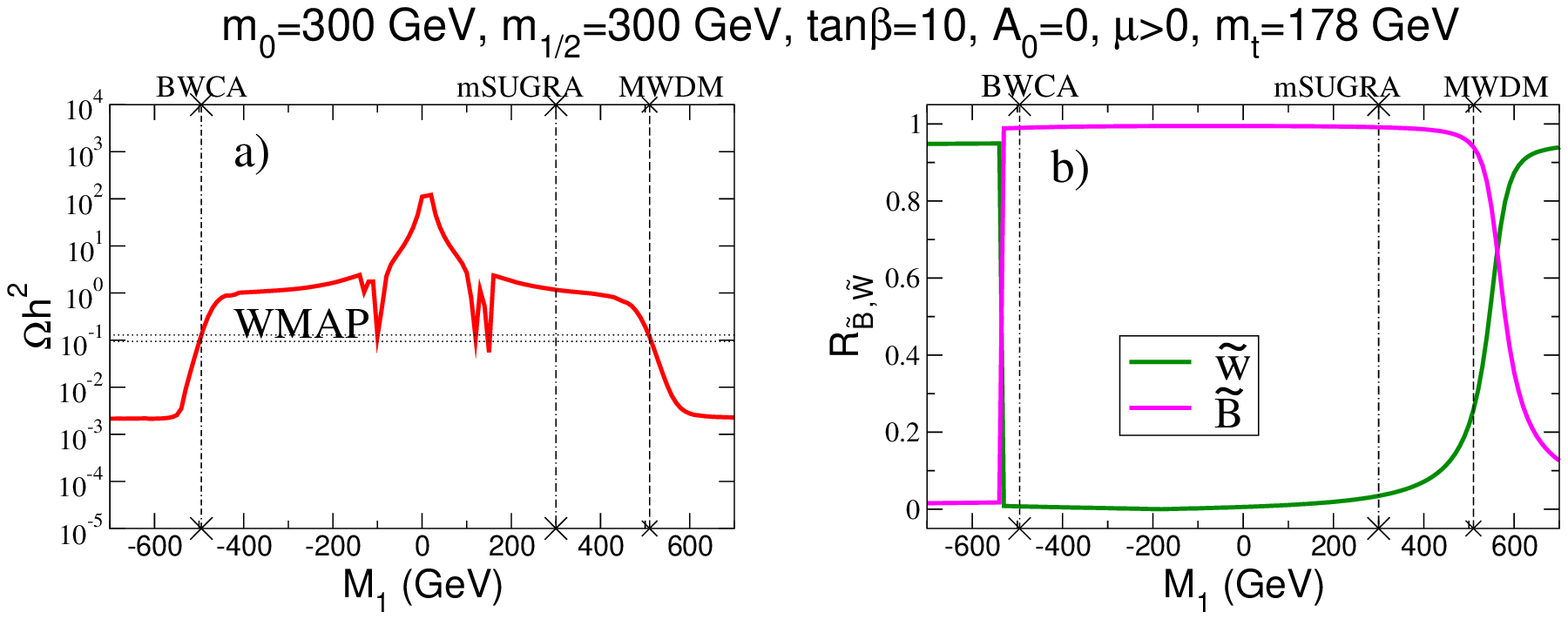,width=7cm,angle=-90} 
\caption{\label{fig:rd_bw1}
A plot of {\it a}) relic density $\Omega_{CDM}h^2$ and 
{\it b}) bino/wino component of the lightest neutralino as a
function of $M_1$ for
$m_0=300$ GeV, $m_{1/2}=300$ GeV, $A_0=0$, $\tan\beta =10$, $\mu >0$
and $m_t=178$ GeV.}}

In frame {\it b}), we show the amplitude, $R_{\tB ,\tw}$, for the
bino/wino content of the lightest neutralino $\tz_1$.  Here, we adopt
the notation of Ref. \cite{bbkt,wss}, wherein the lightest neutralino is
written in terms of its (four component Majorana) Higgsino and gaugino
components as \be
\tz_1=v_1^{(1)}\psi_{h_u^0}+v_2^{(1)}\psi_{h_d^0}+v_3^{(1)}\lambda_3
+v_4^{(1)}\lambda_0 , \ee where $R_{\tw}=|v_3^{(1)}|$ and
$R_{\tB}=|v_4^{(1)}|$.  A striking difference between the positive and
negative $M_1$ portions of frame {\it b}) is the shape of the level
crossings at $R_{\tB}=R_{\tw}=1/\sqrt{2}$: while the transition from a
bino-like to a wino-like LSP is gradual when $M_1> 0$, it is much more
abrupt for negative values of $M_1$.  We had already anticipated this
when we noted that for $M_1({\rm weak})=M_2({\rm weak})$ the mass
eigenstates are the photino, and aside from a small admixture of
higgsinos, the zino, while the corresponding eigenstates were bino-like
and wino-like as long as $|M_1({\rm weak}) - M_2({\rm weak})|$ was
larger than the gaugino-higgsino mixing entries in the neutralino mass
matrix. We also see that, for $M_1 > 0$, the MWDM scenario is realized
for the value $M_1$ where $R_{\tw} < \sin\theta_W \simeq 0.48$ 
as we also expected.
Finally, again as we anticipated, for $M_1 < 0$, the BWCA scenario is
obtained for $|M_1|$ just above the level crossing, when the LSP is
mainly bino-like and close in mass to the wino-like $\tz_1$.
For a bino-like LSP, annihilation to vector boson pairs
is suppressed, and 
the only way to reduce the relic density in the case of large 
negative $M_1$ is by having rather
small $\tz_2-\tz_1$ and $\tw_1-\tz_1$ mass gaps, so that 
chargino and neutralino co-annihilation effects are large\cite{griest_seck}, 
and act to
decrease the relic density. Of course, when $-M_1> M_2$, then the
$\tz_1$ suddenly becomes pure wino-like, and very low relic density
is obtained, as in the case of AMSB models.

Similar results are obtained by keeping $M_1=m_{1/2}$, and varying
$M_2$, as shown in Fig. \ref{fig:rd_bw2}. The MWDM scenario is reached
just to the right of the level crossing where $M_2({\rm weak})$ is
slightly larger than $M_1({\rm weak})$, while the BWCA scenario is
reached for $M_2=-156$ GeV, where $|M_2({\rm weak})|$ is again just above
$|M_1({\rm weak})|$ and the $\tz_1$ is essentially a bino. Again, 
the $\tz_2-\tz_1$ and, as discussed just below, also
the $\tw_1-\tz_1$, mass gaps are very small, so that co-annihilation
plays an essential role in reducing $\Omega_{\tz_1}h^2$ to an acceptable
value.  We see also that there is a region of mainly small negative
$M_2$ values where $m_{\tw_1}<m_{\tz_1}$, where a charged LSP is
obtained. This is, of course, ruled out by the negative results for
searches for charged stable relics from the Big Bang.

In both the MWDM and the BWCA scenarios, $\tz_1, \ \tz_2$ and $\tw_1$
are the lightest sparticles. They will likely have a large impact upon
the phenomenology of these models. Of particular interest is the
ordering of the mass spectrum of these particles. We work this out
in the so-called
large $|\mu|$ approximation, where $|\mu|\gg M_1, M_2$ applicable 
in many models. 
The chargino sector, since it
consists of just two states, is simple and $m_{\tw_1}$ is given by a
relatively simple and well known expression that we do not reproduce
here (see {\it e.g.} Ref. \cite{wss}).  
The neutralino sector is  much more complicated but, in the large
$|\mu|$ approximation,
it is possible to ``integrate out'' the higgsinos, and work
with an effective low scale theory that only includes the neutral bino and
wino as discussed in the Appendix. 
The couplings of winos and binos to higgsinos in the original
theory manifests itself as a mixing between winos and binos in the
effective theory, where this
mixing is suppressed by $1/|\mu|$.  To ${\cal O}(1/\mu^2)$, the {\it
signed} tree level neutralino masses are given (in terms of weak scale
parameters) by,\footnote{We are abusing notation here in that we are
using $m_{\tz_1}$ and $m_{\tz_2}$ to denote the signed neutralino mass,
whereas everywhere else we use the same symbols to denote the physical
(positive) neutralino masses.}
\begin{equation}
(1+\frac{M_W^2}{\mu^2})m_{\tz_{1,2}}= \frac{M_1+M_2
  +a\sec^2\theta_W+M_2\eta \pm \kappa}{1(1+\eta)} + {\cal O}({1\over\mu^3})\;,
\label{highmuneut}
\end{equation}
where 
$$a= {{M_W^2\sin2\beta}\over \mu} \ll |M_1|, \ |M_2|, $$
$$\eta = \frac{M_W^2}{\mu^2}(-1+\tan^2\theta_W)\ll 1,$$ and
\begin{eqnarray*}
\kappa^2&=&
(M_1-M_2)^2+2a(M_1+M_2)\sec^2\theta_W+a^2\sec^4\theta_W\\
&+&2(M_1+M_2)M_2\eta
-4a(M_1+M_2\tan^2\theta_W)-4M_1M_2\eta.
\end{eqnarray*} 
We emphasize
that (\ref{highmuneut}) is valid even when the gaugino masses are
comparable to the gaugino-higgsino mixing terms in the neutralino mass
matrix, and is a useful approximation as long as $|\mu|$ is large.
If $M_1$ is not very close to $M_2$, we can simplify the expressions for
$m_{\tz_{1,2}}$ by
expanding $\kappa \simeq |M_1-M_2|$ plus terms that are suppressed by
powers of $\mu$. Such an expansion then yields,  
\begin{eqnarray}
m_{\tz_1} &=& M_1 +\frac{M_W^2}{\mu}\tan^2\theta_W\sin2\beta
-M_1\frac{M_W^2}{\mu^2}\frac{\sin^2 2\beta\tan^2\theta_W}{M_1-M_2}
+{\cal O}(1/\mu^3),\nonumber \\
m_{\tz_2} &=& M_2+\frac{M_W^2}{\mu}\sin 2\beta -M_2\frac{M_W^2}{\mu^2}
-\frac{M_W^4}{\mu^2}\frac{\tan^2\theta_W}{M_1-M_2}\sin^2 2\beta+{\cal O}(1/\mu^3),
\label{eq:highstates}\\
m_{\tw_1}&=& M_2 + \frac{M_W^2}{\mu}\sin 2\beta -\frac{M_W^2}{\mu^2}M_2
+{\cal O}(1/\mu^3)\nonumber \;,
\end{eqnarray}
where we have assumed $|M_1| < |M_2|$ when we associated the bino-like
state with $\tz_1$. Except for this, (\ref{eq:highstates}) are valid for
all magnitudes and signs of $M_1$ and $M_2$, as long as $|M_1-M_2| \gg
M_W^2/|\mu|$. In particular, we can use these expressions for the BWCA
scenario, but for the MWDM case, we would have to use (\ref{highmuneut})
to get the neutralino masses.

\FIGURE[!htb]{
\epsfig{file=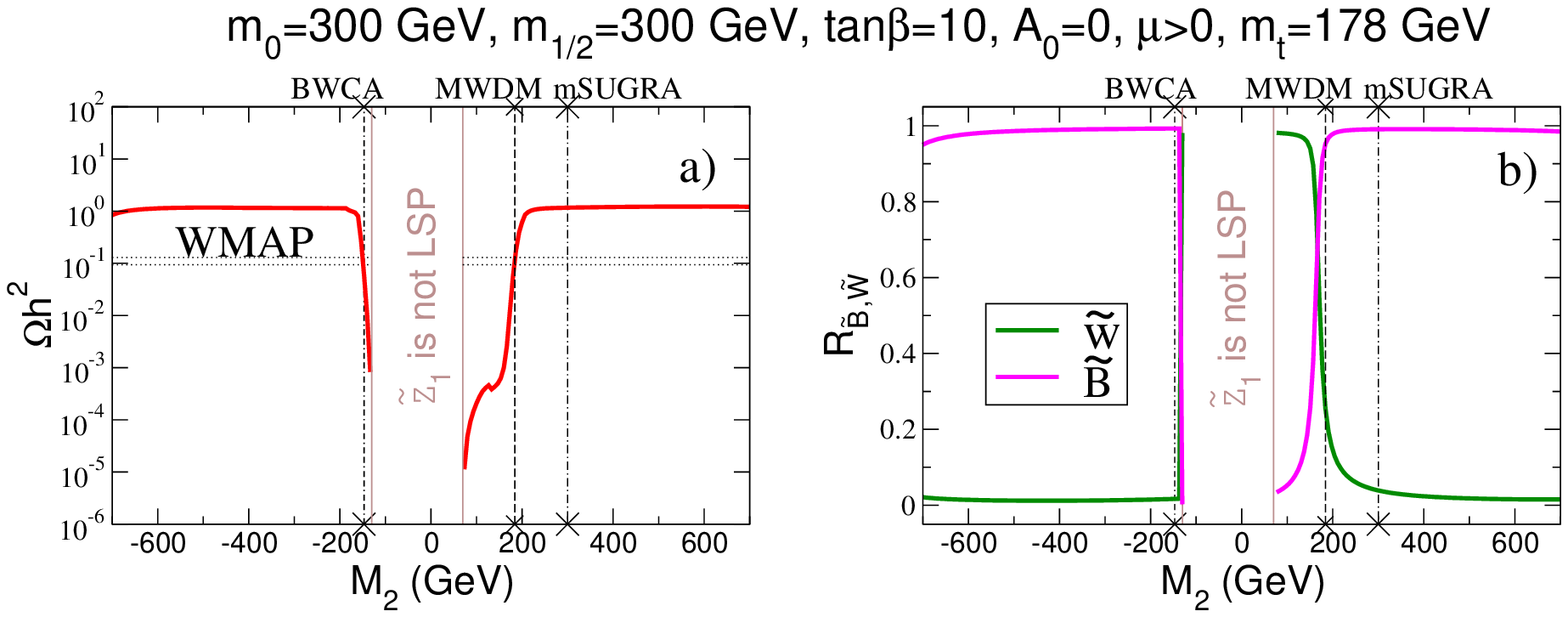,width=7cm,angle=-90} 
\caption{\label{fig:rd_bw2}
A plot of {\it a}) relic density $\Omega_{CDM}h^2$ and 
{\it b}) bino/wino component of the lightest neutralino as a
function of $M_2$ for
$m_0=300$ GeV, $m_{1/2}=300$ GeV, $A_0=0$, $\tan\beta =10$, $\mu >0$
and $m_t=178$ GeV.}}
%

\begin{figure*}[!t]
\begin{center}
\includegraphics[scale=0.5]{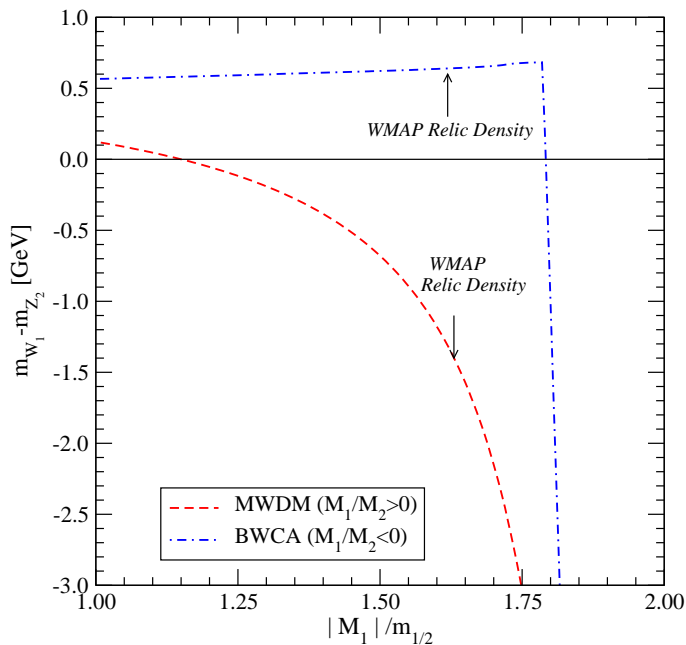}
\end{center}
\caption{The mass splitting between the lightest chargino and the
next-to-lightest neutralino, $m_{\widetilde W_1}-m_{\widetilde Z_2}$ in
the BWCA scenario (dot-dashed blue line, with {\em opposite} $M_1$ and
$M_2$ signs) and in the MWDM scenario (dashed red line, with {\em same}
$M_1$ and $M_2$ signs), as a function of the absolute GUT scale value of
$|M_1|/m_{1/2}$.  The arrows indicate the value of $|M_1|/m_{1/2}$ where
the neutralino thermal relic abundance equals the central WMAP CDM
abundance value.  The other SUSY parameters are as in Fig. 1.}
\label{fig:deltam}
\end{figure*}

The ordering of the mass spectrum of the lightest SUSY particles in the
same sign MWDM and opposite sign BWCA scenarios is different, as can be
seen from Fig.~\ref{fig:deltam}, where we plot the $\widetilde
W_1$-$\widetilde Z_2$ mass splitting as a function of the absolute
GUT-scale value $|M_1|/m_{1/2}$. For the
MWDM case, the lightest chargino is lighter than the next-to-lightest
neutralino, while the opposite holds true in the BWCA case.  As
long as $|M_1({\rm weak})| < |M_2({\rm weak})|$, both $\tw_1$ and
$\tz_2$ are dominantly wino-like, and 
the tree-level mass splitting
between them can be read off from (\ref{eq:highstates}), as long as 
the weak scale values of $M_1$ and $M_2$ are not too close. We then
find that at tree-level \cite{fmrss}, 
%
\begin{equation}
\Delta m_{\rm tree}\equiv m_{\tw_1}-m_{\tz_2} \approx
\frac{m_W^4\tan^2\theta_W}{(M_1({\rm weak})-M_2({\rm
weak})\mu^2)}\sin^22\beta .
\label{eq:split}
\end{equation}
for all combinations of signs of gaugino masses.  For the BWCA case, the
denominator $|M_1({\rm weak})-M_2({\rm weak})|$ is very large, so that the
tree level splitting is negligible compared to the one-loop splitting;
the latter is always positive, hence in the opposite sign BWCA case,
when the LSP is bino-like, the lightest chargino is always heavier than
the next-to-lightest neutralino.  
In contrast, for the same-sign MWDM case, the tree level splitting (over
the range of $M_1$ values where (\ref{eq:split}) is valid) is negative,
and comparable to or larger than the one-loop splitting, so that the
chargino is now usually heavier than $\tz_2$, as may be seen by the
dashed line in Fig.~\ref{fig:deltam}.

Various other
sparticle masses are also affected by varying the gaugino masses, 
since these feed into the soft term evolution via the RGEs. 
In Fig. \ref{mass_M1}, we show the variation of the sparticle 
mass spectrum with respect to the GUT scale ratio 
$M_1/m_{1/2}$ for the same parameters
as in Fig. \ref{fig:rd_bw1}. In the mSUGRA case where $M_1/m_{1/2}=1$, 
there is a
relatively large mass gap between $\tz_2$ and $\tz_1$: 
$m_{\tz_2}-m_{\tz_1}=106.7$ GeV. As $M_1$ varies to
large positive values (the MWDM case), the mass gap shrinks 
to $m_{\tz_2}-m_{\tz_1} =31.9$ GeV. 
As $M_1$ varies to large {\it negative} values, the mass gap also decreases,
this time to just 22.7 GeV in the BWCA scenario.
We also note that as $|M_1|$ increases, the $\te_R$, 
$\tmu_R$ and $\ttau_1$ masses also increase, since $M_1^2$ feeds into their
mass evolution via RGEs. 
This also gives rise to the nearly symmetric behavior versus the sign of
$M_1$ for the mass spectrum of first and second generation sfermions.
As the coefficient appearing in front of $M_1$ in the RGEs is larger 
(and with the same sign) for the right handed sfermions than 
for the left handed ones, one expects, in general, a departure 
from the usual mSUGRA situation where the lightest sleptons are right-handed.
As a matter of fact, whereas in mSUGRA $m_{\te_L}>> m_{\te_R}$ for $m_0
\alt m_{1/2}$, in the 
case of BWCA or MWDM, we find that $m_{\te_L}\sim m_{\te_R}$.
As shown in the figure, the right-handed squark masses also increase
with increasing $|M_1|$, although the relative effect is less dramatic
than the case involving sleptons: 
the dominant driving term in the RGEs is, in this case, 
given by $M_3$ (absent in the case of sleptons), 
hence variations in the GUT value of $M_1$ produce milder effects.
The trilinear SSB $A_t,\ A_b$ and $A_\tau$ parameters have a linear 
dependence on gaugino mass in their RGEs, which means the 
weak scale $A$-parameters will be asymmetric versus the sign of $M_1$.
The $\mu$ parameter is also slightly asymmetric. This gives rise to 
the asymmetric behavior of the third generaton sfermion masses with 
respect to the sign of the gaugino mass.
\FIGURE[!t]{
\epsfig{file=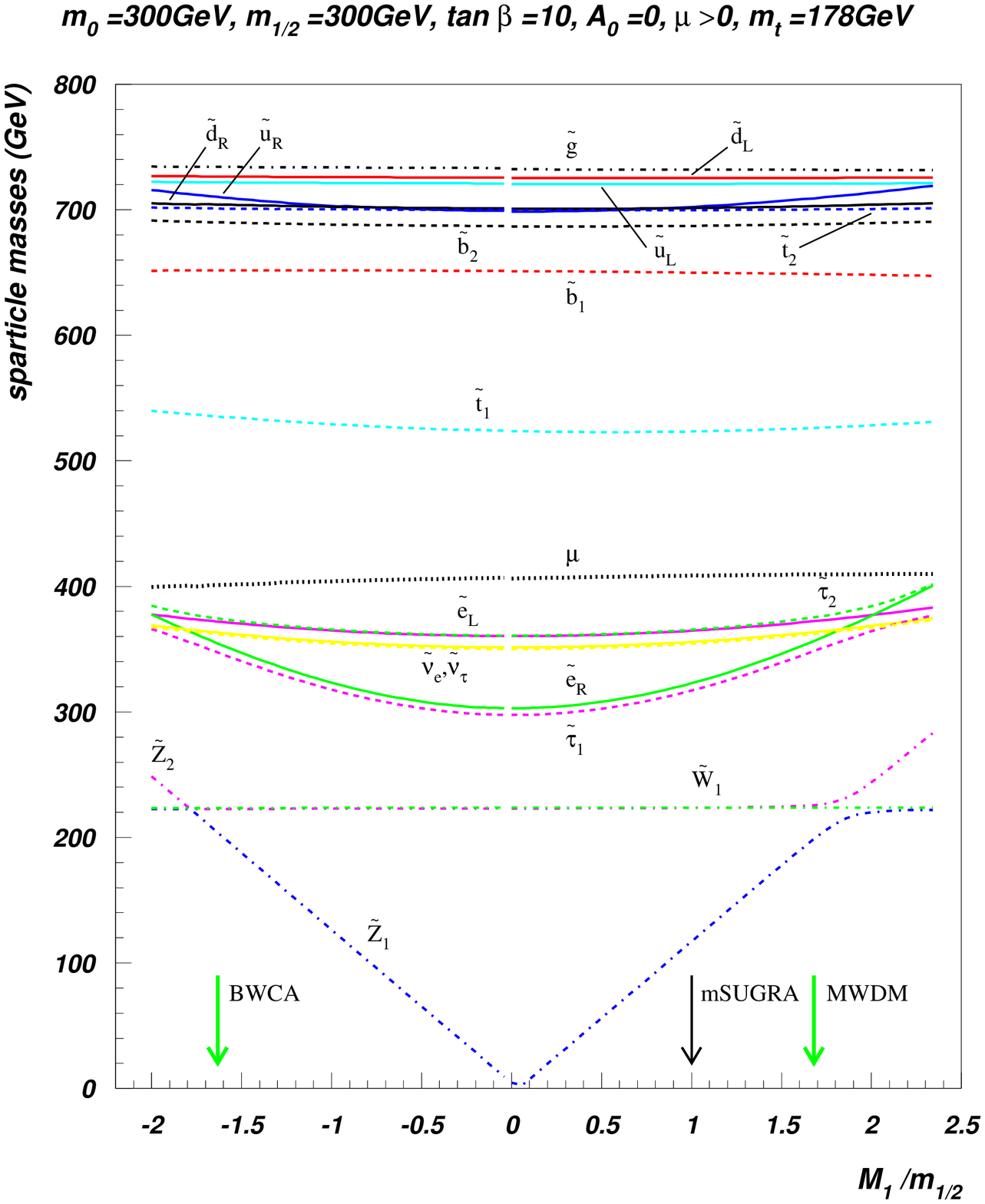,width=10cm} 
\caption{\label{mass_M1}
A plot of various sparticle masses {\it vs.} $M_1/m_{1/2}$ for
$m_0=300$ GeV, $m_{1/2}=300$ GeV, $A_0=0$, $\tan\beta =10$ and $\mu >0$.}}

In Fig. \ref{mass_M2}, we show a plot of sparticle masses for the same
parameters as in Fig. \ref{mass_M1}, but versus $M_2/m_{1/2}$. In this case,
as $|M_2|$ is decreased from its mSUGRA value of 300 GeV, the
$\tw_1$ and $\tz_2$ masses decrease until $\Omega_{\tz_1}h^2$ reaches 0.11
in both the BWCA and MWDM scenarios.
In this case, with decreasing
$|M_2|$, the left- slepton and sneutrino masses also decrease, again
leading to $m_{\te_L}\sim m_{\te_R}$. The left-handed squark masses 
similarly decrease. This increase is more pronounced that in
Fig.~\ref{mass_M1} because the $SU(2)$ gauge coupling is larger than the
$U(1)_Y$ gauge coupling. 
The $SU(2)$ singlet right-handed sfermion masses are not affected, 
with the net result that the mSUGRA $m_{\te_L}>> m_{\te_R}$ 
hierarchy is again altered. 
\FIGURE[!t]{
\epsfig{file=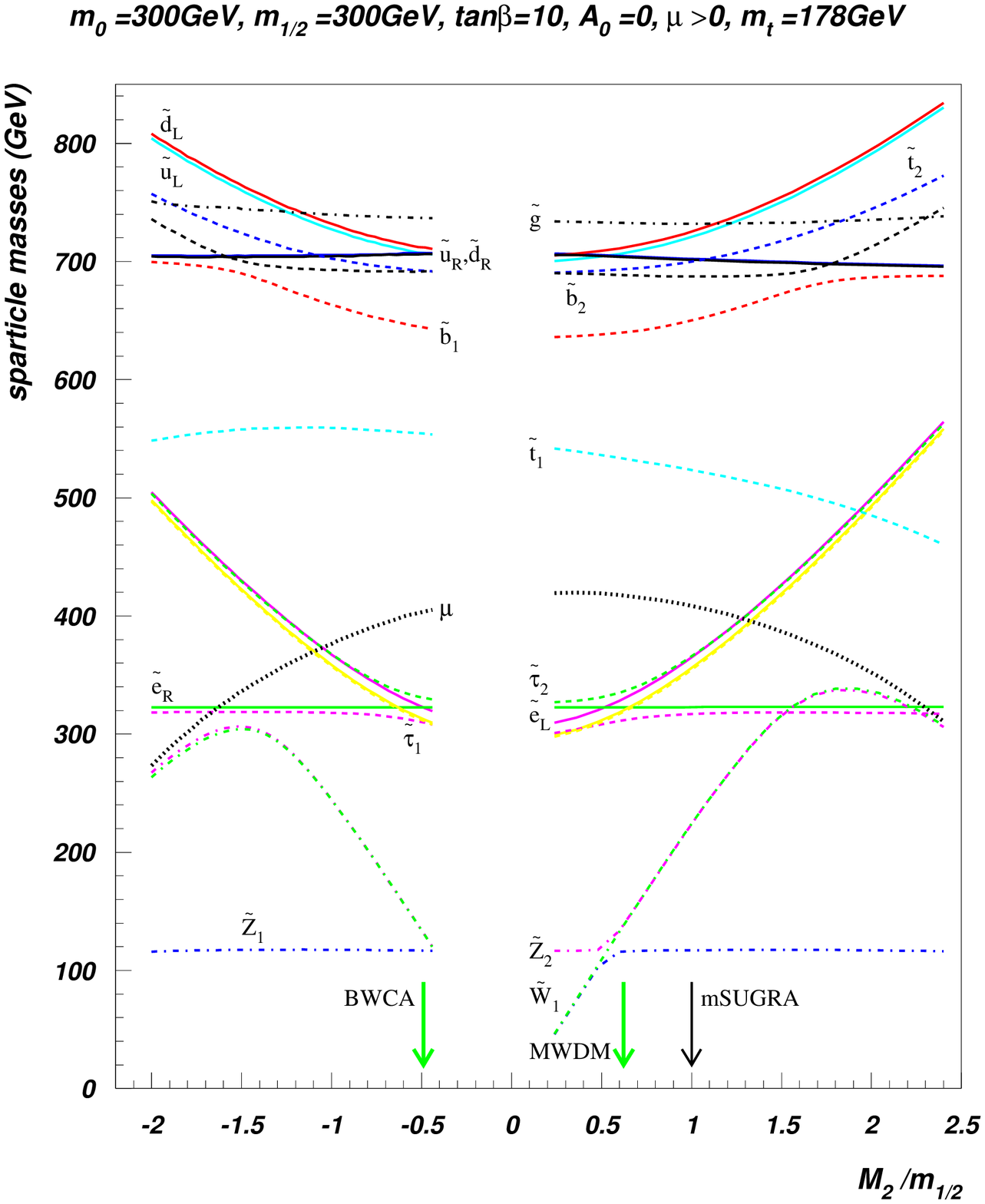,width=10cm} 
\caption{\label{mass_M2}
A plot of various sparticle masses {\it vs.} $M_2/m_{1/2}$ for
$m_0=300$ GeV, $m_{1/2}=300$ GeV, $A_0=0$, $\tan\beta =10$ and $\mu >0$.}}

It should be apparent now that most points in the $m_0\ vs.\ m_{1/2}$
plane can become WMAP allowed by adopting an appropriate {\it negative}
value of either $M_1$ or $M_2$ such that one enters into the BWCA
scenario.  The exception occurs if the WMAP-allowed point is obtained
because the $A$-funnel or stau co-annihilation region is reached
instead.  To illustrate this, we plot in Fig. \ref{planes_r} the ratio
$r_{1}\equiv M_1/m_{1/2}$ in frame {\it a}) or $r_2\equiv M_2/m_{1/2}$
in frame {\it b}) needed to achieve a relic density in accord with the
WMAP central value.  We see in frame {\it a}) that $r_1$ generally
increases as one moves from lower-left to upper-right. The structure in
the upper-left of the plot occurs when $-M_1$ is dialed to such a value
that $2m_{\tz_1}\simeq m_A$, {\it i.e.}  one is entering the $A$-funnel
(even though $\tan\beta$ is relatively low) instead of the BWCA
scenario. These regions will of course have a much larger $\tz_2-\tz_1$
mass gap than points in the BWCA scenario. Like the non-universal mass
scenario \cite{nuhm}, the BWCA scenario allows the $A$ funnel to be
reached for any value of $\tan\beta$, but should be distinguishable from
this because the $\tz_1-\tz_2$ mass gap, for instance, will be quite
different in the two scenarios. In frame {\it b}), the ratio $r_2$ that
gives rise to a WMAP-allowed point is shown to increase as one travels
from lower to higher values of $m_{1/2}$.  In this case, since the value
of $M_2$ hardly changes the value of $m_{\tz_1}$, the $A$-funnel is
never reached, and the BWCA region can be accessed over most of
parameter space, save near the left-hand edge in the stau
co-annihilation region.
\FIGURE[htb]{
\mbox{\hspace{-0.5cm}\epsfig{file=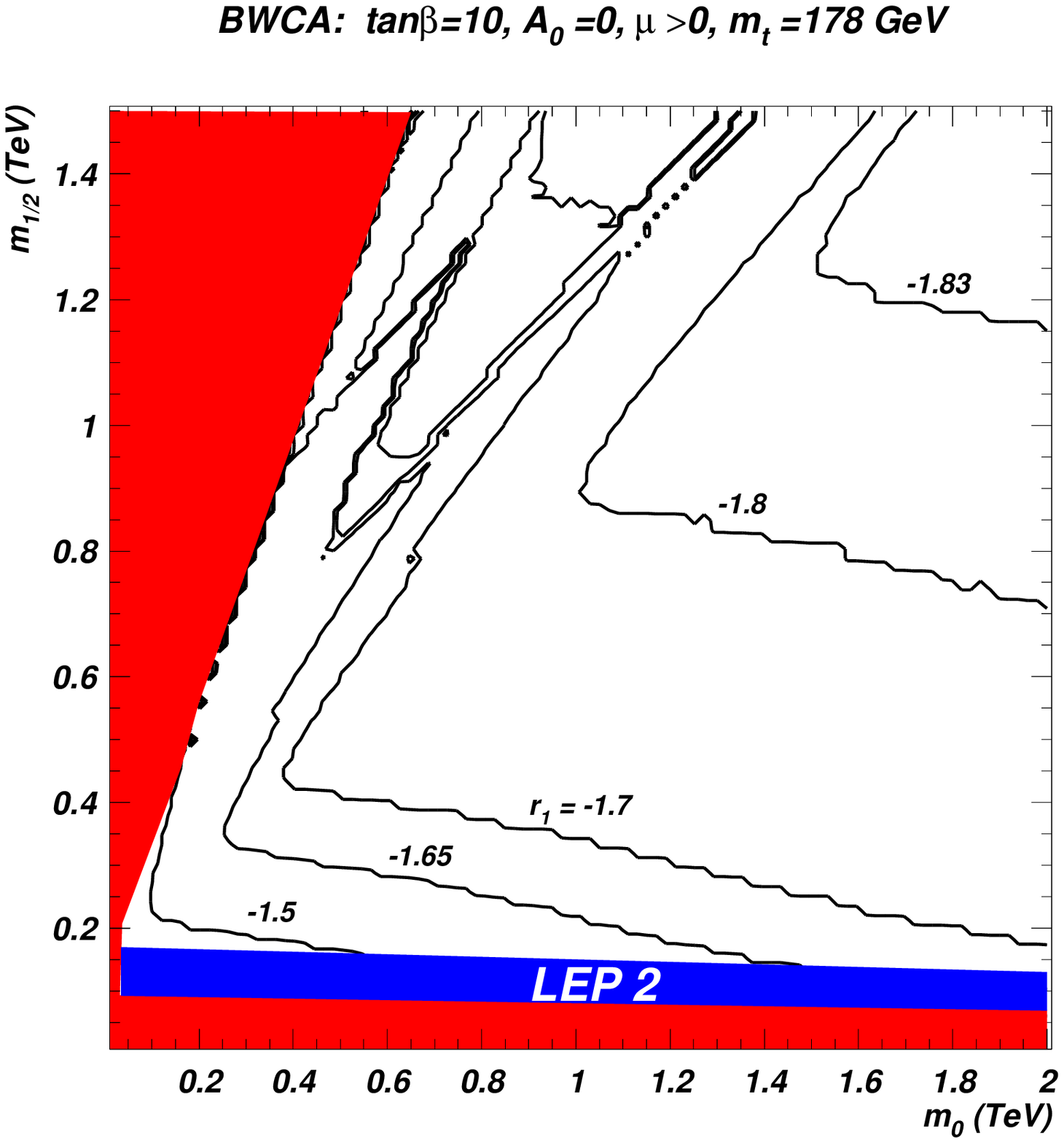,width=8cm}
\epsfig{file=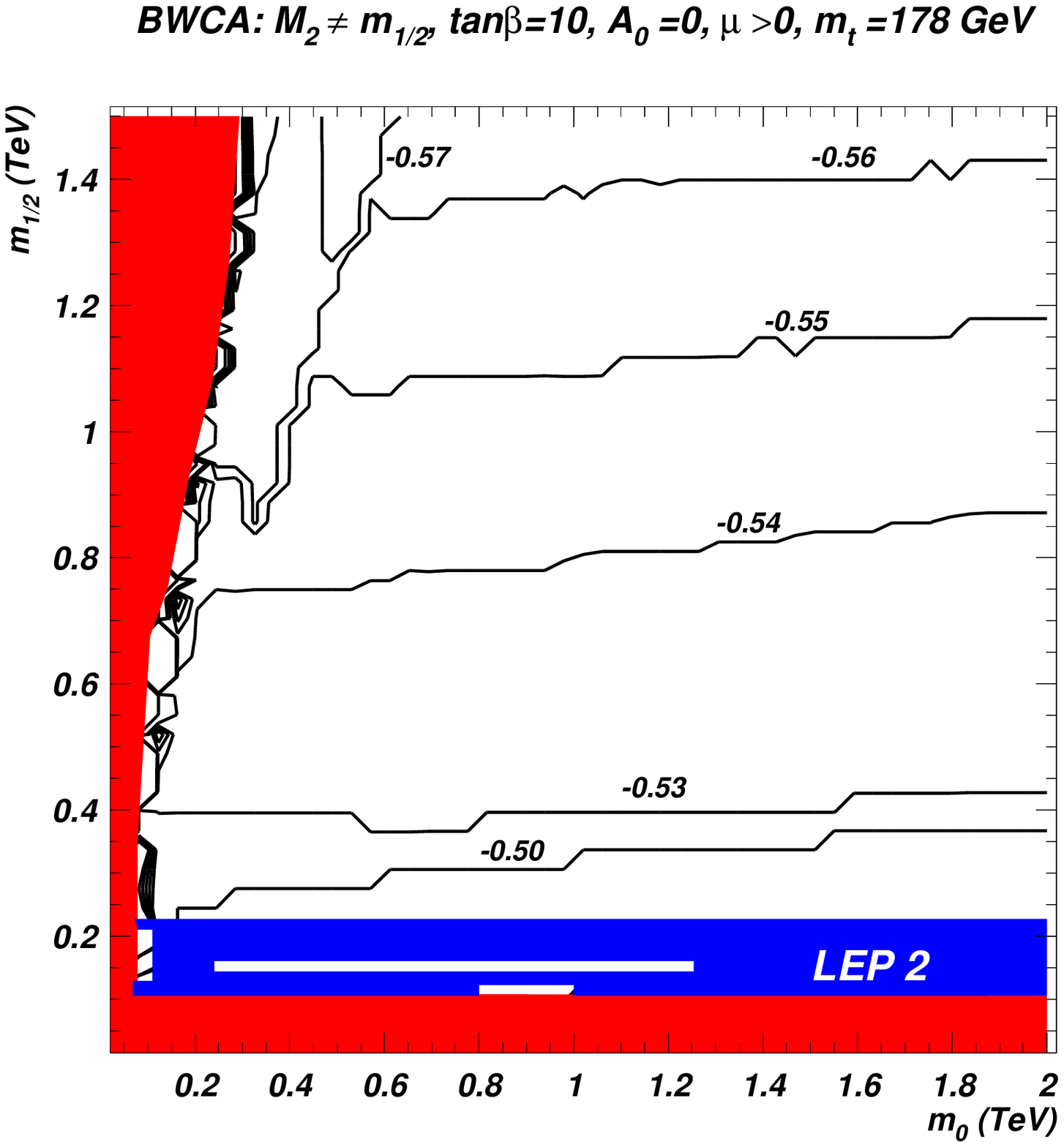,width=8cm}}

\caption{\label{planes_r}
Contours of {\it a}) $r_1$ and {\it b}) $r_2$ in the 
$m_0\ vs.\ m_{1/2}$ plane for
$\tan\beta =10$, $A_0=0$, $\mu >0$. Each point has
$\Omega_{\tz_1}h^2 =0.11$. The origin of the strips in the LEP excluded
region of the right hand frame is discussed in Sec.~\ref{sec:col}.}}

\section{Dark matter in the BWCA scenario}
\label{sec:ddet}

\subsection{Neutralino relic density in the BWCA scenario: a closer look}

\begin{figure*}[!t]
\begin{center}
\includegraphics[scale=0.5]{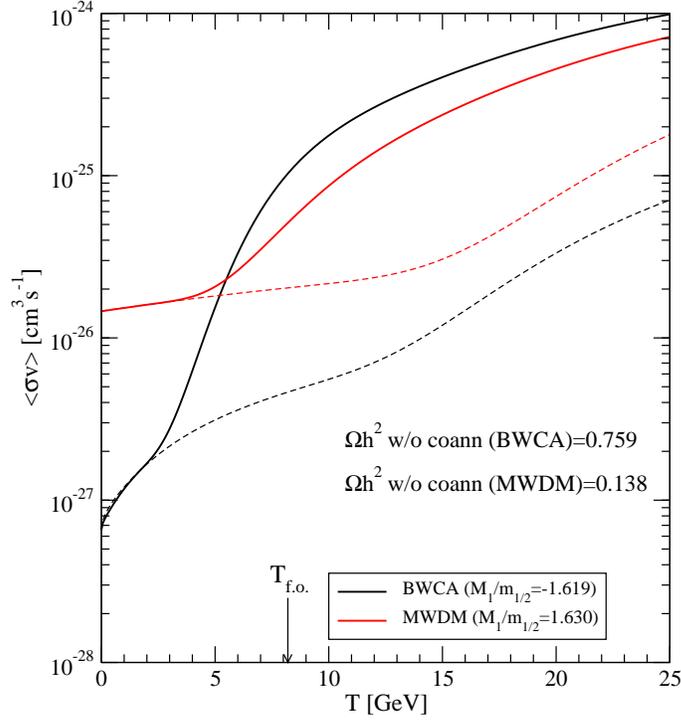}
\end{center}
\caption{The thermally averaged effective pair annihilation cross
section times velocity, $\langle \sigma v\rangle(T)$ as a function of
the temperature, for the BWCA (black lines) and MWDM (red lines)
scenarios. The dotted lines correspond to $\langle \sigma v\rangle(T)$
without the inclusion of coannihilation effects. The arrow indicates the
approximate temperature where freeze-out occurs.}
\label{fig:sv_th_av}
\end{figure*}

In order to better understand the coannihilation mechanisms which drive
the neutralino relic abundance within the WMAP preferred range in the
BWCA vs. the MWDM scenario (with opposite and same signs for $M_1$ and
$M_2$), we adopt the sample point defined by the mSUGRA input parameters
($m_0$=300 GeV, $m_{1/2}$=300 GeV, $\tan\beta$=10, sgn($\mu )>0$,
$A_0$=0, $m_{\rm top}$=178 GeV), and pick the $M_1/m_{1/2}$ values which
give $\Omega_{\widetilde Z_1}h^2=0.11$, {\em i.e.}, respectively,
$M_1/m_{1/2}=-1.619$ and $M_1/m_{1/2}=1.630$.  We plot in
Fig.~\ref{fig:sv_th_av} the thermally averaged cross section including
coannihilations (solid) and without coannihilations (dashed lines) times
the relative velocity as a function of the temperature $T$.  In the same
sign MWDM case, one notices that the very significant wino-component in
the photino-like lightest neutralino gives a non-negligible $s$-wave
contribution via annihilation to $W$ pairs, ({\em i.e.}, a contribution
which goes as $\langle\sigma v\rangle(T)\sim a$), while in the opposite
sign BWCA case the lightest neutralino is a pure bino, and the squark
mediated $t$-channel dominated pair annihilation cross section is
strongly $s$-wave suppressed ($\langle\sigma v\rangle(T)\sim
a^\prime+b^\prime (T/m_{\rm LSP})$, $a^\prime\ll b^\prime$).  As a
result, the pair annihilation cross section at $T=0$ (relevant for
indirect DM detection) is more than one order of magnitude suppressed in
the BWCA case.

A second effect, also traced back to the lack of a significant wino component
in the opposite sign BWCA case, is the role and onset of coannihilations.
To isolate the role of coannihilations, we show by dashed lines
$\langle\sigma v\rangle(T)$ computed without any coannihilation
contribution. In the BWCA case, coannihilations play a much more
important role, as can be understood by looking at the relative size of
the cross sections with and without coannihilations around the lightest
neutralino freeze-out temperature $T_{\rm f.o.}$, indicated by an arrow
in the figure. Furthermore, in the BWCA case, the onset of the
coannihilation regime takes place at lower temperatures, a reflection of
the reduced mass splitting between the coannihilation partners and the
LSP.  At the freeze-out temperature, $\langle\sigma v\rangle(T)$ is
dominated by coannihilations in the opposite sign BWCA case, while the
relative coannihilation contribution for the MWDM case (for which the
$\tz_1$s can annihilate to $WW$) is significantly smaller. Observe also the
indicated relic abundance of the LSP computed without coannihilations,
respectively 0.759 in the BWCA case and 0.138 in the same MWDM.

Coannihilations of particles $A$ and $B$ effectively enter the
neutralino pair annihilation cross section when the center of mass
momentum $p_{\rm c.m.}$ of the neutralino-neutralino system satisfies
the relation
\begin{equation}
s=4p_{\rm c.m.}^2+4m_{\widetilde Z_1}^2 \ge (m_A+m_B)^2.
\end{equation}
A suitable quantity to illustrate the onset of coannihilations is given
by an effective annihilation rate $(\sigma v)_{\rm eff}$, defined as in
Ref. \cite{esug}
\begin{equation}\label{eq:weff}
(\sigma v)_{\rm eff}\equiv\frac{W_{\rm eff}(p_{\rm c.m.})}{4E^2_{\rm
c.m.}}, \qquad E_{\rm c.m.}=\sqrt{p_{\rm c.m.}^2+m_{\widetilde Z_1}^2}
\end{equation}
(we refer the reader to Ref.~\cite{esug} for the definition of the
effective annihilation rate $W_{\rm eff}$, which is essentially the sum
over all (co-)annihilation channels, properly weighted, of the various
annihilation rates per unit volume and unit time) and such that
\begin{equation}
\lim_{p_{\rm c.m.}\rightarrow 0}(\sigma v)_{\rm eff} =\langle\sigma
v\rangle(T=0) .
\end{equation}
The temperature dependence of $\langle\sigma v\rangle(T)$ is factored
out in the weight function $\kappa(p_{\rm c.m.},T)$ \cite{esug}, so that
\begin{equation}
\langle\sigma v\rangle(T)=\int_0^\infty{\rm d}p_{\rm c.m.}\frac{W_{\rm
eff}(p_{\rm c.m.})}{4E^2_{\rm c.m.}}\kappa(p_{\rm c.m.},T) .
\end{equation}

\begin{figure*}[!t]
\begin{center}
\includegraphics[scale=0.5]{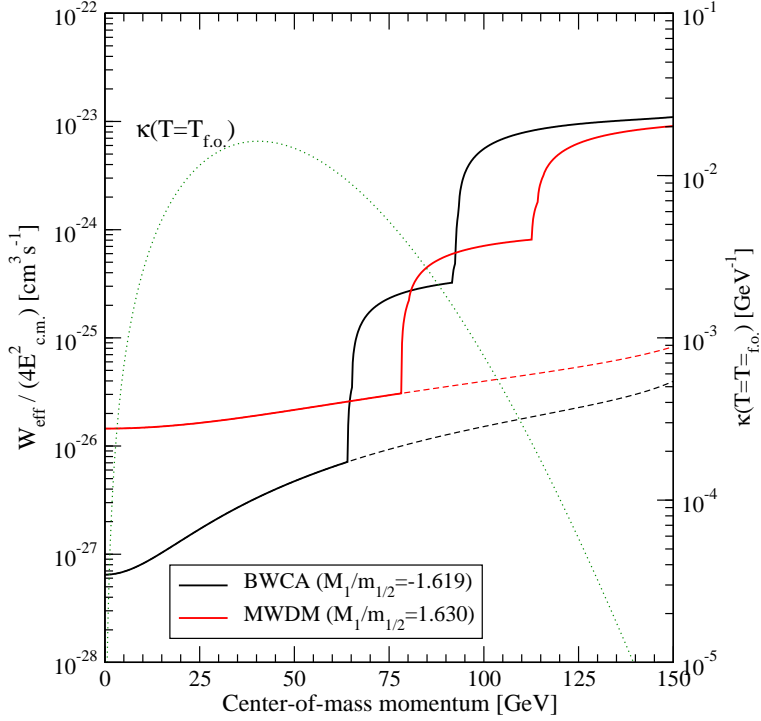}
\end{center}
\caption{The effective $(\sigma v)_{\rm eff}$ term, as defined in
Eq.~(\protect\ref{eq:weff}), as a function of the center of mass
momentum $p_{\rm eff}$, for the BWCA (black lines) and MWDM (red lines)
scenarios, with (solid lines) and without (dotted lines)
coannihilations. The green dotted line indicates the ``{\em weight
function}'' $\kappa(p_{\rm c.m.},T=T_{\rm f.o.})$.}
\label{fig:weff}
\end{figure*}

In Fig.~\ref{fig:weff}, we plot the two effective annihilation rates
$(\sigma v)_{\rm eff}$ for the two BWCA and MWDM sample cases at the
mSUGRA point given above, with the correct WMAP relic abundance.  The
coannihilation thresholds are shifted to larger $p_{\rm c.m.}$ values in
the same sign case (the mass splitting of coannihilating partners is
increased).  The $\widetilde W_1$-$\widetilde Z_2$ mass splitting is too
small to resolve the separate contributions, and the two bumps
correspond to the onset of $\widetilde Z_1$-$\widetilde W_1,\widetilde
Z_2$ coannihilations and the onset of (co-)annihilations amongst
$\widetilde W_1$ and $\widetilde Z_2$.  When the bino-wino mixing is
suppressed, the second processes contribute much more than the first
processes to $(\sigma v)_{\rm eff}$. However, as shown by the weight
function $\kappa(p_{\rm c.m.},T=T_{\rm f.o.})$, $(\sigma v)_{\rm eff}$
is largely sampled in a $p_{\rm c.m.}$ range where the $\widetilde
W_1,\widetilde Z_2$ (co-)annihilations are not kinematically accessible.
This picture, however, depends on the LSP mass : had we picked a larger
value for $m_{\tz_1}$, the mass splitting between $\tz_1$ and $\tz_2,\
\tw_2$ needed to obtain a sufficiently low relic abundance would have
been smaller, so that the coannihilation bumps in $W_{eff}/4E^2$ would
have occurred at smaller center-of-mass momenta, where the sampling
function $\kappa$ has not yet fallen to very small values. In this case,
the role of $\tw_1-\tz_2$ coannihilations would have been enhanced even
in the MWDM case.

\begin{figure*}[!t]
\begin{center}
\includegraphics[scale=0.5]{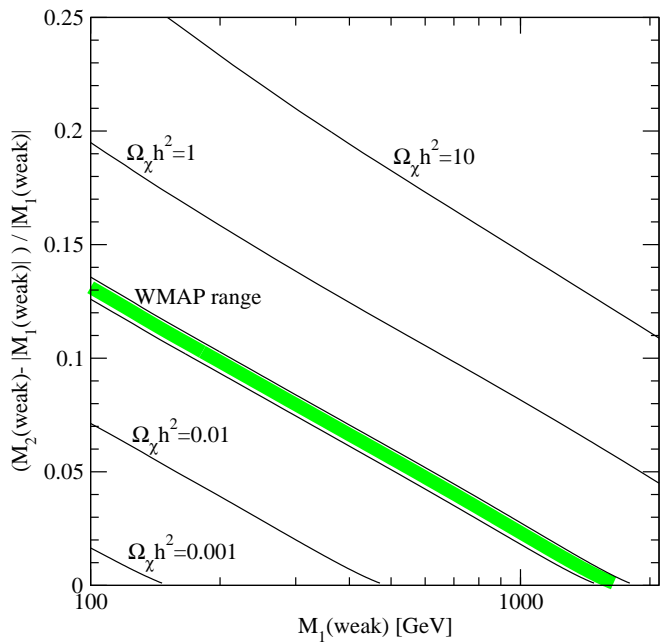}
\end{center}
\caption{Iso-level curves at given values of the neutralino relic abundance in the $M_1,(M_2-|M_1|)/|M_1|$ plane. The green strip indicates the WMAP 2-$\sigma$ range. We assumed here $\mu,m_{\widetilde S},m_A=100\cdot M_2$, $m_{\widetilde S}$ being all sfermion masses.}
\label{fig:omega}
\end{figure*}

In the limit in which sfermions are heavy, 
and thus the bino annihilation 
cross section is extremely suppressed, and in which
$\mu\gg M_1(\rm weak),\ M_2(\rm weak)$ 
(in the remainder of this paragraph, we will
implicitly use $M_{1,2}=M_{1,2}(\rm weak)$), 
{\em i.e.}, in the pure gaugino limit, the relic abundance should only
depend on {\it i}.) the LSP mass scale and on {\it ii.})  the LSP-wino
system splitting relative to the LSP mass. Since a pure wino-like system
has a relic abundance which goes like \cite{Profumo:2005xd}
\begin{equation}
\Omega_{\rm pure \ winos}h^2\simeq c\cdot\left(\frac{M_2}
{1\ {\rm TeV}}\right)^\gamma,\qquad c\simeq0.024,\ \gamma\simeq 1.9 ,
\end{equation}
the relic abundance of a coannihilating bino will be given by that of
the pure wino system, rescaled by the exponential factor
$\exp[-((M_2-M_1)/M_1)\cdot \widetilde x)]$, with
$\widetilde x\simeq M_1/T_{\rm f.o.}$, and rescaled by the new parasite
bino degrees of freedom\footnote{The degrees of freedom should be as
well weighted according to the mass splitting, but this is a higher
order effect.}.  The wino system carries 2+4 degrees of freedom, while
the bino 2, hence we expect an enhancement of the relic abundance for a
coannihilating bino of a factor $(4/3)^2$ \cite{Profumo:2004wk}. The
relic abundance should then take the form
\begin{equation}
\Omega_{\rm bino+wino}h^2\approx a\cdot\left(\frac{M_2}{1\ {\rm TeV}}\right)^\gamma
\cdot\exp\left[\frac{M_2-M_1}
{M_1}\cdot \widetilde x\right],\quad\quad a\approx c(4/3)^2\label{eq:oh2} .
\end{equation}
Since $M_2\simeq M_1$, this allows in principle to define a strip in the $M_1,\frac{M_2-M_1}{M_1}$ 
plane of WMAP preferred relic abundance. We show the result in Fig.~\ref{fig:omega}, 
taking $\mu,m_{\widetilde S},m_A=100\cdot M_2$, $m_{\widetilde S}$ being all sfermion masses. 
Indeed, the iso-level curves for the relic abundance show the functional form of Eq.~(\ref{eq:oh2}).

\subsection{Direct and indirect detection of neutralino CDM}

In this section, we turn to consequences of the BWCA scenario
for direct and indirect detection of neutralino dark matter\cite{eigen}. 
We adopt the DarkSUSY code\cite{darksusy}, 
interfaced to Isajet, for the computation of the various rates, and resort 
to the Adiabatically Contracted 
N03 Halo model\cite{n03} for the dark matter distribution 
in the Milky Way\footnote{For a comparison of the implications of 
different halo model choices for indirect DM detection rates, see
{\it e.g.} Refs. \cite{bo,bbko,antimatter,nuhm}.}.
We evaluate the following neutralino DM detection rates:
\begin{itemize}
\item Direct neutralino detection via underground cryogenic
detectors\cite{direct}.  Here, we compute the spin independent
neutralino-proton scattering cross section, and compare it to expected
sensitivities\cite{bbbo} for Stage 2 detectors (CDMS2\cite{cdms2},
Edelweiss2\cite{edelweiss}, CRESST2\cite{cresst}, ZEPLIN2\cite{zeplin})
and for Stage 3, ton-size detectors (XENON\cite{xenon},
GERDA\cite{gerda}, ZEPLIN4\cite{zeplin4} and WARP\cite{warp}).  We
take here as benchmark experimental reaches of Stage 2 and Stage 3
detectors the projected sensitivities of, respectively, CDMS2 and XENON
1-ton at the corresponding neutralino mass.

\item Indirect detection of neutralinos via neutralino annihilation to
neutrinos in the core of the Sun\cite{neut_tel}. 
Here, we present rates for detection of $\nu_\mu \to \mu$ conversions
at Antares\cite{antares} or IceCube\cite{icecube}. 
The reference experimental sensitivity we use is that of IceCube, 
with a muon energy threshold of 25 GeV, corresponding to a flux 
of about 40 muons per ${\rm km}^2$ per year. 
\item Indirect detection of neutralinos via neutralino annihilations in the
galactic center leading to gamma rays\cite{gammas}, 
as searched for by EGRET\cite{egret}, and 
in the future by GLAST\cite{glast}. 
We evaluate the integrated continuum $\gamma$ ray flux above a 
$E_\gamma=1$ GeV threshold, and assume a GLAST sensitivity 
of 1.0$\times10^{-10}\ {\rm cm}^{-2}{\rm s}^{-1}$.
\item Indirect detection of neutralinos via neutralino annihilations in the
galactic halo leading to cosmic antiparticles, including
positrons\cite{positron} (HEAT\cite{heat}, Pamela\cite{pamela} 
and AMS-02\cite{ams}), antiprotons\cite{pbar} (BESS\cite{bess}, 
Pamela, AMS-02) and anti-deuterons ($\bar{D}$s) (BESS\cite{bessdbar}, 
AMS-02, GAPS\cite{gaps}). 
For positrons and antiprotons we evaluate the averaged differential 
antiparticle flux in a projected energy bin centered at a kinetic 
energy of 20 GeV, where we expect an optimal statistics and 
signal-to-background ratio at space-borne antiparticle 
detectors\cite{antimatter,statistical}. We take the  
experimental sensitivity that of the Pamela experiment 
after three years of data-taking as our benchmark.
Finally, the average differential antideuteron flux has been 
computed in the $0.1<T_{\bar D}<0.4$ GeV range, 
where $T_{\bar D}$ stands for the antideuteron kinetic energy per nucleon, 
and compared to the estimated GAPS sensitivity\cite{gaps} (see Ref.~\cite{baerprofumo} for an updated discussion of the role of antideuteron searches in DM indirect detection).
\end{itemize}

In Fig. \ref{dmrates1}, we show various direct and indirect DM detection
rates for $m_0=m_{1/2}=300$ GeV, with $A_0=0$, $\tan\beta =10$ and $\mu
>0$, while $M_1$ is allowed to vary.  The $M_1$ value corresponding to
the mSUGRA model is denoted by a dot-dashed vertical line, while the
BWCA and MWDM scenarios with $\Omega_{\tz_1}h^2=0.11$ are denoted by
dash-dash-dot and dashed vertical lines, respectively. The dotted lines
correspond to the sensitivity level of each of these experiments; {\it
i.e.}, the signal is observable only when the model prediction is higher
than the corresponding dotted line. While the minimum sensitivity for
the direct detection rates in frames {\it b}) -- {\it f}) refers to the minimum
magnitude of the signal that is detectable (and hence independent of the
LSP mass), the smallest detectable cross section shown by the dotted
curves in frame {\it a})
depends on the value of $m_{\tz_1}$.

In frame {\it a}), we plot the spin-independent neutralino-proton
scattering cross section. 
We see that as $M_1$ is decreased, and becomes increasingly negative, the 
neutralino-proton scattering cross section plummets to values
in the $10^{-12}$ pb range, far below the sensitivity of any planned detector.
The drop-off is due to increasing negative interference amongst the 
contributing Feynman diagrams.
\FIGURE[htb]{
\mbox{\hspace{-1cm}
\epsfig{file=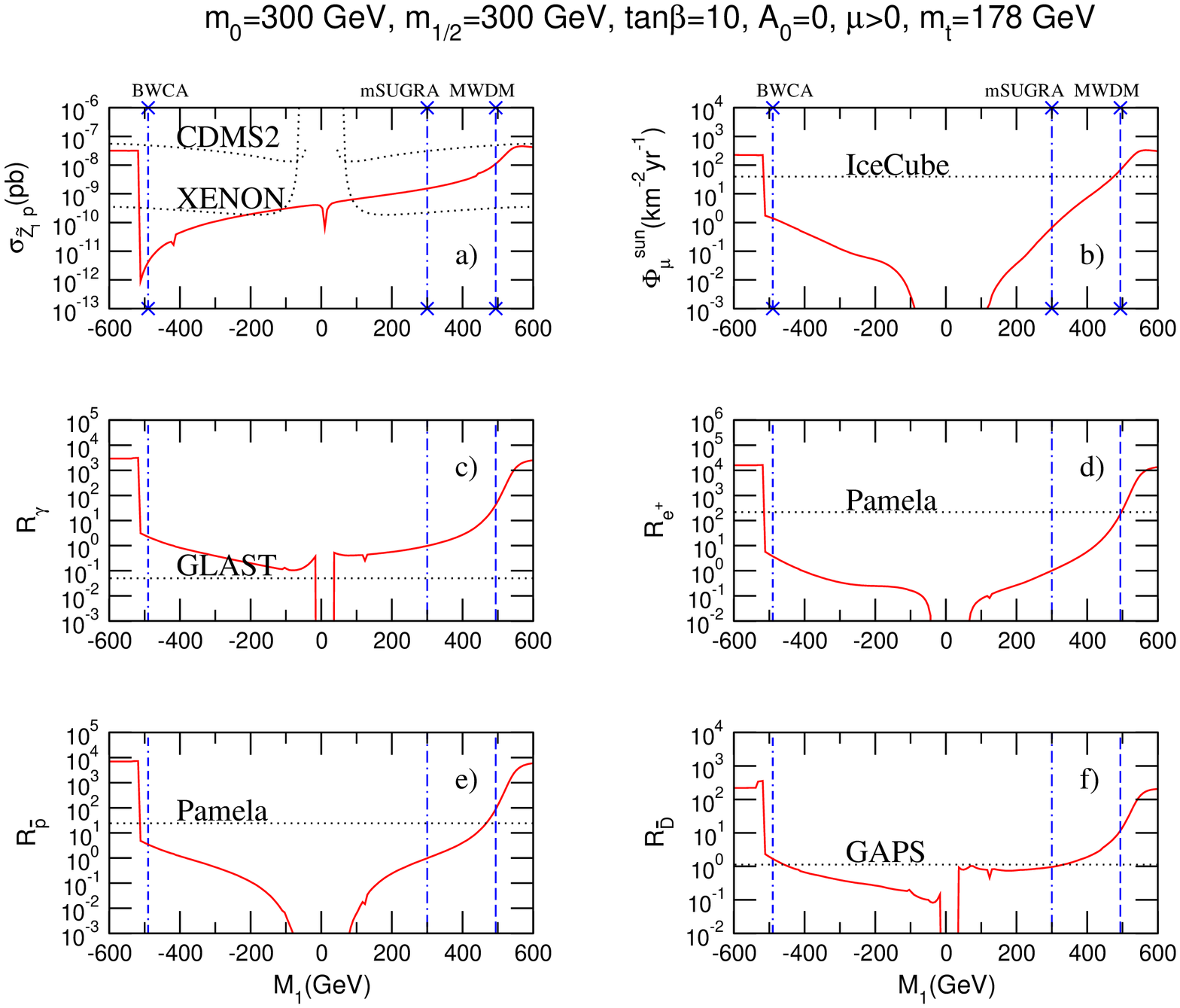,width=10cm,angle=-90} }
\caption{\label{dmrates1}
Rates for direct and indirect detection of neutralino dark matter
vs. $M_1$ for $m_0=m_{1/2}=300$ GeV, with
$\tan\beta =10$, $A_0=0$, $\mu >0$. 
Frames {\it c}) -{\it f}) show the ratio of indirect detection rates 
compared to the mSUGRA model. 
In this plot, we adopt the adiabatically contracted 
N03 distribution for halo dark matter.}}

In frame {\it b}), we show the flux of muons from neutralino 
pair annihilations  in the core of the Sun.
The muon flux is below the reach of IceCube in the mSUGRA case, and it
remains below IceCube observability in the BWCA case.
The rate for
neutralino annihilation in the sun or earth is given by
\be
\Gamma_A ={1\over 2}C\tanh^2(t_{\odot}/\tau),
\ee
where $C$ is the capture rate, $A$ is the total annihilation rate times
relative velocity per volume, $t_\odot$ is the present age of the solar
system and $\tau =1/\sqrt{CA}$ is the equilibration time. 
For small $C$, as shown in frame {\it a}), the equilibration time becomes 
large, so that $\Gamma_A\sim\frac{1}{2}C^2 A t^2$,
and is hence sensitive to the neutralino annihilation cross section
times relative velocity, 
unlike cases where the neutralino-nucleon scattering cross section is large.
The muon flux jumps to observable levels at more negative values of $M_1$, 
but this is only because the $\tz_1$ suddenly becomes wino-like,
so that the relic density becomes too low.

In frames {\it c}), {\it d}), {\it e}) and {\it f}) we show the flux of
photons, positrons, antiprotons and antideuterons, respectively.  The
results here are plotted as ratios of fluxes normalized to the mSUGRA
point, in order to give results that are approximately halo-model
independent.  (We do show the above described expected experimental
reach lines as obtained by using the Adiabatically Contracted N03 Halo
model\cite{n03}.)  The rates for indirect detection via observation of
halo annihilation remnants are typically low in the BWCA scenario, since
the bino annihilation cross sections are $s$-wave suppressed. Observable
results are indicated for $\gamma$ rays by GLAST, but this is due in
part to the very favorable N03 halo distribution which is assumed.

In Fig. \ref{dmrates2}, we show the same direct and indirect DM
detection rates as in Fig. \ref{dmrates1}, except this time versus $M_2$
instead of $M_1$. In frame {\it a}), the neutralino-nucleon scattering
rates do not have negative interference, and can remain at observable,
although not enhanced, levels.  The rates for detection of BWCA DM at
IceCube are relatively low, as are rates for anti-matter detection by
Pamela in frames {\it d}) and {\it e}).  The rates for $\gamma$
detection by GLAST in frame {\it c}) are similar to those from the
mSUGRA case, while the rate for antideuteron detection by GAPS is just
barely observable for BWCA DM in frame {\it f}).
\FIGURE[htb]{
\mbox{\hspace{-1cm}
\epsfig{file=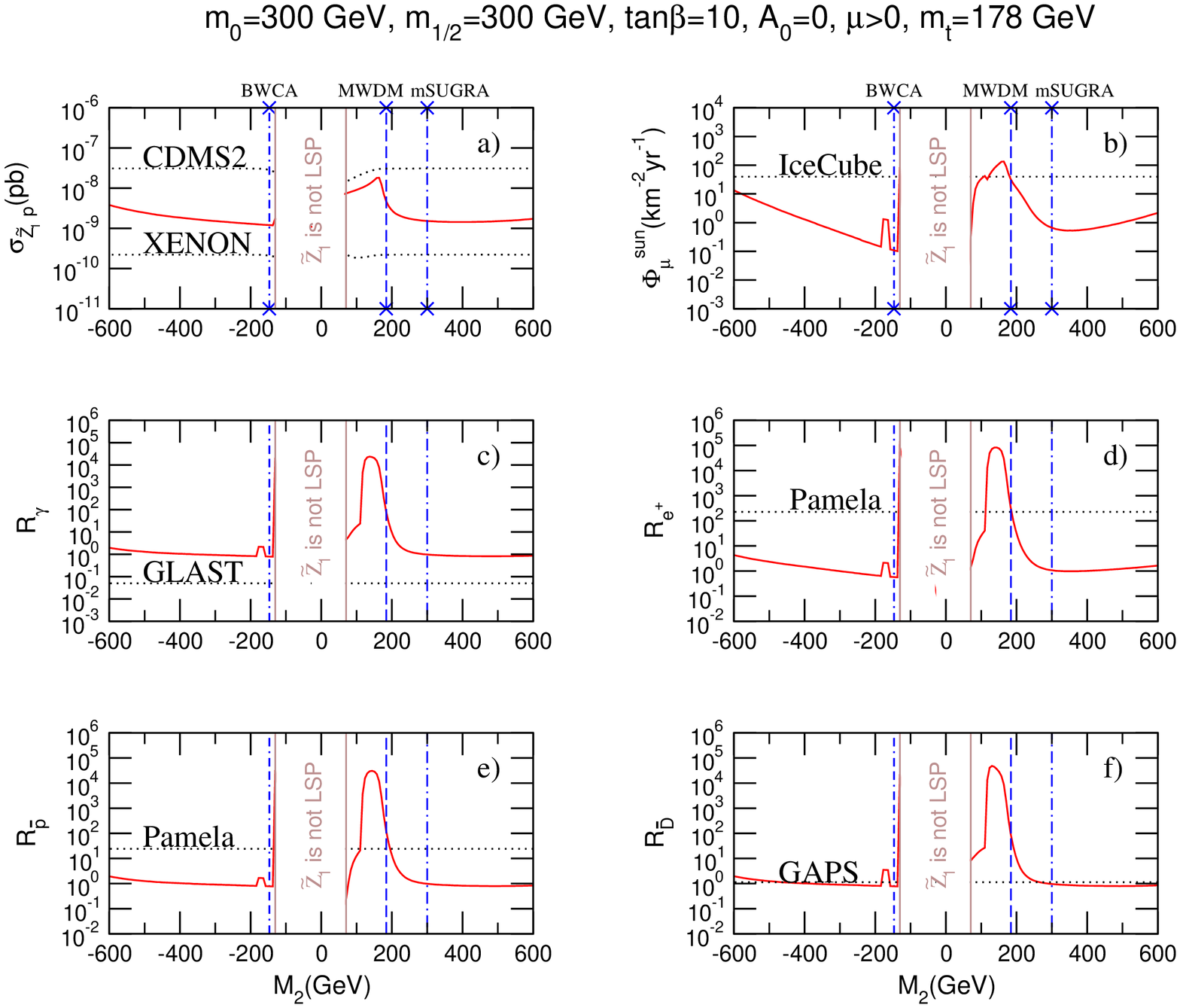,width=10cm,angle=-90} }
\caption{\label{dmrates2}
Rates for direct and indirect detection of neutralino dark matter
vs. $M_2$ for $m_0=m_{1/2}=300$ GeV, with
$\tan\beta =10$, $A_0=0$, $\mu >0$. 
Frames {\it c}) -{\it f}) show the ratio of indirect detection rates 
compared to the mSUGRA model. 
In this plot, we adopt the adiabatically contracted
N03 distribution for halo dark matter.}}

Overall, prospects for direct or indirect detection of BWCA dark matter
are generally at or below levels expected in the mSUGRA model.  For this
reason, we do not present direct and indirect detection rates in the
$m_0\ vs.\ m_{1/2}$ plane. We also mention that the situation is sharply
different in the MWDM scenarios where the corresponding rates are
generally larger than in the mSUGRA model. Thus, a detection of a signal
in the XENON or Pamela experiments could serve to discriminate between
these scenarios, especially if we already have some information of the
SUSY spectrum from collider experiments.

\section{Neutralino radiative decay in the BWCA scenario}
\label{sec:z2z1g}

The loop-induced radiative decay width for $\tz_2\to \tz_1\gamma$ has
been calculated in Ref. \cite{width,hw}. A thorough numerical
analysis\cite{am,bk} has shown that the radiative decay can be large and
even dominant in certain regions of MSSM parameter space.  A necessary
(but not sufficient) condition for this is that all tree level
two body decay modes of $\tz_2$ be kinematically forbidden. In this
case, $\tz_2$ usually decays via $\tz_2 \to \tz_1 f{\bar f}$, where $f$
is a light SM fermion. However, if $\tz_1$ and $\tz_2$ are close in
mass, the formally higher order two body radiative decay becomes
competitive with the three body decays $\tz_2\to f\bar{f}\tz_1$. This is
because $\Gamma(\tz_2 \to \tz_1\gamma) \propto
(1-m_{\tz_1}/m_{\tz_2})^3$, while $\Gamma(\tz_2\to \tz_1 f\bar{f})
\propto (1-m_{\tz_1}/m_{\tz_2})^5$. 

As we have seen, in both the MWDM and BWCA scenarios $m_{\tz_2}-m_{\tz_1}$
is small, so that we may expect that the branching fraction for radiative
decays may be enhanced.  Moreover, in both cases, the couplings of the
neutralinos to the $Z$ boson, which occur only via the higgsino
components of the neutralino are strongly suppressed so that virtual
$Z$ boson exchange contribution to three body decay amplitudes
is correspondingly suppressed. However, vector
boson-gaugino loops essentially also decouple from the radiative decay in the
BWCA case because the bino does not couple to these. These do not,
however, decouple in the MWDM case since both the photino and the zino
couple to $W^\pm \tw^\mp$ system. The branching fraction for the
radiative decay is thus a result of a complicated interplay between the
kinematic and dynamic effects discussed above.



In Fig. \ref{rd_bf1}, we show in upper frame {\it a}) the 
neutralino relic density, and in {\it b}) the $BF(\tz_2\to\tz_1\gamma )$
for the same parameters as in Fig. \ref{fig:rd_bw1}, 
versus GUT scale gaugino mass 
$M_1$. The radiative branching fraction at this point in the mSUGRA
model is just $\sim 10^{-5}$. As $M_1$ climbs to 490 GeV, in the MWDM
case, the branching fraction has climbed to $\sim 5\%$.
When $M_1$ varies to large negative values in the BWCA case, the branching 
fraction has climbed to $10\%$. In this case, we may expect that
a considerable fraction of SUSY events at colliders 
to contain hard isolated photons via the decay of $\tz_2$ that is
produced either directly, or via cascade decays of heavier sparticles. 
In the two lower frames, we show the same figures except for large 
$m_0=1$ TeV. In this case, the sfermion loops mediating the 
$\tz_2\to\tz_1\gamma$ decay become suppressed, and the branching fraction
is much smaller, reaching $\sim 8\%$ in the MWDM case, and just
$1\%$ in the BWCA case.
\FIGURE[!t]{
\epsfig{file=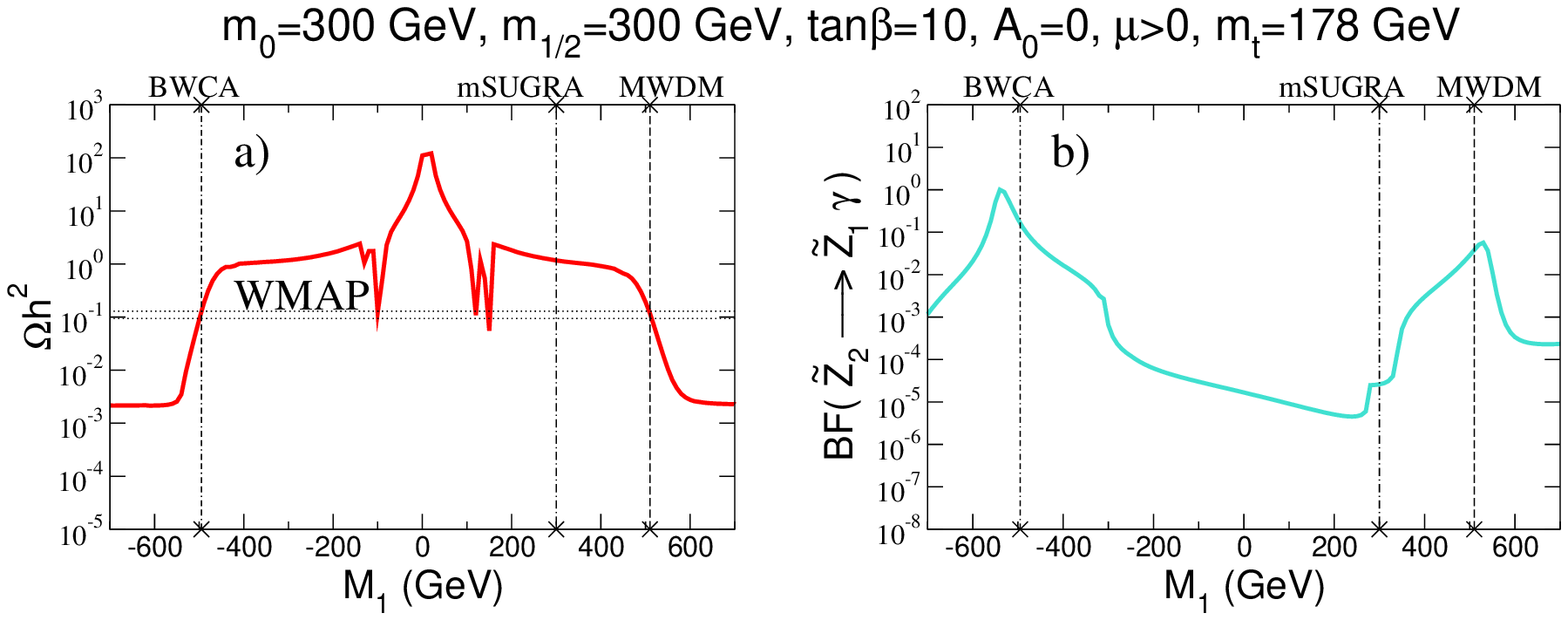,width=7cm,angle=-90}\\
 \epsfig{file=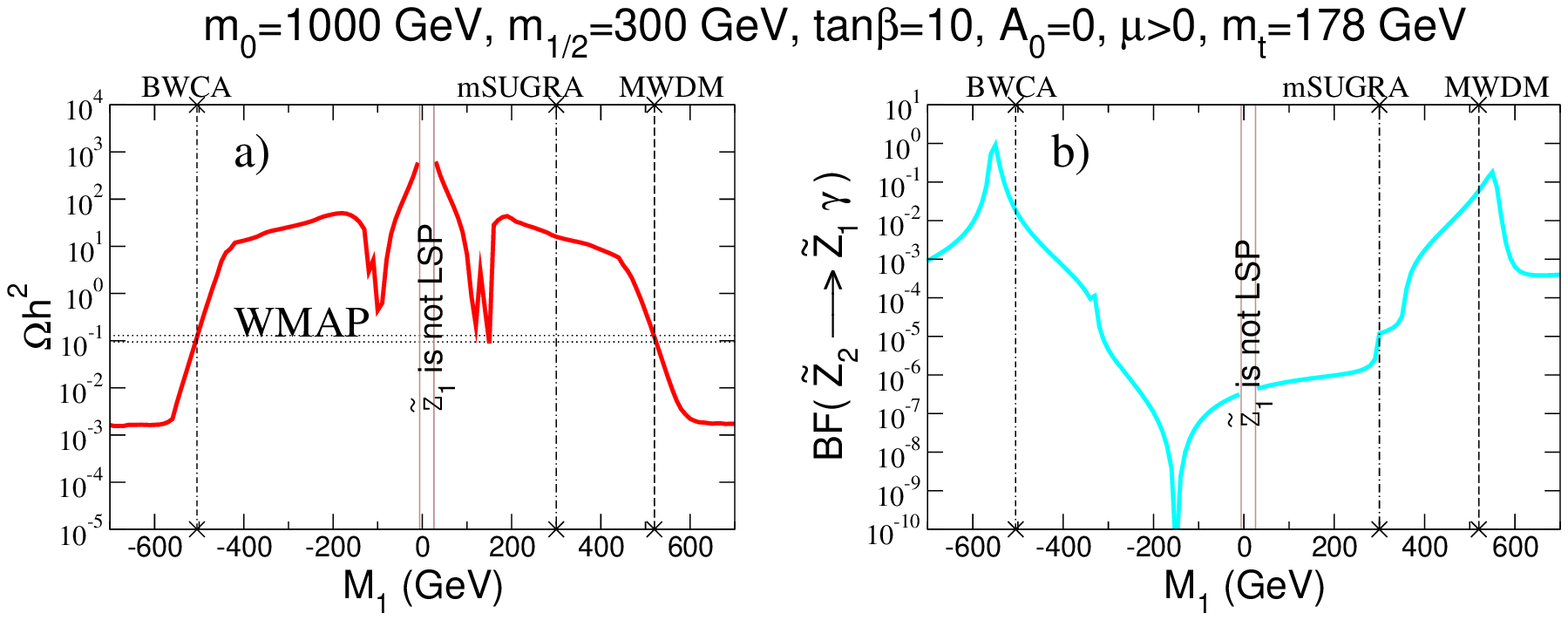,width=7cm,angle=-90}
\caption{\label{rd_bf1}
Upper: A plot of {\it a}) relic density $\Omega_{CDM}h^2$ and 
{\it b}) $BF(\tz_2\to\tz_1\gamma$) as a
function of $M_1$ for
$m_0=300$ GeV, $m_{1/2}=300$ GeV, $A_0=0$, $\tan\beta =10$, $\mu >0$
and $m_t=178$ GeV. Lower: Same plot for $m_0=1000$ GeV.
}}
\FIGURE[!t]{
\epsfig{file=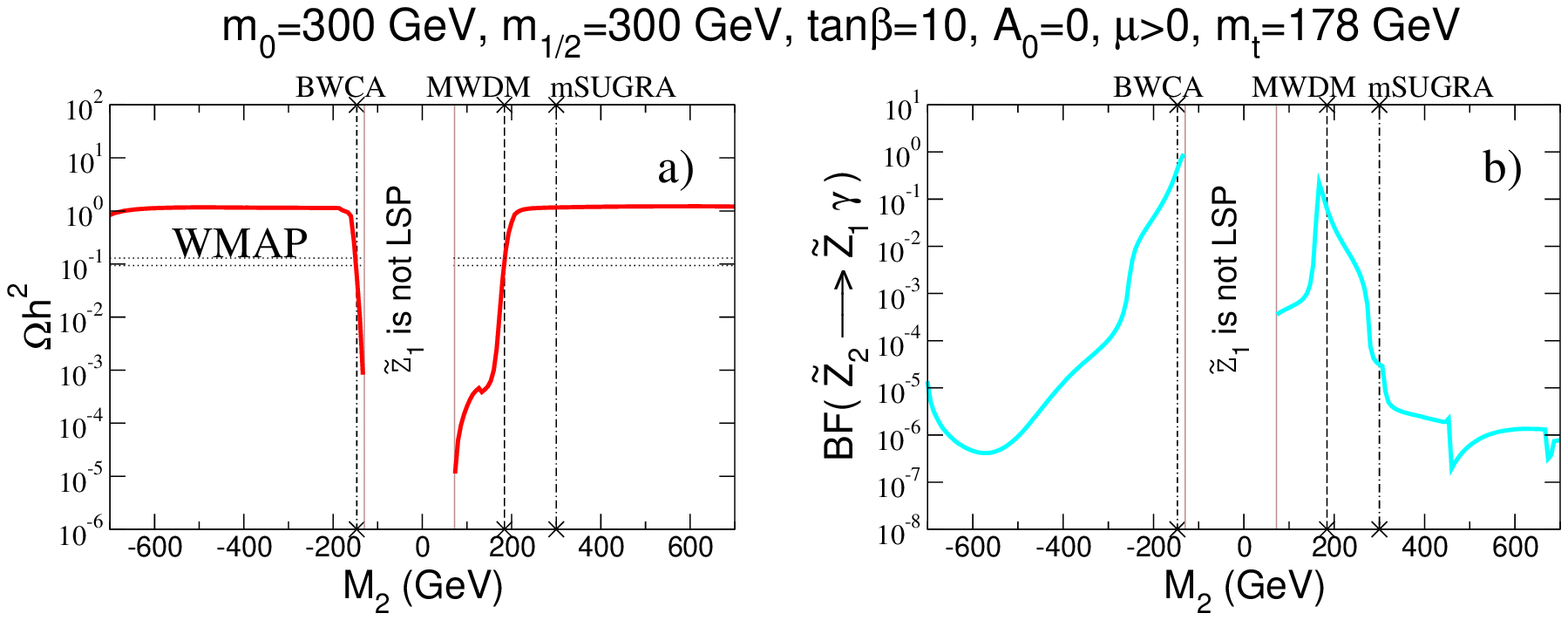,width=7cm,angle=-90}\\
\epsfig{file=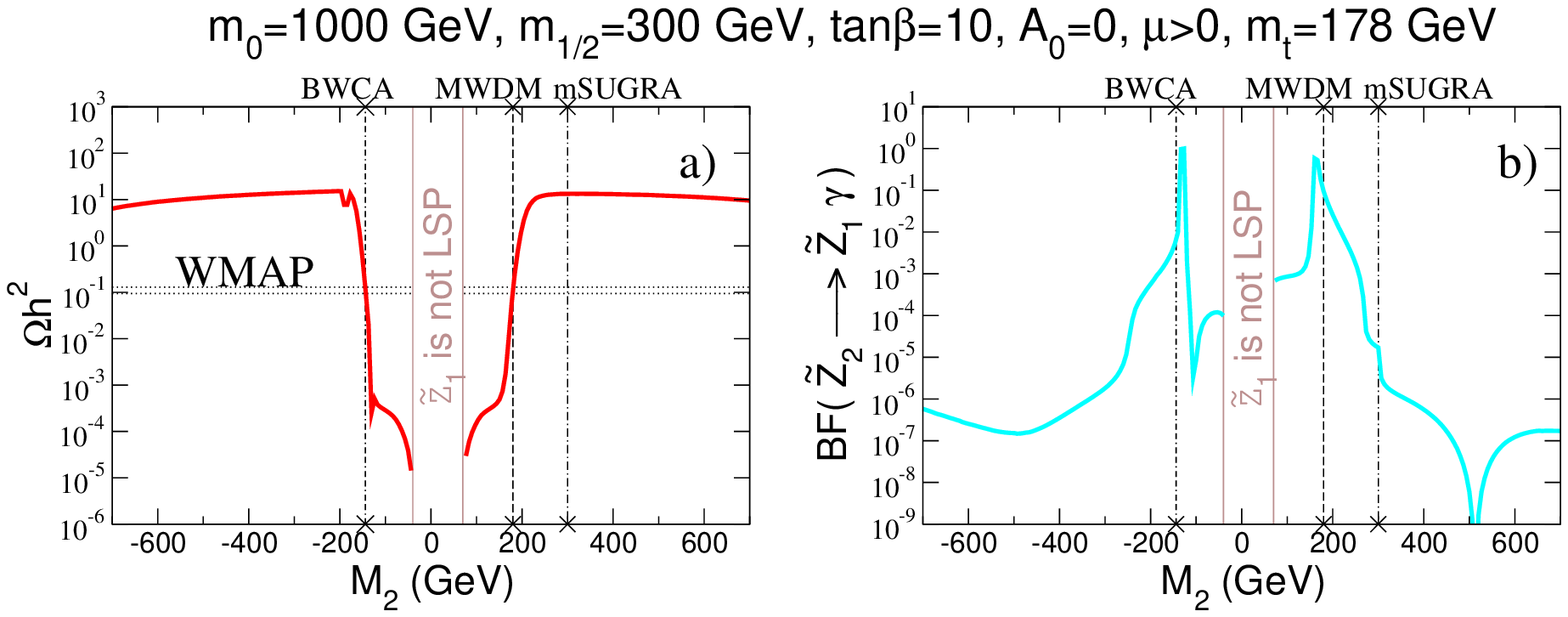,width=7cm,angle=-90}
\caption{\label{rd_bf2}
Upper: A plot of {\it a}) relic density $\Omega_{CDM}h^2$ and 
{\it b}) $BF(\tz_2\to\tz_1\gamma$) as a
function of $M_2$ for
$m_0=300$ GeV, $m_{1/2}=300$ GeV, $A_0=0$, $\tan\beta =10$, $\mu >0$
and $m_t=178$ GeV. Lower: Same plot for $m_0=1000$ GeV.
}}
A similar situation is illustrated in Fig. \ref{rd_bf2}, where now we
plot versus variable $M_2$, while keeping $M_1=m_{1/2}=300$ GeV. In the
upper frames for $m_0=300$ GeV, we see that while
$BF(\tz_2\to\tz_1\gamma )$ reaches $5\%$ in the MWDM case, it reaches
25\% in the BWCA scenario.  In the lower frames, we see that for $m_0=1$
TeV, the branching fraction reaches $8\%$ for the case of MWDM and 0.8\%
for BWCA dark matter. We note that in both figures $B(\tz_2 \to
\tz_1\gamma)$ attains a higher value at its peak when $M_1/M_2$ is
negative because it is in this case that the sparticle masses get really
close (see the level crossings in Figs.~\ref{mass_M1} and
\ref{mass_M2}).
\FIGURE[!t]{
\epsfig{file=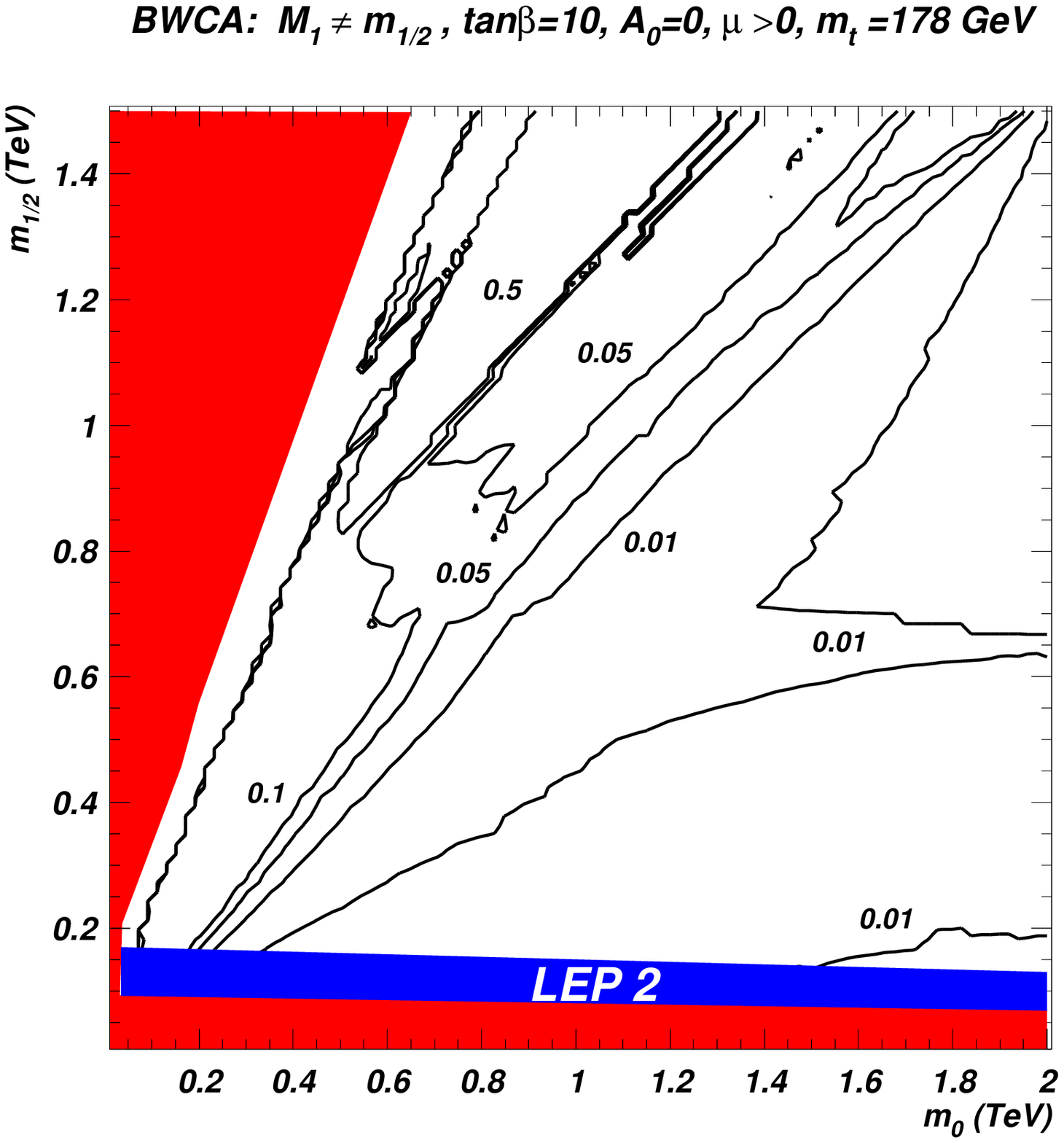,width=7cm},
\epsfig{file=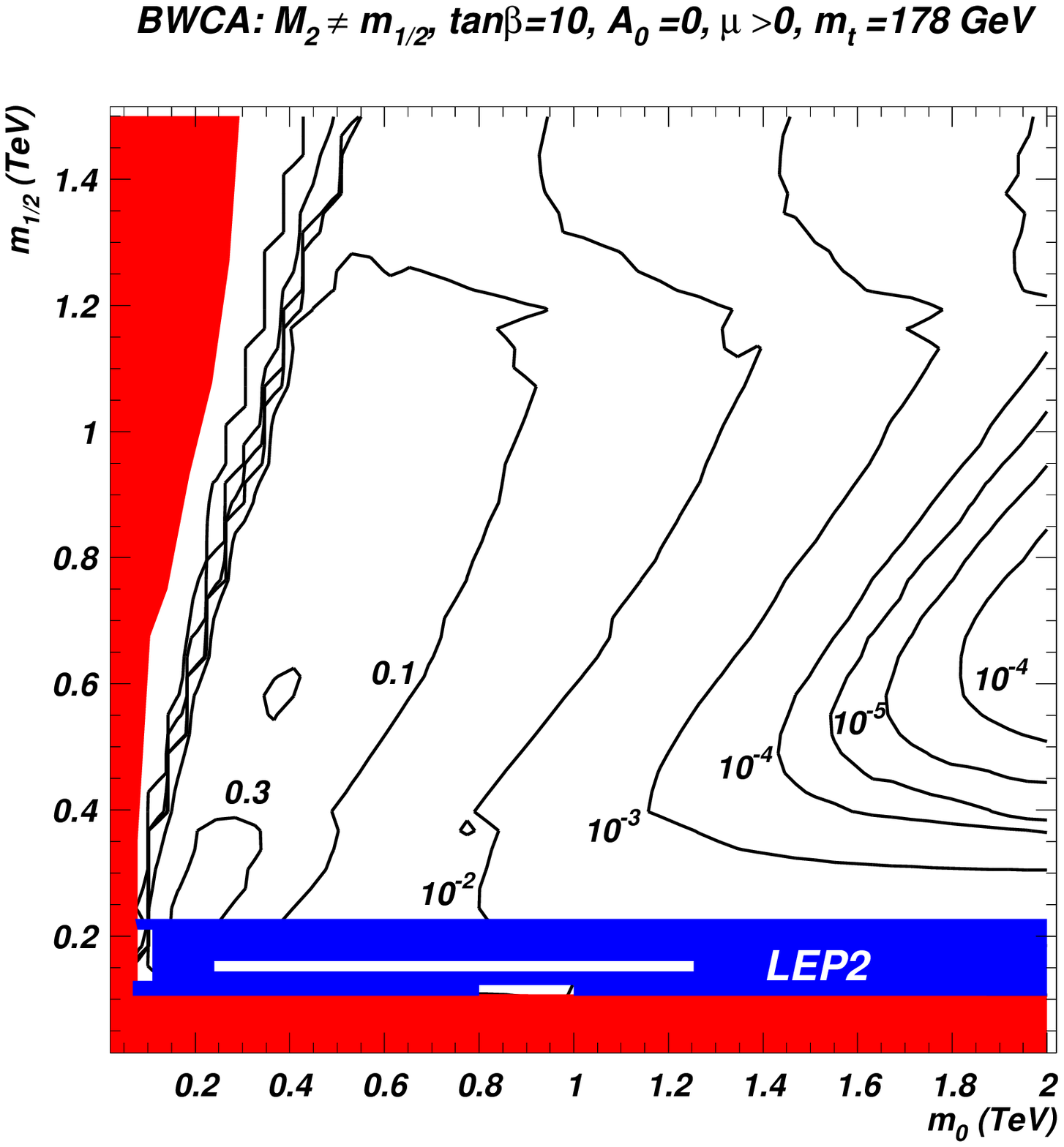,width=7cm}\\
\epsfig{file=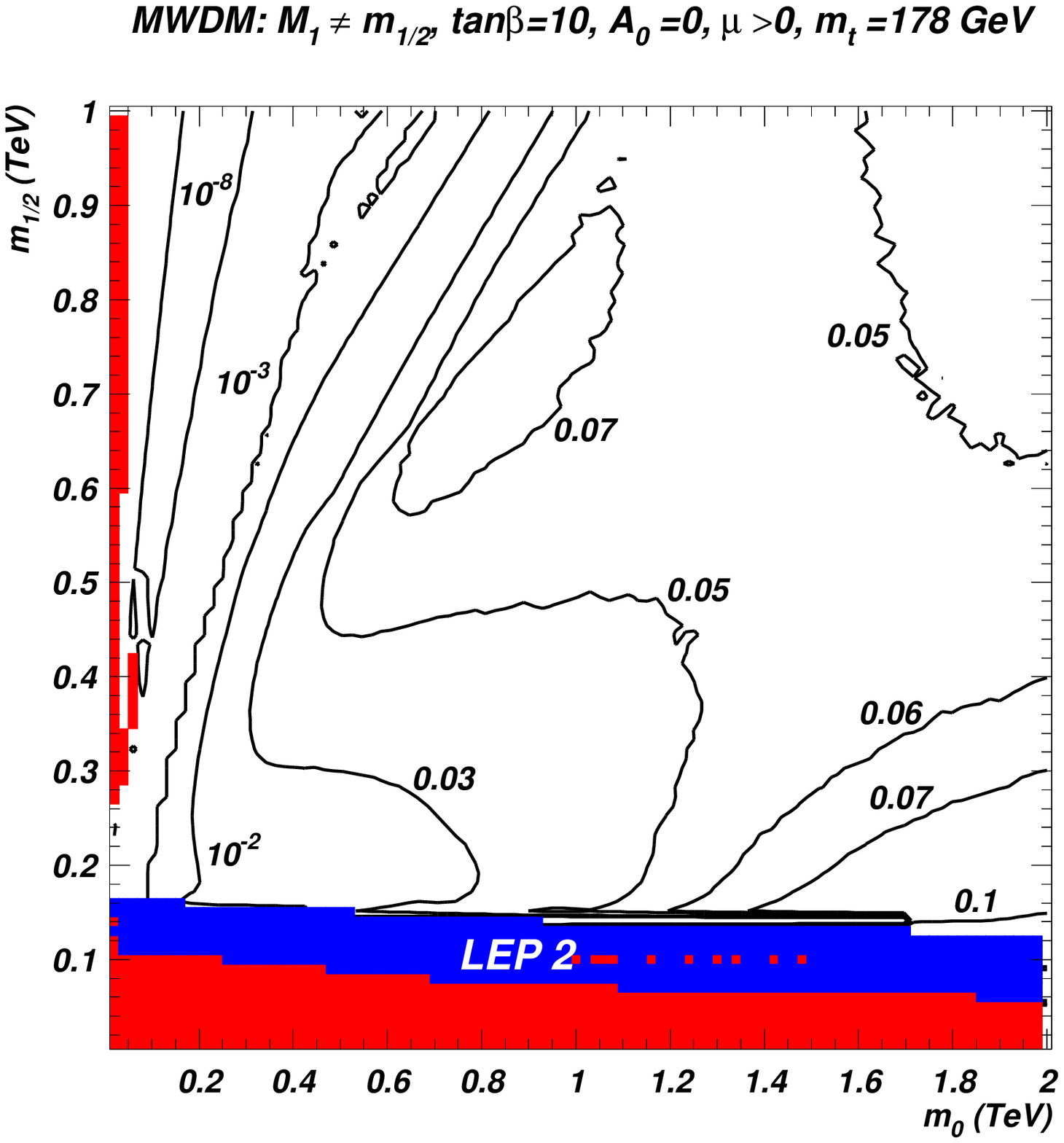,width=7cm},
\epsfig{file=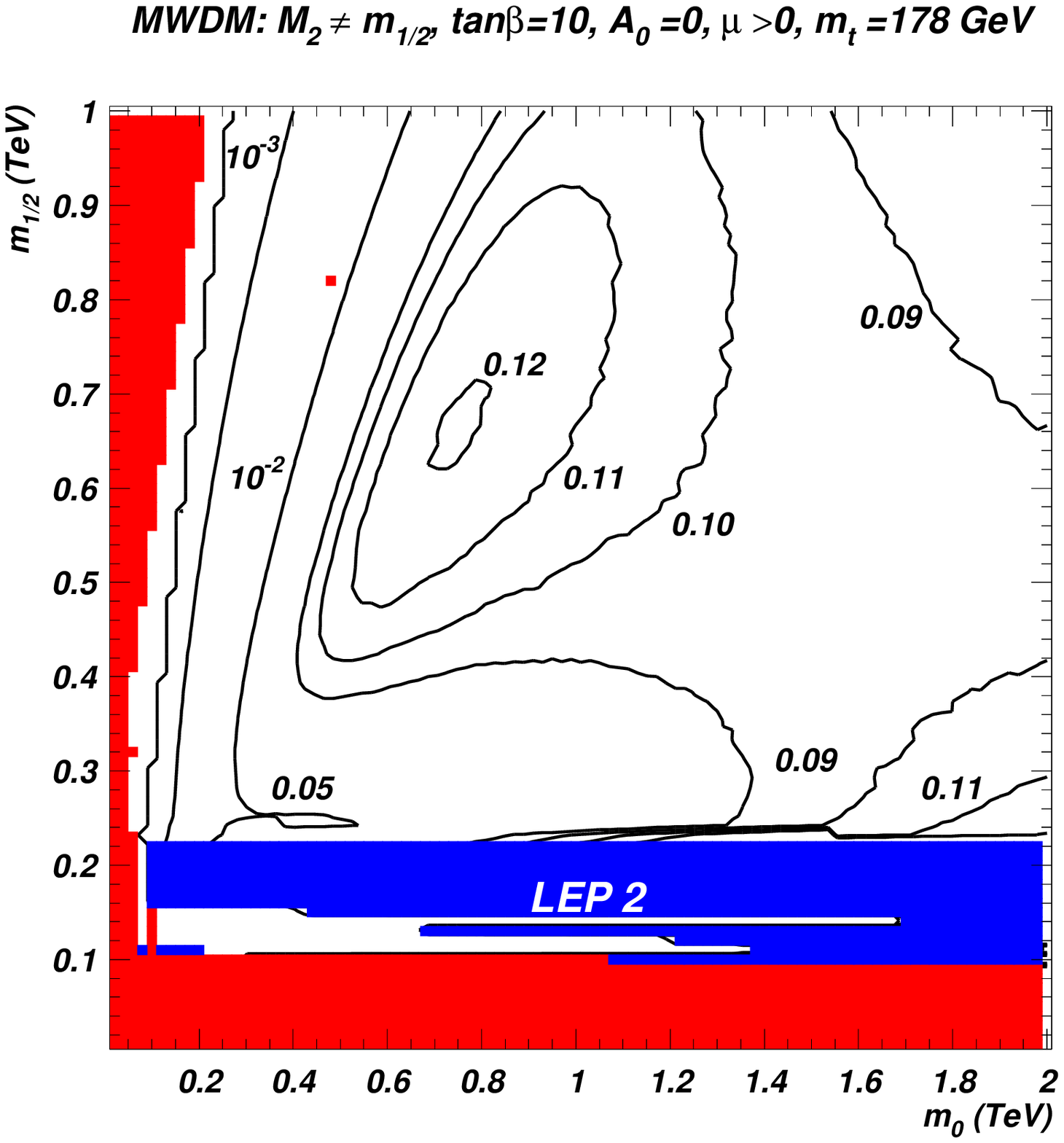,width=7cm}
\caption{\label{bfplane}
Contours of $BF(\tz_2\to\tz_1\gamma )$ in 
the $m_0\ vs.\ m_{1/2}$ plane for
$\tan\beta =10$, $A_0=0$, $\mu >0$, $m_t=178$ GeV.
In {\it a})., we have adjusted $M_1$ everywhere to negative values
so that $\Omega_{\tz_1}h^2=0.11$ in the BWCA scenario. 
In {\it b})., we have adjusted $M_2$ everywhere to negative values
so that $\Omega_{\tz_1}h^2=0.11$ in the BWCA scenario. 
In {\it c}). ({\it d}).), we have adjusted 
$M_1$ and ($M_2$) everywhere to positive values
so that $\Omega_{\tz_1}h^2=0.11$ in the MWDM scenario. 
}}

In Fig. \ref{bfplane}, we plot contours of $BF(\tz_2\to\tz_1\gamma )$ in
the $m_0\ vs.\ m_{1/2}$ plane for $A_0=0$, $\tan\beta =10$, $\mu >0$ and
$m_t=178$ GeV. At every point in the plane of frame {\it a}), we have
adjusted $M_1$ to a negative value chosen so that
$\Omega_{\tz_1}h^2=0.11$ in the BWCA scenario. We see that the
$BF(\tz_2\to\tz_1\gamma )$ exceeds 50\% around $m_0\sim 600-1000$ GeV
for $m_{1/2}\sim 1-1.4$ TeV. The branching fraction remains large at all
$m_{1/2}$ values, but diminishes for $m_0\agt 500-1000$ GeV.  In
Fig. \ref{bfplane}{\it b}), we plot the same contours except that at
every point in the plane, we have instead adjusted $M_2$ to negative
values so that $\Omega_{\tz_1}h^2=0.11$. In this case, we see again that
the branching fraction can become larger than 0.3 for low $m_0$ values.
In frames {\it c}) and {\it d}) once again we show contours of
$B(\tz_2\to\tz_1\gamma)$ but for the case of MWDM, where $M_1$ and
$M_2$ are dialed to positive values.  While the radiative branching
fractions never reach much beyond the 10\% level for MWDM, they maintain
a significant rate out to large values of $m_0$. This is because in the
MWDM case the radiative loop decays are dominated by $W$-chargino
exchange, whereas in the BWCA case the radiative loops are dominated by
fermion-sfermion exchange.

\section{BWCA dark matter at colliders}
\label{sec:col}

\subsection{$\tz_2-\tz_1$ mass gap in BWCA scenario}
\label{ssec:massgap}

An important question is whether collider experiments would be able to 
distinguish the case of BWCA dark matter from other forms of neutralino DM
such as MHDM as occur in the mSUGRA model or MWDM. We have seen 
from the plots of sparticle mass spectra that the squark and
gluino masses vary only slightly with changing $M_1$ or $M_2$. However, the
chargino and neutralino masses change considerably, and in fact
rather small mass gaps $m_{\tw_1}-m_{\tz_1}$ and $m_{\tz_2}-m_{\tz_1}$ are 
in general expected in both BWCA and MWDM scenarios,
as compared to the case of models
that incorporate gaugino mass unification close to the GUT scale.

In Fig. \ref{z21gap}, we show contours of 
the mass gap $m_{\tz_2}-m_{\tz_1}$ in the $m_0\ vs.\ m_{1/2}$ plane for 
$A_0=0$, $\tan\beta =10$ and $\mu >0$ for
{\it a}) the mSUGRA model, {\it b}) the case of BWCA DM where $-M_1$ is
raised at every point until $\Omega_{\tz_1}h^2\to 0.11$ and {\it c})
the case of BWCA DM where $-M_2$ is lowered until $\Omega_{\tz_1}h^2\to 0.11$.
In the case of the mSUGRA model, most of the parameter space has
$m_{\tz_2}-m_{\tz_1}>90$ GeV, which means that $\tz_2\to \tz_1 Z^0$ decay is
allowed. When this decay is allowed, its branching fraction is always large, 
unless it competes with other two-body decays such as $\tz_2\to \tz_1 h$
or $\tz_2\to \bar{f}\tilde f$ or $f{\bar{\tilde f}}$ (where $f$ is 
a SM fermion). In the case of BWCA DM in frames {\it b}) and {\it c}), 
we see that (aside from the left-most portion of frame {\it b}), 
which is not a region of BWCA), 
the mass gap is much smaller, so that 
two-body tree level decays of $\tz_2$ and $\tw_1$ 
are closed and three-body decays are dominant.
If $\tz_2$'s are produced at large rates either directly or via gluino
or squark cascade decays\cite{cascade}, it should be possible to
identify opposite sign/ same flavor dilepton pairs from their decays, to
reconstruct their invariant mass, and extract the upper edge of the
invariant mass distribution\cite{mlldist,frank}.

\FIGURE[!t]{
\epsfig{file=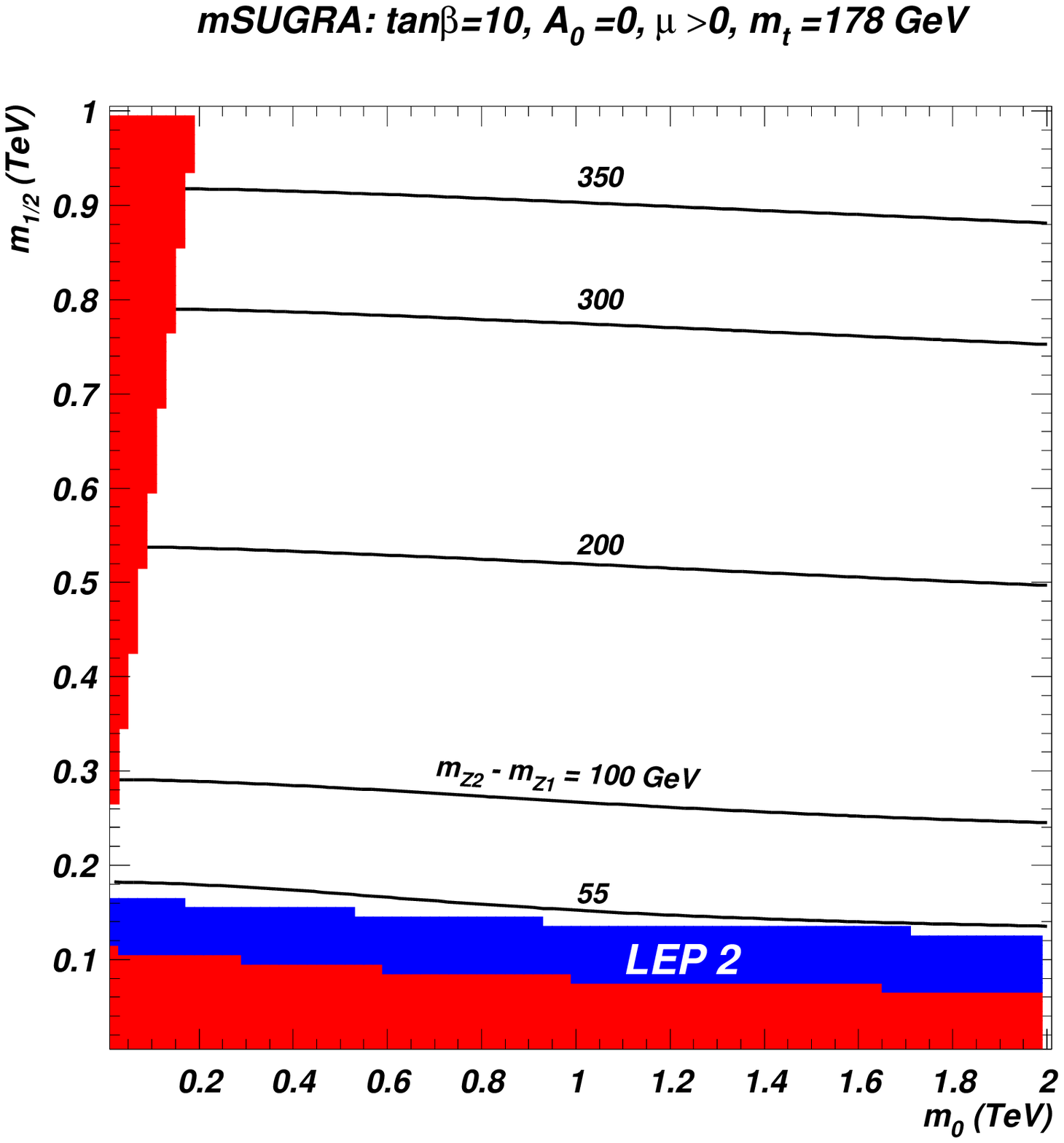,width=7.5cm} 
\mbox{\hspace*{-1cm}\epsfig{file=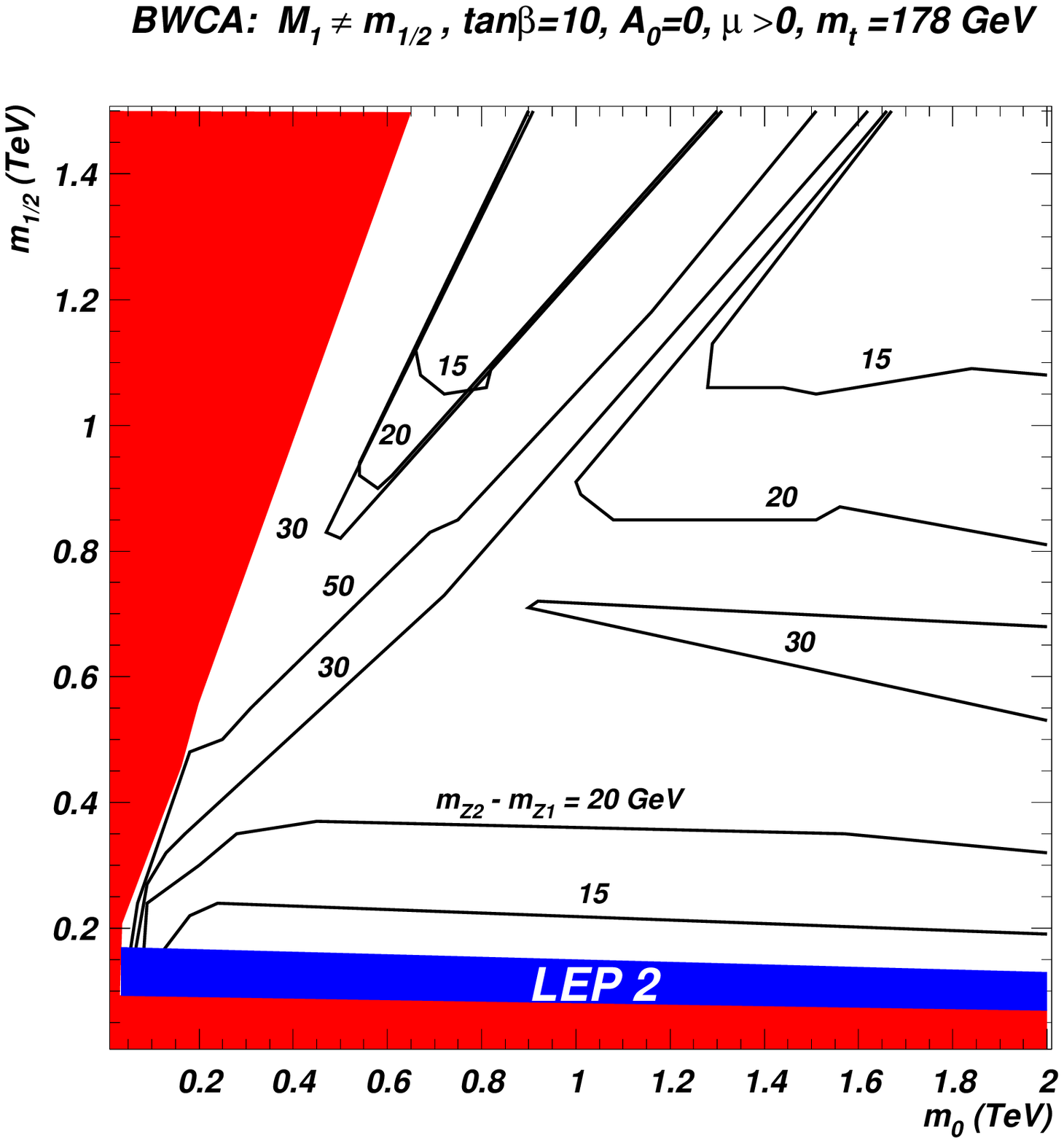,width=7.5cm} 
\epsfig{file=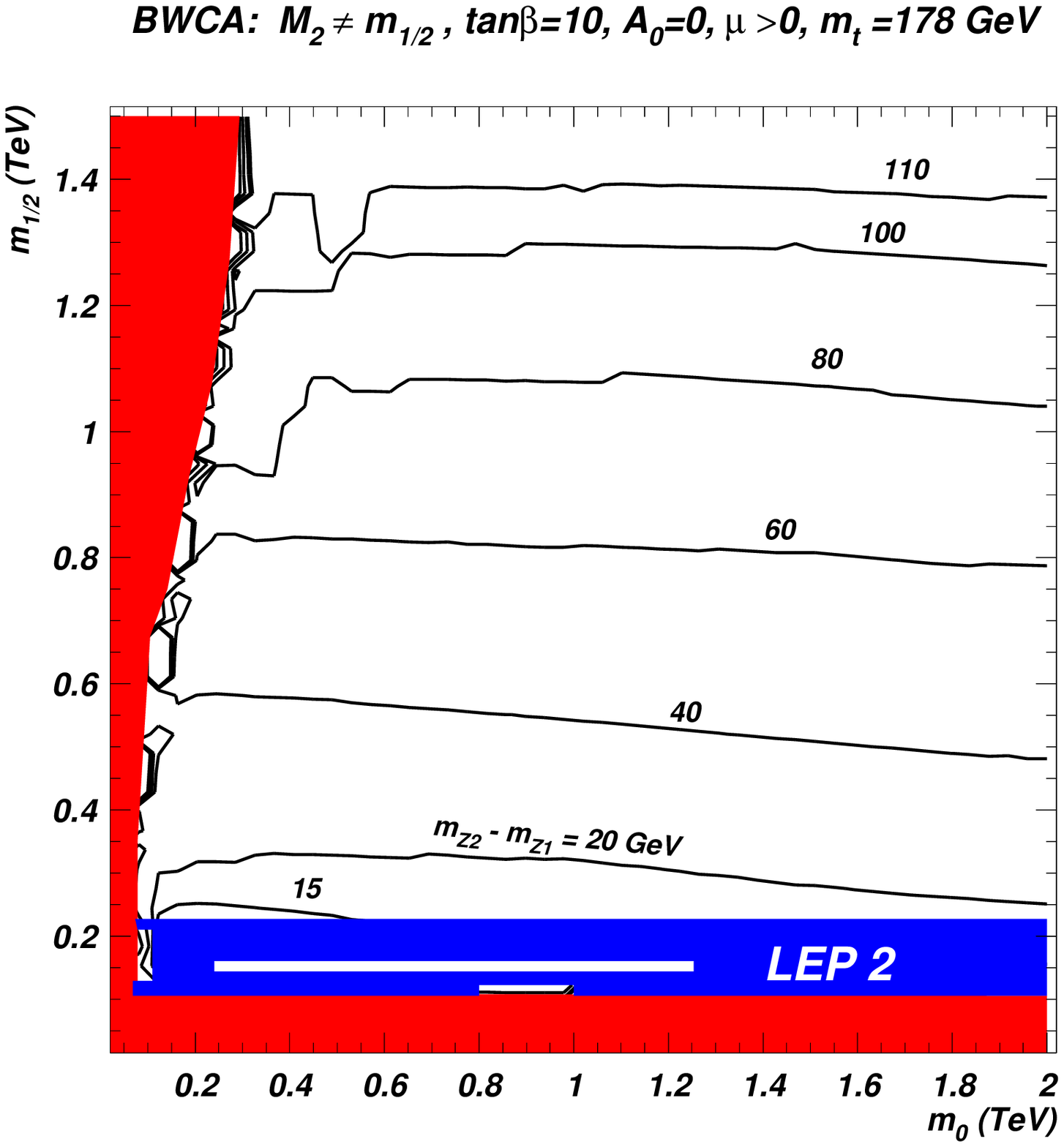,width=7.5cm} }
\caption{\label{z21gap}
Contours of $m_{\tz_2}-m_{\tz_1}$ mass gap in 
the $m_0\ vs.\ m_{1/2}$ plane for
$\tan\beta =10$, $A_0=0$, $\mu >0$ and
{\it a}) mSUGRA model, {\it b}) $-M_1>m_{1/2}$ BWCA  and
{\it c}) $-M_2<m_{1/2}$ BWCA.}}

Finally, we note one curious feature of Fig. \ref{z21gap}{\it c} that
was referred to in the caption of Fig.~\ref{planes_r}.
Within the blue shaded LEP2 excluded region there appear three allowed strips.
The lower horizontal strip at $m_{1/2}\sim 100$ GeV corresponds to the
neutralino $Z$-annihilation funnel, where $M_2$ does not have to be dialed to
low values, since $Z$ annihilation already reduces the relic density.
However, in this case where $M_2$ is negative, terms in the two-loop gaugino
mass RGE conspire to yield a weak scale $|M_2|$ value which is somewhat 
larger than its mSUGRA counterpart. This means that while 
$2m_{\tz_1}\sim M_Z$, at the same time $m_{\tw_1}$ is just above 
$103.5$ GeV, and is thus LEP2 allowed.
A similar situation occurs for the longer allowed strip at 
$m_{1/2}\sim 150$ GeV, except in this case it is the $h$ resonance
which reduces the relic density. Finally, at very low $m_0\sim 80$ GeV,
bulk annihilation via light sleptons reduces the relic density,
while the negative $M_2$ pushed $m_{\tw_1}$ above the LEP2 limit.  
The corresponding strips do not appear in frame {\it b}) because here,
we have only scanned $M_1$ values {\it above} $m_{1/2}$, as required to
get agreement with WMAP: as we can see from Fig.~\ref{fig:rd_bw1} that the
$Z$ and $h$ resonances appear for $|M_1| < m_{1/2}$.

\subsection{Fermilab Tevatron}
\label{ssec:tev}

In the mSUGRA model, the best reach for SUSY at the Fermilab Tevatron occurs
in the clean trilepton channel\cite{trilep,new3l}. 
We examined the clean trilepton
signal rate for case study point BWCA3. For this point, 
the total SUSY particle production cross section was $\sim 440$ fb for
$\sqrt{s}=2$ TeV $p\bar{p}$ collisions. Using the soft trilepton cuts SC2 of 
Ref. \cite{new3l} (three isolated leptons with 
$p_T(\ell_1,\ell_2,\ell_3 )>11,7,5$ GeV, $|\eta (\ell_{1,2/3})|<1,2$, 
$\eslt >25$ GeV, $m(\ell\bar{\ell})>20$ GeV, $m_T(\ell,\eslt)<60$ GeV
plus a $Z$ mass veto), we find a surviving signal cross section of only 0.035 fb,
well below observability. The main problem here is the 
$m(\ell\bar{\ell})>20$ GeV cut. This cut is essential to remove background 
from virtual photons in $q\bar{q}'\to\ell\nu_\ell\ell'\bar{\ell}'$
electroweak production. However, the cut also kills much of the SUSY signal
in the BWCA case, since the $\ell\bar{\ell}$ invariant mass is 
constrained to be $<m_{\tz_2}-m_{\tz_1}$, which 
is already quite small.

Another possibility is to search for events at the Tevatron containing
isolated high $E_T$ photons from radiative $\tz_2$ decays.  We examined
the isolated photon inclusive channel ($E_T(\gamma )>8$ GeV with
$\sum_{cone}^{R=0.4} E_T<0.1 E_T(\gamma )$ and $|\eta (\gamma)|<1.1$),
the photon plus lepton channel ($p_T(\ell )>15$ GeV with
$\sum_{cone}^{R=0.4} E_T<5$ GeV and $\eslt >20$ GeV), and photon plus
$\eslt >20$ GeV channels for the BWCA3 case study (see
Table~\ref{tab:bwca}). In the latter two channels, we also vetoed jets.
The signal rates were 43, 2.7 and 13.5 fb respectively, while
backgrounds from $W\gamma$ ($Z\gamma$) production were 6911 (1803), 2500
(6.2) and 394 (253) fb, respectively.  In light of these results, it
appears very difficult to use the isolated photon
channels in the BWCA scenario to identify  SUSY 
at the Tevatron.

\subsection{CERN LHC}
\label{ssec:lhc}

If the $R$-parity conserving MSSM is a good description of nature
at the weak scale, then
multi-jet plus multi-lepton plus $\eslt$ events should occur at
large rates at the CERN LHC, provided that 
$m_{\tg}\alt 2-3$ TeV\cite{susycascade}.
The LHC reach for SUSY in the mSUGRA model has been calculated in
Ref. \cite{susylhc,bbbkt}.
The ultimate mSUGRA reach results, coming from
the $\eslt +$ jets channel, should also 
apply qualitatively to the BWCA case, since the values of 
$m_{\tg}$ and $m_{\tq}$ change little in going from mSUGRA to BWCA,
and the $\eslt +$ jets reach mainly depend on these masses.

The reach in other channels such as multileptons plus jets and isolated
photons plus jets may change substantially in the BWCA scenario.  The
reach of the LHC in the ($m_0,\ m_{1/2}$) plane of the mSUGRA model was
recently re-assessed in Ref.~\cite{bbbkt} .  The search strategy was
based on the detection of gluino and squark cascade decay products,
namely multiple high transverse momentum jets and/or leptons and/or
photons plus large missing transverse energy.  Here, we use Isajet
7.72$^\prime$~\cite{isajet} for the simulation of signal and background
events at the LHC.  The event and detector simulation was performed
along the lines established in Ref. \cite{bbbkt}, where details on cuts
and detector resolution along with our definitions of jets and isolated
leptons and photons may be found.

\FIGURE[!h]{
\epsfig{file=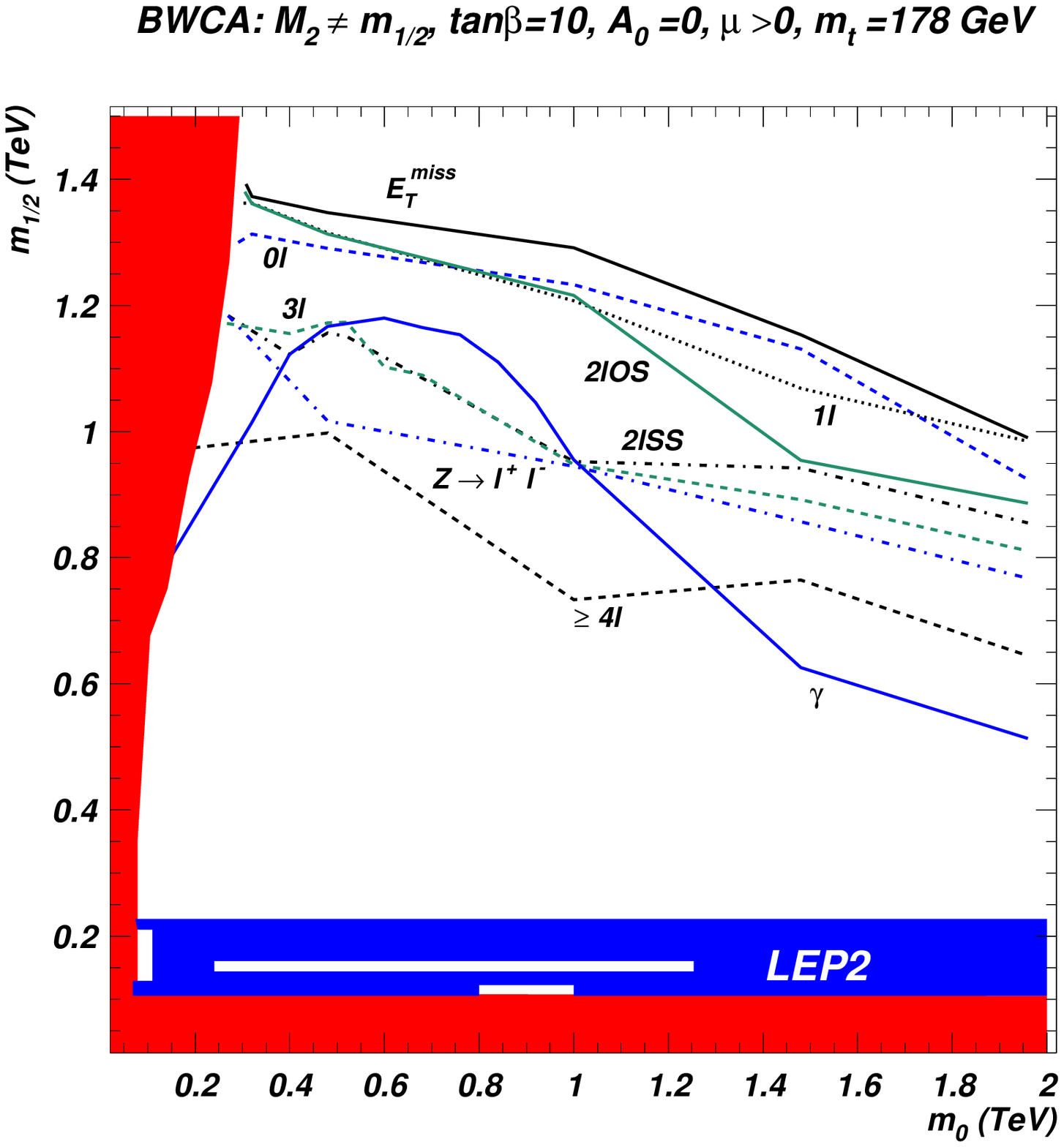,width=8.5cm} 
\caption{\label{fig:lhc}
Reach contours of the CERN LHC for 100 fb$^{-1}$ of integrated
luminosity for various signal topologies in 
the $m_0\ vs.\ m_{1/2}$ plane for
$\tan\beta =10$, $A_0=0$, $\mu >0$ and for $-M_2<m_{1/2}$ 
such that $\Omega_{\tz_1}h^2 =0.11$ at every point in the plane.}}
We plot the reach of the LHC in Fig.~\ref{fig:lhc} for 
the BWCA case using the 
procedure described in \cite{bbbkt}. All events had to 
pass the pre-cuts, which impose the requirement that $E_T^{\rm miss}>200$ GeV 
and there are at 
least 2 jets with $p_T^{\rm jet}>40$ GeV. 
We then optimized the cuts using the strategy in Ref. \cite{bbbkt} --
generally speaking, harder jet $E_T$ and $\eslt$ 
cuts apply for heavier sparticles, and softer cuts apply for 
lighter sparticles. 
The events are divided into several classes, 
characterized by the number of leptons or the presence of
an isolated photon in the final state. 
The $5\sigma$ discovery reach for 100 $fb^{-1}$ 
of integrated luminosity is shown for the various channels.
The ultimate reach in the $\eslt +$ jets channel ranges from 
$m_{1/2}\sim 1.4$ TeV at low $m_0$ to $m_{1/2}\sim 1$ TeV at $m_0=2$ TeV.
These results are similar to those obtained in the mSUGRA model.
The reach contour in the isolated photon plus jets plus $\eslt$ channel,
however,
has greatly increased in the BWCA scenario compared to the mSUGRA case.
In mSUGRA, the reach contour varies from $m_{1/2}:850\rightarrow 500$ GeV
as $m_0: .1\rightarrow 2$ TeV. In the BWCA case, where $\tz_2\to\tz_1\gamma$
at a large rate,
the photon reach contour reaches a maximum of $m_{1/2}=1.2$ TeV for 
$m_0\sim 500$ GeV. Thus, at CERN LHC, the BWCA scenario will be signalled
by multijet plus isolated multilepton plus $\eslt$ events, but with 
a {\it large content of hard isolated photons} as well, at least for the case
where $m_0\alt 1$ TeV.\footnote{Gauge mediated SUSY breaking (GMSB) models in
  which the next-to-lightest SUSY particle (NLSP) decays to a photon and a
  gravitino also have photons in SUSY events. It is unlikely that these
  models will be confused with the BWCA scenario because not only is the
  sparticle mass spectrum quite different but the photons in the GMSB
  scenario would typically have much larger energy because the gravitino
is essentially massless. Moreover, unless the NLSP has many decay modes,
the multiplicity of photons in the GMSB case would be much larger since
every SUSY event would contain two photons. }
For larger values of $m_0>1$ TeV, the $\tz_2\to\tz_1\gamma$ branching 
fraction drops, and the reach projections become similar to the case of
mSUGRA. We have also checked that the reach using the $\gamma + n$ lepton
channel is smaller than the reach via the corresponding $n$ lepton
channel, primarily because the signal becomes too small to pass our 10
event/100~fb$^{-1}$ requirement. 

For SUSY searches at the CERN LHC, Hinchliffe {\it et al.} have 
pointed out\cite{frank} that an approximate value of $m_{\tq}$ or
$m_{\tg}$ can be gained by extracting the maximum in the
$M_{eff}$ distribution, where 
$M_{eff}=\eslt +E_T(jet\ 1)+E_T(jet\ 2)+E_T(jet\ 3)+E_T(jet\ 4)$.
Their analysis will carry over to the BWCA scenario, as well as in 
models with gaugino mass unification, so that the approximate
mass scale of strongly interacting sparticles will be known soon 
after a supersymmetry signal has been established.  

In mSUGRA, a dilepton mass edge should be
visible in SUSY signal events only if $m_{1/2}\alt 250$ GeV 
or if $\tz_2\to \tell\bar{\ell},\ \bar{\tell}\ell$ decays are allowed.
In the case of BWCA DM, as with MWDM, 
the dilepton mass edge should be visible over almost all 
parameter space. We illustrate the situation for four case studies
listed in Table \ref{tab:bwca}.\footnote{In this study, 
a toy detector simulation 
is employed with calorimeter cell size 
$\Delta\eta\times\Delta\phi=0.05\times 0.05$ and $-5<\eta<5$. The hadronic 
energy resolution is taken to be $80\%/\sqrt{E}$ for $|\eta|<2.6$ and 
$100\%/\sqrt{E}$ for $|\eta|>2.6$. The electromagnetic energy resolution 
is assumed to be $3\%/\sqrt{E}$. We use a UA1-like jet finding algorithm 
with jet cone size $R=0.5$ and $p_T^{\rm jet}>25$ GeV. 
We also require that 
$|\eta_{\ell}|<2.5$ and $|\eta_j|<3$.
Leptons ($e$s or $\mu$s) have to also satisfy $p_T^{\rm lepton} \ge 10$~GeV.
Leptons are considered 
isolated if the visible activity within the cone $\Delta R<0.3$ is 
$\Sigma E_T^{\rm cells}<2$ GeV. The strict isolation criterion helps reduce 
multi-lepton background from heavy quark (especially $t\bar t$)
production. } 
The first case, labeled mSUGRA,
has $m_0=m_{1/2}=300$ GeV, with $A_0=0$, $\tan\beta =10$ and $\mu >0$.
In this case, $\tg\tg$, $\tg\tq$ and $\tq\tq$ production occurs with a 
combined cross section of about 12 pb, 
while the total SUSY cross section is around
13.4 pb (the additional 1.4 pb comes mainly from -ino pair production and
-ino-squark or -ino-gluino associated production). 
The case of BWCA1, with $M_1=-480$ GeV, has similar rates of 
sparticle pair production. The case of BWCA2, 
with lighter chargino and neutralino masses, has a total production cross 
section of 19.2 pb, wherein strongly interacting sparticles are 
pair produced at similar rates as in mSUGRA or BWCA1, but -ino pairs are 
produced at a much larger rate $\sim 6.1$ pb.
We also show the case of BWCA3, which is similar to that of BWCA2
except that $\mu <0$, which gives a better fit to $(g-2)_\mu $
measurements.
%
\begin{table}
\begin{tabular}{lcccc}
\hline
parameter & mSUGRA & BWCA1 & BWCA2 & BWCA3 \\
\hline
$M_1$ & 300 & -480 & 300 & 300 \\
$M_2$ & 300 & 300 & -156 & -170 \\
$\mu$ & 409.2 & 401.3 & 402.3 & -401.6 \\
$m_{\tg}$ & 732.1 &  733.4 & 736.3 & 736.8 \\
$m_{\tu_L}$ & 713.9 & 715.3 & 701.9 & 703.5 \\
$m_{\tst_1}$ & 523.4 & 535.0 & 554.8 & 566.4 \\
$m_{\tb_1}$ & 650.0 & 651.8 & 645.1 & 646.1 \\
$m_{\te_L}$ & 364.7 & 371.6 & 324.5 & 327.7 \\
$m_{\te_R}$ & 322.8 & 352.7 & 322.6 & 322.6 \\
$m_{\tw_2}$ & 432.9 & 426.0 & 419.6 & 421.3 \\
$m_{\tw_1}$ & 223.9 & 223.4 & 138.5 & 141.7 \\
$m_{\tz_4}$ & 433.7 & 425.0 & 415.2 & 419.7 \\
$m_{\tz_3}$ & 414.8 & 409.4 & 414.0 & 410.0 \\ 
$m_{\tz_2}$ & 223.7 & 222.7 & 138.6 & 141.4 \\ 
$m_{\tz_1}$ & 117.0 & 200.0 & 116.8 & 118.8 \\ 
$m_A$       & 538.7 & 537.1 & 508.4 & 508.5 \\
$m_{H^+}$   & 548.0 & 546.4 & 517.9 & 518.0 \\
$m_h$       & 115.7 & 115.3 & 114.0 & 112.7 \\
$\Omega_{\tz_1}h^2$& 1.1 & 0.11 & 0.10 & 0.12 \\
$BF(b\to s\gamma)$ & $3.2\times 10^{-4}$ & $3.3\times 10^{-4}$ &
$3.7\times 10^{-4}$ & $4.6\times 10^{-4}$ \\
$\Delta a_\mu    $ & $12.1 \times  10^{-10}$ & $9.9 \times  10^{-10}$ & 
$-13.8\times 10^{-10}$ & $13.1\times 10^{-10}$ \\ 
$\sigma_{sc} (\tz_1p )$ & 
$1.5\times 10^{-9}\ {\rm pb}$ & $4.7\times 10^{-12}\ {\rm pb}$ & 
$1.2\times 10^{-9}\ {\rm pb}$ & $8.0\times 10^{-11}\ {\rm pb}$\\
$BF(\tz_2\to\tz_1\gamma )$ & $2.7\times  10^{-5}$ &  0.096 & 0.25 & 0.25 \\
\hline
\end{tabular}
\caption{Masses and parameters in~GeV units
for mSUGRA and three BWCA scenarios. In each case,
$m_0=m_{1/2} =300$ GeV, $A_0=0$, $\tan\beta =10$ and $m_t=178$ GeV.
The third BWCA case has $\mu <0$, while the first two have $\mu >0$.
}
\label{tab:bwca}
\end{table}

We have generated 50K 
LHC SUSY events for each of these cases using Isajet 7.72$^\prime$, 
and passed them through 
a toy detector simulation as described above. 
Since gluino and squark  masses of the three case studies 
are similar to those 
of LHC point 5 of the study of Hinchliffe {\it et al.}\cite{frank}, 
we adopt the same 
overall signal selection cuts which efficiently select the SUSY signal
while essentially eliminating SM backgrounds:
$\eslt >max(100\ {\rm GeV}, 0.2 M_{eff})$,
at least four jets with $E_T>50$ GeV, where the hardest jet has
$E_T>100$ GeV, transverse sphericity $S_T>0.2$ and $M_{eff}>800$ GeV.

\FIGURE[!ht]{
\epsfig{file=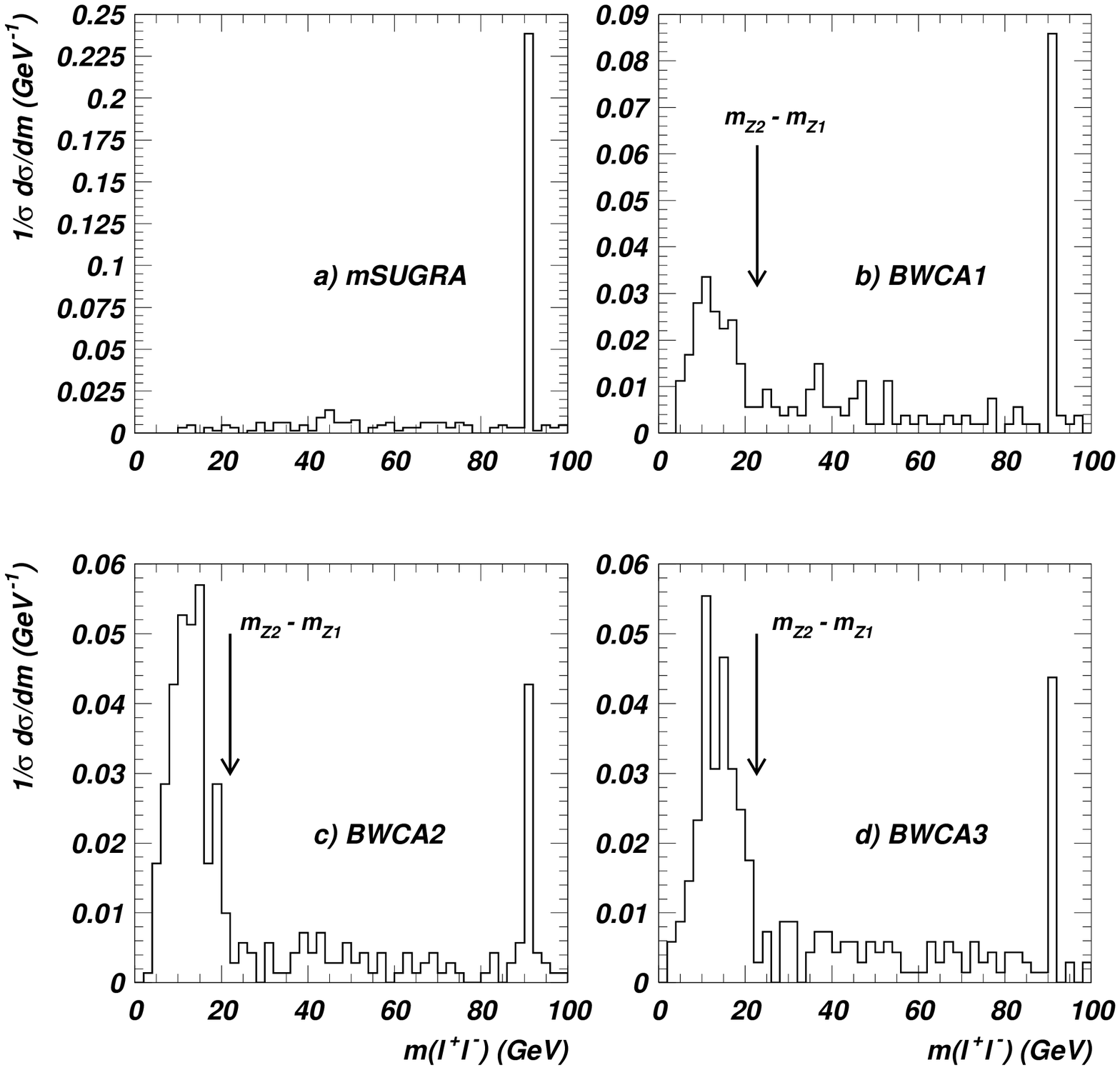,width=13cm} 
\caption{\label{fig:mll}
Distribution of same flavor/opposite sign dileptons from SUSY 
events at the CERN LHC 
from {\it a}) mSUGRA, {\it b}) BWCA1, {\it c}) BWCA2 
and {\it d}) BWCA3 cases 
as in Table \ref{tab:bwca}.}}
In these events, we require at least two isolated leptons, and then plot 
the invariant mass of all same flavor/opposite sign dileptons. 
The results are shown in Fig. \ref{fig:mll}. In the case of the mSUGRA model,
frame {\it a}), 
there is a sharp peak at $m(\ell^+\ell^- )\sim M_Z$, which comes 
from $\tz_2\to \tz_1 Z^0$ decays where $\tz_2$ is produced in the gluino
and squark cascade decays. In the case of BWCA1 in frame {\it b}), 
we again see a $Z^0$ peak,
although here the $Z^0$s arise from $\tz_3$, $\tz_4$ and $\tw_2$ decays.
We also see the continuum distribution in 
$m(\ell^+\ell^- )<m_{\tz_2}-m_{\tz_1}=22.7$ GeV. The cross section 
plotted here is $\sim 0.035$ pb, which would correspond to 
3.5K events in 100 fb$^{-1}$ of integrated luminosity (the sample 
shown in the figure contains just 135 events). 
In frame {\it c})-- with a cross section of $\sim 0.07$ pb 
(but just 187 actual entries)-- we see again the $Z^0$ peak, 
but also we see again the 
$m(\ell^+\ell^- )<m_{\tz_2}-m_{\tz_1}=21.8$ GeV continuum.
In both these BWCA cases, the $m_{\tz_2}-m_{\tz_1}$ mass edge should 
be easily measurable. It should also be obvious that it is inconsistent 
with models based on gaugino mass unification, in that the
projected ratios $M_1:M_2:M_3$ will not be in the order $1:\sim 2:\sim 7$ as in
mSUGRA. Although the $\tz_2 -\tz_1$ mass edge will be directly measurable, 
the absolute neutralino and chargino masses will be difficult 
to extract at the LHC.
In frame {\it d}), we show the spectrum from BWCA3, which is similar to
the case of BWCA2.

\subsection{Linear $e^+e^-$ collider}
\label{ssec:ilc}

\FIGURE[!t]{
\epsfig{file=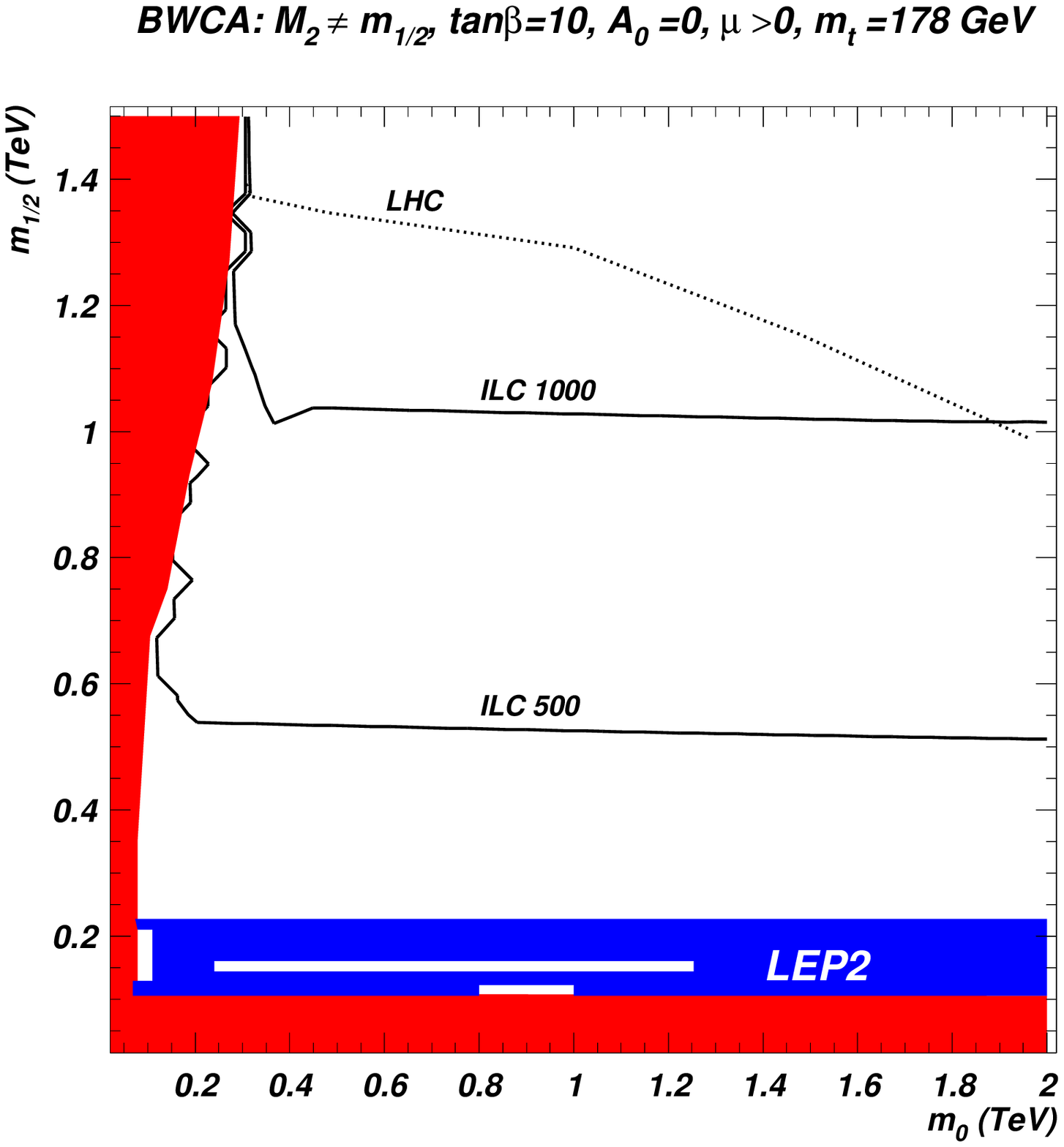,width=8cm} 
\caption{\label{fig:ilcreach}
Reach of the ILC and the LHC for SUSY in the BWCA DM scenario
where $|M_2|$ is lowered until $\Omega_{\tz_1}h^2=0.11$ at every
point in the $m_0\ vs.\ m_{1/2}$ plane. We assume 100 fb$^{-1}$ of
integrated luminosity for each collider, and show the ILC reach for
$\sqrt{s}=500$ GeV and 1000 GeV. }}
The reach of the CERN LHC for supersymmetric matter is determined mainly by 
$m_{\tq}$ and $m_{\tg}$, which depend on $m_0$ and $m_{1/2}$.
In contrast, the reach of the ILC for SUSY is largely determined by 
whether or not the reactions
$e^+e^-\to \tw_1^+\tw_1^-$ or $e^+e^-\to \tell^+\tell^-$ are kinematically 
accessible\cite{nlc}. 
For instance, chargino pair production is expected to be visible if 
$\sqrt{s}>2m_{\tw_1}$.
The value of $m_{\tw_1}$ depends mainly on $M_2$ and $\mu$. 
Thus, in the BWCA case where $M_2=m_{1/2}$ but $M_1$ is variable, 
the reach of the ILC in the $m_0\ vs.\ m_{1/2}$ 
plane will be similar to the case of the mSUGRA model.
However, in the BWCA case where $M_1=m_{1/2}$, with variable $M_2$,
the reach of the ILC will be enhanced compared to the mSUGRA case, 
since $|M_2|$ is typically much smaller for a given
set of $m_0$ and $m_{1/2}$ values.
The situation is illustrated in Fig. \ref{fig:ilcreach} where we show the
ultimate reach of the LHC and the ILC in the $m_0\ vs.\ m_{1/2}$ plane for 
$\tan\beta =10$, $A_0=0$, $\mu >0$ and $m_t=178$ GeV. We have dialed
$-M_2$ at every point to give $\Omega_{\tz_1}h^2 =0.11$, in accord with 
the WMAP observation. We have assumed 100 fb$^{-1}$ of integrated luminosity
for both LHC and ILC. The reach of ILC with $\sqrt{s}=500$ GeV
(denoted by ILC 500) extends to $m_{1/2}\sim 500$ GeV, while the 
corresponding reach in the mSUGRA model with gaugino mass unification 
extends to $m_{1/2}\sim 320$ GeV\cite{nlc}. 
The reach of ILC with $\sqrt{s}=1000$ GeV
extends to $m_{1/2}\sim 1$ TeV, compared with the mSUGRA value of
$m_{1/2}\sim 600$ GeV. In fact, we see that for $m_0\agt 1.9$ TeV, the 
ILC1000 reach begins to exceed that of the LHC. In this region, 
$m_{\tw_1}\sim 500$ GeV, while $m_{\tg}\sim 2300$ GeV and 
$m_{\tq}\sim 2675$ GeV. 

At a $\sqrt{s}=500$ GeV ILC, the new physics reactions for the four
case studies shown in Table \ref{tab:bwca} would 
include $Zh$, $\tw_1^+\tw_1^-$,
$\tz_1\tz_2$ and $\tz_2\tz_2$ production. It was shown in Ref. \cite{nlc}
that even in the case of a small $\tw_1 -\tz_1$ mass gap 
chargino pair production events could still be identified 
above SM backgrounds using cuts specially designed to pick out
low visible energy acollinear signal events over backgrounds from
$e^+e^-$ and $\gamma\gamma$ processes. 
The chargino and neutralino masses can be inferred from the resultant dijet
distribution in 
$\tw_1^+\tw_1^-\to (\bar{\ell}\nu_\ell\tz_1 )+(q\bar{q}\tz_1 )$
events\cite{jlc,bmt,nlc}. 
These measurements should allow the absolute mass scale of the 
sparticles to be pinned down, and will complement the $\tz_2 -\tz_1$ 
mass gap measurement from the CERN LHC. 
The combination of $m_{\tz_2}$, $m_{\tw_1}$,
$m_{\tz_1}$ and $m_{\tz_2} -m_{\tz_1}$ measurements 
will point to whether or not gaugino
mass unification is realized in nature.

\FIGURE[!ht]{
\epsfig{file=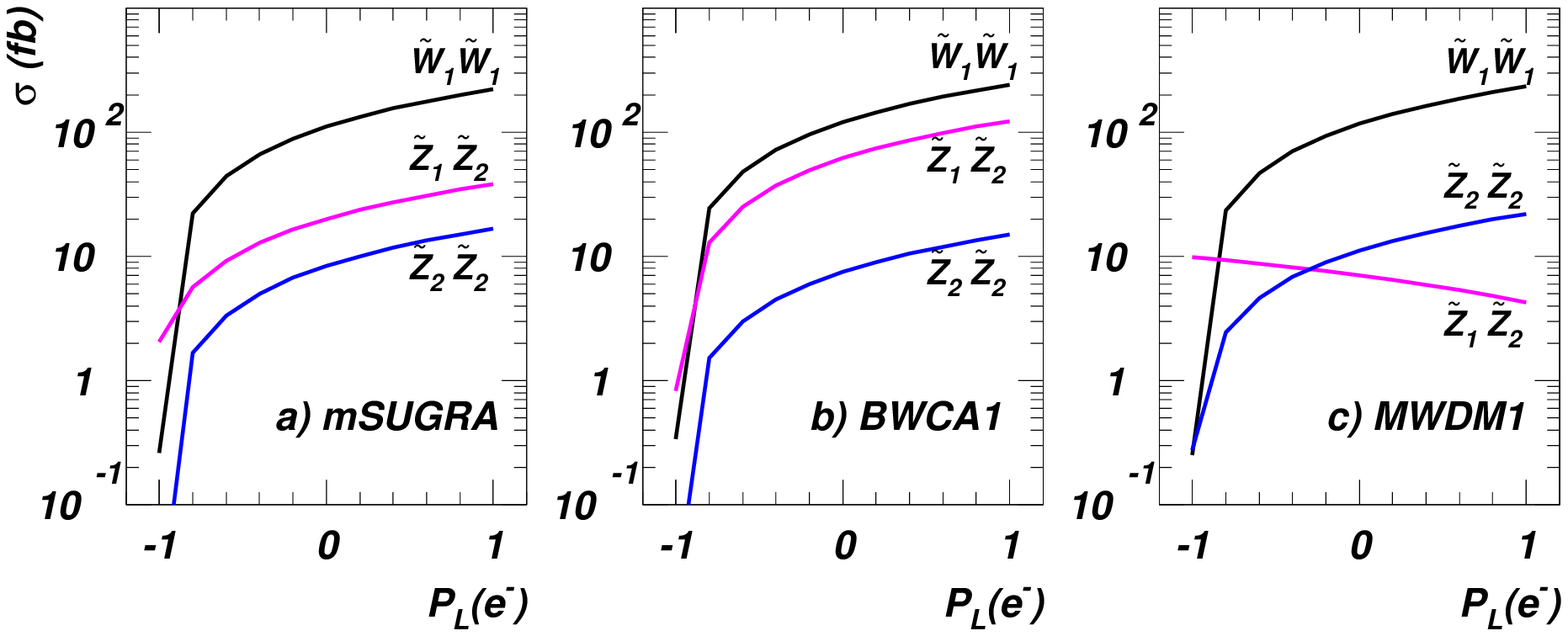,width=14cm} 
\caption{\label{fig:ilc} Cross sections for $e^+e^-\to \tw_1^+\tw_1^-$,
$\tz_1\tz_2$ and $\tz_2\tz_2$ processes versus electron beam
polarization $P_L(e^-)$ for an electron-positron linear collider
operating at $\sqrt{s}=500$ GeV for {\it a}) mSUGRA, {\it b}) BWCA1 and
{\it c}) MWDM1 from Ref.~\cite{mwdm}. The positron beam is taken to be
unpolarized.}}

The dependence of the cross sections for $\tw_1^+\tw_1^-$, $\tz_1\tz_2$
and $\tz_2\tz_2$ production on the longitudinal polarization of the
electron beam provides an additional tool at the ILC.  This dependence
is illustrated in Fig.~\ref{fig:ilc} for the mSUGRA model in
Table~\ref{tab:bwca} in frame {\it a}), for case BWCA1 in frame {\it b})
and for case MWDM1, which is point 1 of Ref. \cite{mwdm}. In all three
cases, the chargino and the second lightest neutralino have dominant
wino components so that the magnitudes and the polarization depence of
$\sigma(\tw_1\tw_1)$ and $\sigma(\tz_2\tz_2)$ are qualitatively
similar to one another and to the corresponding dependence in the mSUGRA
model\cite{bmt}. A minor difference is that, for $P_L=-1$,
$\sigma(\tz_2\tz_2)$ does not fall to quite as small values for the
MWDM1 case because its hypercharge gaugino component always remains
significant, as we have already discussed. The polarization-dependence
and/or the magnitude of the $e^+e^- \to \tz_1\tz_2$ process are,
however, quite different in the three cases. In the limit that $\tz_1$
is a bino and $\tz_2$ is the wino (a good approximation in the mSUGRA
case and an even better approximation in the BWCA case), $t$-channel
selectron exchange dominates the amplitude.\footnote{Recall that $Z$
couples to neutralinos only via their suppressed higgsino components.}
Recall, however, that for $P_L(e^-)=1$ ($P_L(e^-)=-1$) only $\te_L$
($\te_R$) exchange is possible. Since the chargino is always wino-like,
its couplings to $\te_R$ are strongly suppressed, accounting for the
behaviour of the cross sections in the first two frames.  For the MWDM1
case in frame {\it c}), we would naively expect that $\tz_1$ would be
photino-like and $\tz_2$ would be zino-like. It turns out, however, that
(for the specific MWDM1 parameters) the difference $M_1({\rm weak}) -
M_2({\rm weak})$ is sufficiently large, and results in a flip of the
relative sign between the gaugino components of both $\tz_1$ and
$\tz_2$, while roughly preserving the magnitude. This strongly
suppresses the $e\te_L\tz_1$ coupling, and hence, the $\te_L$ exchange
amplitude, resulting in the relatively flat and even decreasing 
polarization-dependence of
the cross section.\footnote{When both $\te_L$ and $\te_R$ exchanges are
dynamically suppressed, the effect of the usually small $Z$ exchange
contribution (which leads to a more or less flat polarization
dependence) may also be significant.}

A very striking feature of the figure is the very large cross section
for $\tz_1\tz_2$ production for a left polarized electron beam for the
BWCA1 case in frame {\it b}), as compared with the mSUGRA case. In both
cases, since $\tz_1$ and $\tz_2$ couple to (a not-so-heavy) $\te_L$
mainly via
their large hypercharge and $SU(2)$ gaugino components, respectively,
$\te_L$ exchange completely dominates $\tz_1\tz_2$ production at
$P_L=1$. Moreover, the magnitudes of the couplings, as well as selectron
masses, are very comparable in the two scenarios. The reason for the
difference in the cross sections lies in the relative sign between the
$\tz_1$ and $\tz_2$ eigenvalues (not physical masses) of the neutralino
mass matrix. Since $M_1/M_2 < 0$, we expect this sign to flip in the
BWCA case as compared with the mSUGRA or the MWDM cases, where $M_1/M_2$
is positive. This is relevant because when we square the 
$t$-channel amplitude and sum over the neutralino spins, there is one term
that is proportional to the product $m_{\tz_1}m_{\tz_2}$ (see {\it e.g.}
Eq.~(8f) of Ref.~\cite{bbkmt}, where the last term that includes the
factor $(-1)^{\theta_i+\theta_j}$ is the one we are referring to). This
term always flips sign between any scenario with positive $M_1/M_2$ {\it vs.}
negative $M_1/M_2$. As a result, what was a significant cancellation in
the mSUGRA case in frame {\it a}) of Fig.~\ref{fig:ilc} turns into a sum
in frame {\it b}), accounting for the factor $\sim 2.5$ increase in
$\sigma(\tz_1\tz_2)$ at $P_L=1$. 

We examine this potential enhancement of the $e^+e^-\to\tz_1\tz_2$
production cross section at a $\sqrt{s}=0.5$ TeV linear collider in
Fig. \ref{sigz2z1}.  In frame {\it a}), we show contours of $\sigma
(e^+e^-\to \tz_1\tz_2)$ in fb in the $m_0\ vs.\ m_{1/2}$ plane for
$A_0=0$, $\tan\beta =10$ and $\mu >0$, in the case of the mSUGRA
model. The light (yellow) shaded region is where either chargino
pair production or selectron pair production is kinematically
accessible at a $\sqrt{s}=0.5$ TeV ILC.  We see that the $\tz_1\tz_2$ 
cross section only
exceeds the 100 fb level at the very lowest values of $m_0$ and
$m_{1/2}$, in the lower left corner. The cross section at the $\sim 10$
fb level gives some additional reach of an ILC for SUSY beyond the range
where chargino pair production or slepton pair production are
possible\cite{bmt,nlc}. In frame {\it b}), we show the same situation
for the BWCA scenario where $M_1$ has been set to negative values
everywhere in the $m_0\ vs.\ m_{1/2}$ plane such that the WMAP value
$\Omega_{\tz_1}h^2=0.11$ is fulfilled. In this case, the
$\tz_1\tz_2$ production cross section is in general increased
everywhere, and exceeds 100 fb in a much larger region. Of course, the
region of reaction kinematic accessibility has decreased somewhat due to
the increase in $m_{\tz_1}$ for a given $m_0$ and $m_{1/2}$ value, so
that contours of accessibility reach only up to $m_{1/2}\sim 350$ GeV
(as opposed to $420$ GeV in the mSUGRA case). Also, the cross section
falls off at large values of $m_0$ since the amplitude is suppressed
with increasing selectron mass. In frame {\it c}), we
show the same cross section in the $m_0\ vs.\ m_{1/2}$ plane, except here
$M_2$ is taken negative, and is decreased at each point 
in absolute value until the
WMAP value is obtained. In this case, the kinematically
accessible region increases since for a given $m_0$ and $m_{1/2}$ value
$m_{\tz_2}$ is lowered compared to the mSUGRA case, while $m_{\tz_1}$
stays fixed. Again, a rather large region appears with $\sigma >10-100$
fb, so that $\tz_1\tz_2$ production should be robust over much of
parameter space in the BWCA scenario as long as $\tz_1\tz_2$ production
is kinematically accessible and $m_0$ is not too large. An unexpectedly
large value of $\sigma (\tz_1\tz_2)/\sigma (\tw_1\tw_1 )$ may, therefore,
be an indication that $M_1/M_2 <  0$ and that the selectrons are not too
heavy.

\FIGURE[!t]{
\epsfig{file=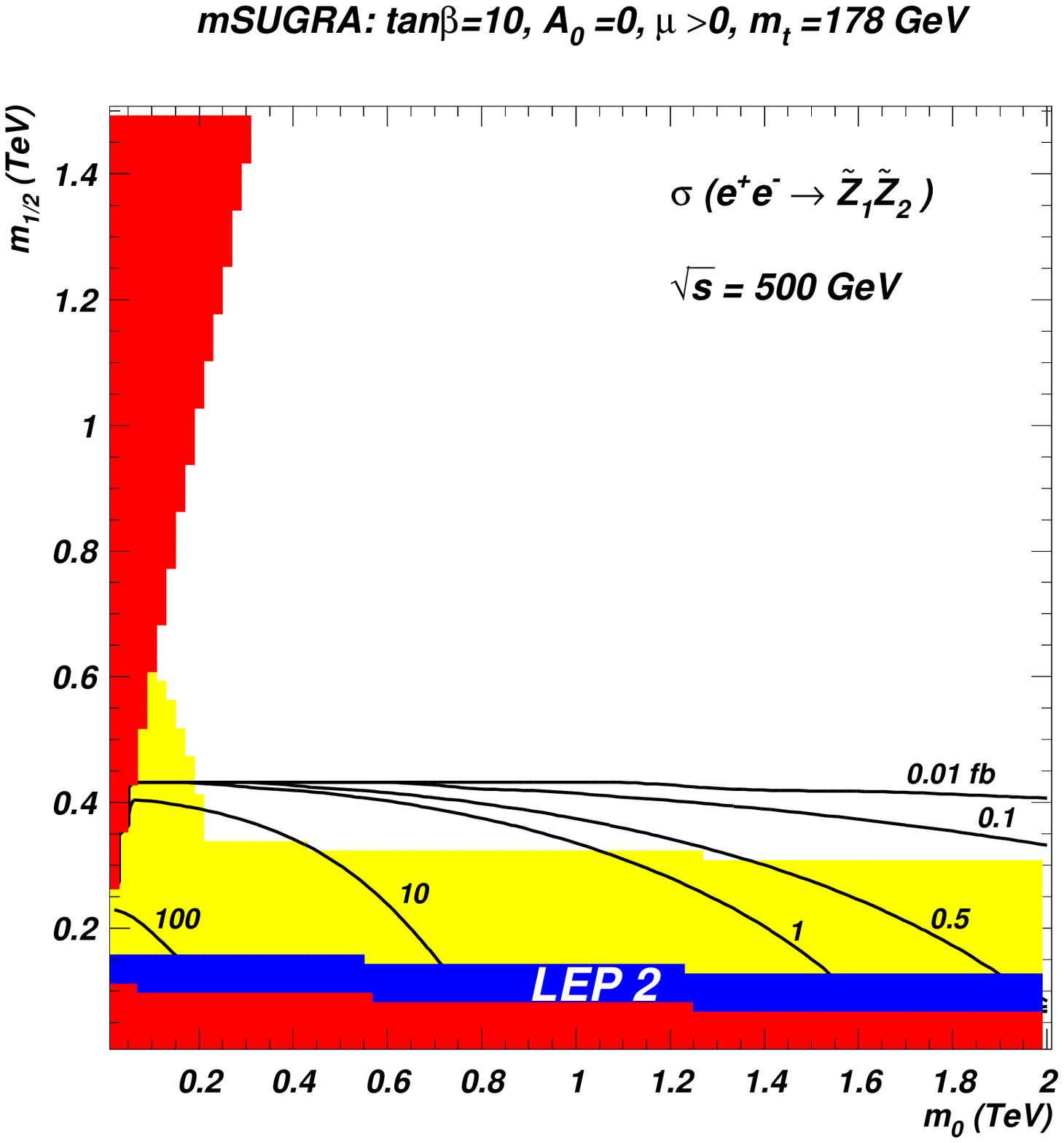,width=7.5cm} 
\mbox{\hspace*{-1cm}\epsfig{file=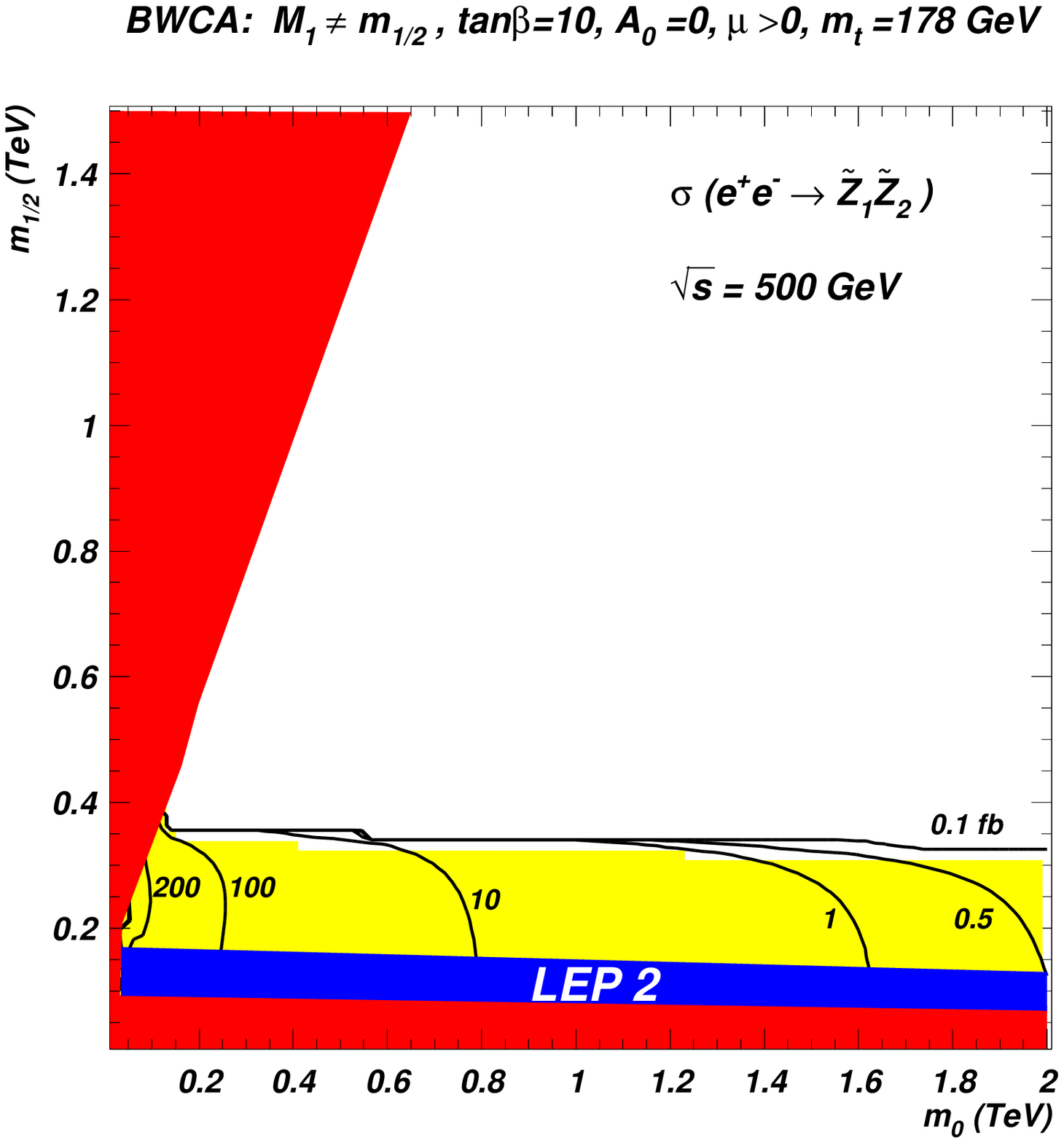,width=7.5cm} 
\epsfig{file=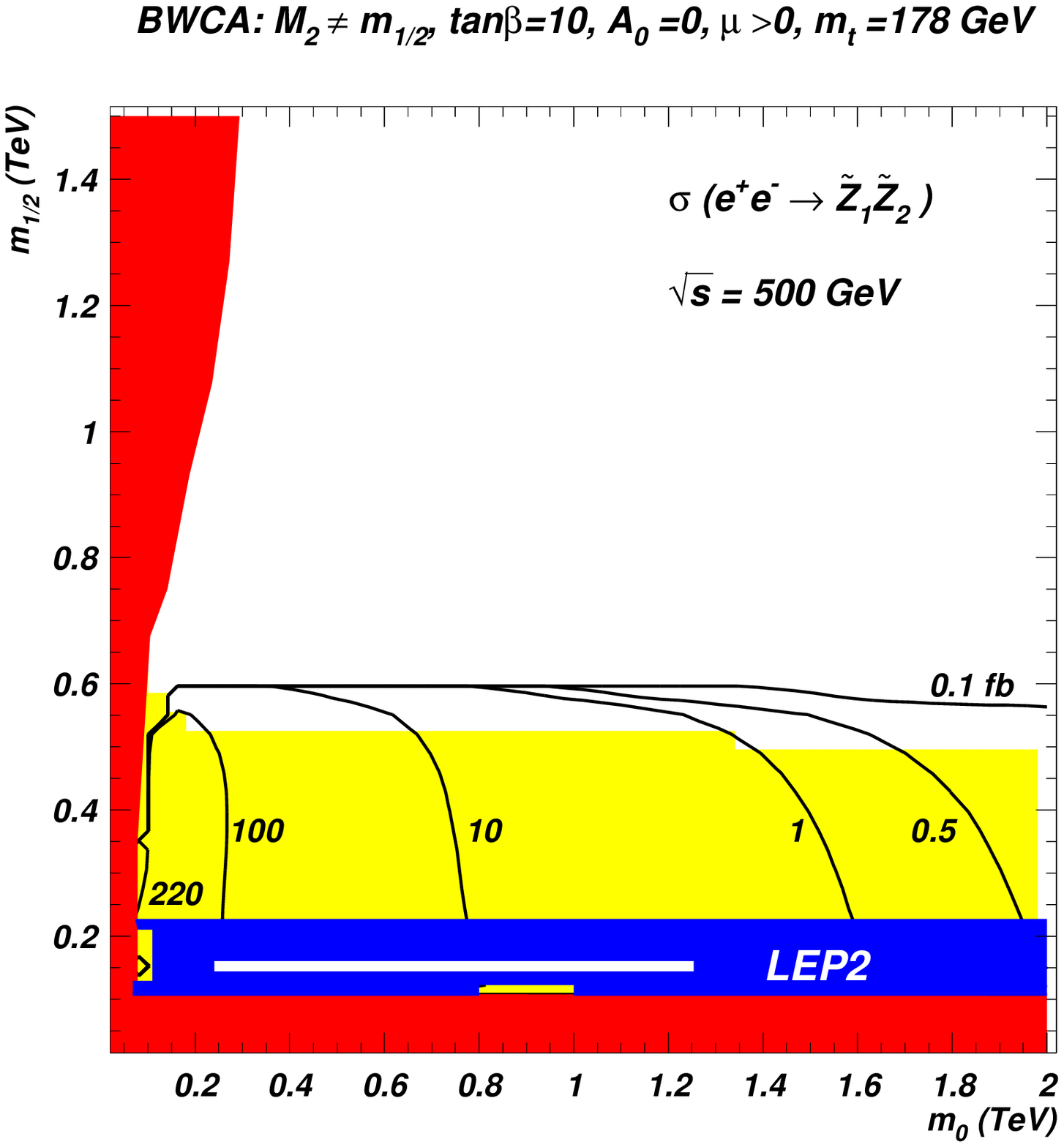,width=7.5cm} }
\caption{\label{sigz2z1}
Contours of $\sigma (e^+e^-\to \tz_1\tz_2 )$ in fb for a $\sqrt{s}=0.5$ TeV
ILC with unpolarized beams in the $m_0\ vs.\ m_{1/2}$ plane for
$\tan\beta =10$, $A_0=0$, $\mu >0$ and
{\it a}) mSUGRA model, {\it b}) $-M_1>m_{1/2}$ BWCA  and
{\it c}) $-M_2<m_{1/2}$ BWCA.
The yellow shaded regions show where $\tw_1^+\tw_1^-$ and/or $\tell^+\tell^-$
should also be kinematically accessible to a $\sqrt{s}=0.5$ TeV ILC.
}}

The rather large $e^+e^-\to \tz_1\tz_2$ cross section at the ILC potentially
leads to another distinguishing feature of SUSY events at the ILC in the BWCA
model: at least for
$m_0\alt 1$ TeV, we should expect a large rate for events
with one or more isolated photons
due to the enhanced branching fraction for $\tz_2\to\tz_1\gamma$
decays. Thus, from $\tz_1\tz_2$ production, we may expect $\gamma
+\emiss$ events, while from $\tz_2\tz_2$ production, we may expect
$\gamma\gamma +\emiss$, $\gamma+jets+\emiss$ events,
$\gamma+\ell\bar{\ell}+\emiss$ events and $\gamma +\emiss$.  Since the
$\gamma$ arises from a two body decay, the endpoints of the $E(\gamma )$
distribution will be functions of $\sqrt{s},\ m_{\tz_2}$ and
$m_{\tz_1}$. A measurement of the endpoints of this distribution will
thus allow an independent measurement of the two lighter neutralino
masses.  The $E(\gamma )$ distribution is shown in Fig. \ref{fig:egam}
for BWCA2 at a $\sqrt{s}=0.5$ TeV ILC.
\FIGURE[!t]{
\epsfig{file=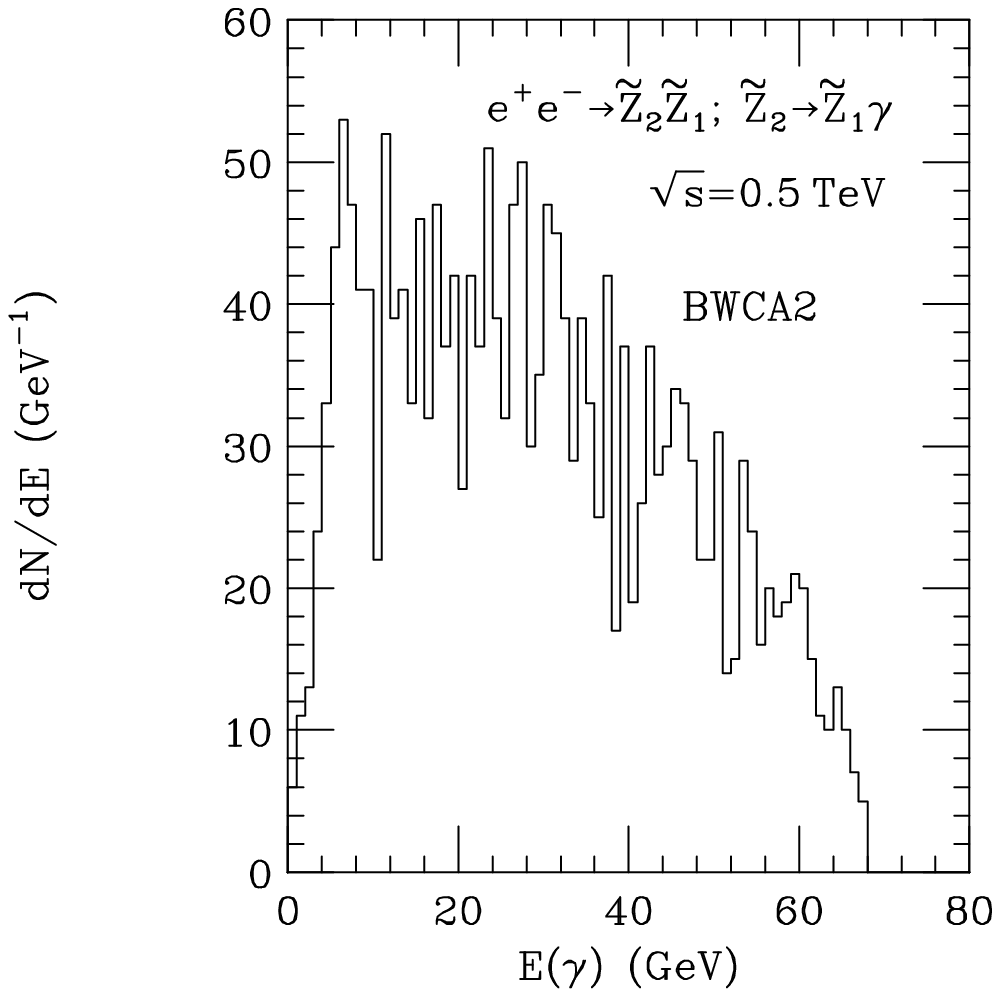,width=8cm,angle=90} 
\caption{\label{fig:egam}
Distribution in $E(\gamma )$ from 
$e^+e^-\to \tz_2\tz_1$ with $\tz_2\to \gamma \tz_1$ for
point BWCA2 in $e^+e^-$ collisions at $\sqrt{s}=0.5$ TeV.}}

\section{Conclusions}
\label{sec:conclude}

In this paper, we have considered the phenomenological implications of
neutralino dark matter in the bino-wino co-annihilation (BWCA) scenario,
and compared these to the case of mixed wino dark matter which
has a qualitatively similar spectrum.
The BWCA scenario arises in models with non-universal gaugino masses
where, at the weak scale, $|M_1|\sim |M_2|$, but where the two gauginos
masses have {\it opposite signs}. The same sign case gives rise to mixed
wino dark matter, while in the BWCA case, there is little mixing so that
the $\tz_1$ remains nearly a pure bino, while $\tz_2$ remains nearly a
pure wino.  The scenario can be brought into accord with the WMAP
$\Omega_{\tz_1}h^2\sim 0.11$ constraint by arranging for $m_{\tw_1}\sim
m_{\tz_2}\sim m_{\tz_1}$, so that bino-wino co-annihilation is the
dominant neutralino annihilation mechanism in the early universe.

Since co-annihilation processes are the dominant mechanism 
for the  annihilation of relic neutralinos from the Big Bang,
it is expected that indirect dark matter detection rates (which depend on the
$\tz_1\tz_1$ pair-annihilation cross section) will typically be quite
low in the BWCA scenario. Direct detection of neutralinos may be possible
for BWCA DM at stage 3 detectors in some portions of parameter space. 
In contrast, for the MWDM case, neutralino annihilation
cross sections are enhanced relative to mSUGRA so that it is more likely 
that an indirect signal for DM will be seen. 
If we already have some infomation about the sparticle spectrum
from the LHC, it may be that results from direct and  indirect 
DM detection experiments may serve to discriminate between these scenarios.

The small mass gap expected between
$m_{\tz_2}$ and $m_{\tz_1}$ is a hallmark of both the BWCA 
and the MWDM scenarios, and leads to a variety of interesting
consequences for collider experiments. 
In teh BWCA case, if $m_0\alt 1$ TeV, then the
radiative decay $\tz_2\to\tz_1\gamma$ is greatly enhanced, leading to
the production of isolated photons at the Tevatron, LHC and ILC. While
extraction of the isolated photon signal above SM background looks
difficult at the Tevatron, they should yield observable signals at the
CERN LHC. The LHC should also be able to extract the
$m_{\tz_2}-m_{\tz_1}$ mass difference from gluino and squark cascade
decay events which contain isolated opposite sign/same flavor dilepton
pairs.  By comparing the neutralino mass difference to measurements of the
gluino mass, we should readily be able to 
infer that the weak scale gaugino masses are incompatible
with expectation from models with a universal gaugino mass at the high scale.
At the ILC, again small
$m_{\tw_1}-m_{\tz_1}$ and $m_{\tz_2}-m_{\tz_1}$ mass gaps should be
measurable, as in the MWDM and MHDM scenarios. It should also be
possible to extract the weak scale values of the gaugino mass parameters.
 However, in the case of
BWCA, $\tz_1\tz_2$ production should occur at enhanced rates relative to
mSUGRA and MWDM, with a production cross section which increases with
$P_L(e^- )$.  In constrast to MHDM,
the $\tz_3$ and $\tz_4$ states will typically occur 
at much higher mass scales associated with  large values for the $|\mu |$
parameter.

\acknowledgments

We thank K.~Melnikov for the idea of using effective theories
in quantum mechanics, as discussed in the appendix. 
This research was supported in part by the U.S. Department of Energy
grant numbers DE-FG02-97ER41022, DE-FG03-94ER40833. 
T. K. was supported in part by the 
U.S. Department of Energy under contract No. DE-AC02-98CH10886.

\setcounter{section}{0}
\section*{Appendix: Effective Theories in Quantum Mechanics}

Consider a situation where the total Hamiltonian can be split into two pieces,
$$H=H_0+V,$$ so that the spectrum of $H_0$ is hierarchical, {\it i.e.}, it
consists of ``low energy'' states $|l_i\rangle$, possibly
interacting with one another 
via interactions included in $H_0$, and ``high energy''
states, $|h_a\rangle$ (again possibly interacting with themselves via
interactions included in $H_0$). Without loss of generality,
{\it all} interactions between the states 
$\left\{ |l_i\rangle\right\}$ and the states $\left\{ |h_a\rangle\right\}$ are
encapsulated in $V$. In other words, $\langle h_a|H_0|l_i\rangle=0$, and 
$\langle l_i|V|l_j\rangle = \langle h_a|V|h_b\rangle=0$. 

We seek a description of low energy states in terms of an effective
Hamiltonian $H_{\rm eff}$, that acts only on the low energy states
$\left\{|l_i\rangle\right\}$, assuming that we are at energies that are
too low to excite the high energy states. Note that even if $H_0$ is
completely diagonal in the low energy sector, we would 
expect that scattering of low energy states (with energy $E \ll M_a$) would
occur via their interactions with the high energy sector, with an
amplitude suppressed by $E/M_a$. Here, we show how this comes about
and obtain an expression for $H_{\rm eff}$. 

We {\it define} $H_{\rm eff}$ by matching the low energy Green's
functions of the full theory with those of the effective theory, 
\begin{equation}
\langle l_i| \frac{1}{H_0+V-z}|l_j\rangle \equiv \langle l_i|
  \frac{1}{H_{\rm eff}-z}|l_j\rangle, \ \ \ {\rm if} \ |z| \ll M_a.
\label{app:def}
\end{equation}
Here, $z$ is the complex argument of the Green's function. 
Expanding the left hand side of (\ref{app:def}), we obtain 
\begin{eqnarray}
\langle l_i|\frac{1}{H_{\rm eff}-z}|l_j\rangle &=& 
\langle l_i|\frac{1}{H_0-z}|l_j\rangle
+\langle l_i|\frac{1}{H_0-z} (-V) \frac{1}{H_0-z}|l_j\rangle \nonumber \\
&+&\langle l_i|\frac{1}{H_0-z} (-V)\frac{1}{H_0-z}(-V)
\frac{1}{H_0-z}|l_j\rangle+ \cdots .
\label{app:exp}
\end{eqnarray}
Clearly, if $V=0$ (no interactions between low and high energy states),
$H_{\rm eff}$ is just $H_0$ restricted to the low energy subspace, {\it
  i.e.} $H_{\rm eff}=H_0 {\cal P}_{\rm low}$, where ${\cal P}_{\rm low}$
is the projector on to the low energy subspace. It is also clear that
terms with an odd number of $(-V)$ factors on the right hand side of
(\ref{app:exp}) are all zero because $V$ only connects states in the low energy
sector with those in the high energy sector, whereas $H_0$  only connects
low energy (high energy) states with one another.
The third term on the right hand
side of (\ref{app:exp})
can be written as
\begin{eqnarray*}
\left(\frac{1}{H_0-z}\right)_{ii'}(-V)_{i'a}\left(\frac{1}{H_0-z}
\right)_{ab}(-V)_{bj'}\left(\frac{1}{H_0-z}\right)_{j'j},
\end{eqnarray*}
with the repeated indices $i', a, b$ and $j'$ all summed up.  We now
defined an induced effective potential $\Vind$ by,
\begin{equation}
-\left(\Vind\right)_{i'j'} =
 (-V)_{i'a}\left(\frac{1}{H_0-z}\right)_{ab}(-V)_{bj'}, 
\label{app:vind}
\end{equation}
in terms of which this term can be written as,
\begin{eqnarray*}
\left(\frac{1}{H_0-z}\right)_{ii'}\left(-\Vind\right)_{i'j'}\left(\frac{1}{H_0-z}\right)_{j'j}. 
\end{eqnarray*}
In exactly the same manner, the term with four factors of $(-V)$ will
end up as one with two factors of $\left(-\Vind\right)$, {\it etc.} so
that we have a geometric series that can be summed to give
\begin{equation}
\langle l_i|\frac{1}{H_{\rm eff}-z}|l_j\rangle = \langle
l_i|\frac{1}{H_0+\Vind-z}|l_j\rangle, 
\label{app:form}
\end{equation} 
showing that the interactions between the low and high sectors 
effectively induce an
additional ``potential'' in the low sector. 

Up to now, our considerations (though formal) have been ``exact'' in the
sense that we have not made any low energy approximation. This shows up
in the fact that the ``potential'' $\Vind$ defined in (\ref{app:vind})
depends on the ``energy'' $z$. If $|z|$ is small compared with the
high energy scale associated with the spectrum of $H_0$, we can ignore
it in the evaluation of the matrix elements of $(H_0-z)^{-1}$ between
high energy states, and approximate $\Vind$ by, 
\begin{eqnarray}
\left(\Vind\right)_{ij} &=& -\sum_{a,b}
\left[V_{ia}\left(\frac{1}{H_0}\right)_{ab}V_{bj} + V_{ia}{\cal
  O}(\frac{|z|}{M_a^2})V_{bj}+\cdots \right] \nonumber \\ 
&\simeq& -\sum_{a,b}
V_{ia}\left(\frac{1}{H_0}\right)_{ab}V_{bj},
\label{app:approx1}
\end{eqnarray}
where, in the last step, we have made the ``low energy approximation''
and obtained what is a conventional potential (independent of
$z$). Assuming that the matrix elements $V_{ia}$ have magnitudes
corresponding to the low energy scale, we see that the 
$\left(\frac{1}{H_0}\right)_{ab}$ term supresses the low energy matrix
elements of $\Vind$ by $|z|/M_a$ as expected. 

This expression is particularly useful in two cases.
\begin{enumerate}
\item If $H_0{\cal P}_{\rm low}$ is diagonal, then all interactions
  arise only from $\Vind$ and this analysis is essential to obtain any
  scattering in the low energy sector.
\item If the low energy sector has an approximate symmetry that is
  violated only by its interactions with the high mass sector, or
  even just by interactions solely within the high mass sector, we can
  use $\Vind$ to study these symmetry violations. 
\end{enumerate}

In the analysis up to now, we have retained just the leading correction
in powers of $M_a$. It is straightforward to retain the ${\cal
O}(1/M_a^2)$ term in the expansion of the Green's function $(H_{\rm
eff}-z)^{-1}$. The energy eigenvalues $E$ in the low energy theory are
given by those values of $z$ where the corresponding Green's function
develops a singularity,{\it i.e.}, where $$H_{\rm eff} |\psi\rangle =
E|\psi\rangle.$$ Assuming, for simplicity, that 
$H_0$ is diagonal
in the high mass sector and retaining terms to ${\cal O}(1/M_a^2)$, 
we  find that these eigenvalues are given by
the {\it generalized} matrix eigenvalue equation,
\begin{equation}
\sum_j[(H_0)_{ij} - \sum_a V_{ia}\frac{1}{M_a}V_{aj}]c_j
=E\sum_j[\delta_{ij}+\sum_aV_{ia}\frac{1}{M_a^2}V_{aj}]c_j. 
\end{equation}
The positive operator on the right hand side serves as a metric in the
low energy subspace. 

This formalism can be directly used to obtain the eigenvalues
(\ref{eq:highstates}) of the neutralino mas matrix in the large $|\mu|$
limit. In this case, the low energy sector comprises of the neutral wino
and the bino, and the generalized eigenvalue equation takes the form,
\begin{equation}
\left(\begin{array}{cc} M_2+a & - a\tan\theta_W \\ -a\tan\theta_W &
  M_1+a\tan^2\theta_W \end{array}\right)\left(\begin{array}{c} c_1 \\
  c_2\end{array}\right) = 
E\left(\begin{array}{cc} 1+\frac{M_W^2}{\mu^2} & - \frac{M_W^2}{\mu^2}\tan\theta_W \\ -\frac{M_W^2}{\mu^2}\tan\theta_W &
  1+\frac{M_W^2}{\mu^2}\tan^2\theta_W \end{array}\right)\left(\begin{array}{c} c_1 \\
  c_2\end{array}\right),
\end{equation}
with $a\equiv \frac{M_W^2}{\mu}\sin2\beta$ much smaller than $M_1$ and
$M_2$ in the large $|\mu|$ limit. 
	
%

\end{document}